\documentclass[aps,prd,nofootinbib,twocolumn,superscriptaddress,preprintnumbers,balancelastpage,longbibliography]{revtex4-1}

\usepackage{appendix}
\usepackage[utf8]{inputenc}
\usepackage{amsmath,amssymb,mathtools,bm}
\usepackage{graphicx, color, hepunits}
\usepackage[dvipsnames]{xcolor}
\usepackage{float}
\usepackage{multirow}
\usepackage{tikz-feynman} 
 \usepackage{hyperref}
\hypersetup{
    colorlinks=true,       
    linkcolor=blue,        
    citecolor=blue,        
    filecolor=magenta,     
    urlcolor=blue          
}
\usepackage[utf8]{inputenc}
\usepackage[english]{babel}

\usepackage{tensor}

\newcommand{\es}[2] {\begin{equation} \label{#1} \begin{split} #2 \end{split} \end{equation}}

\begin{document}

\title{A Search for Dark Matter Lines at the Galactic Center with 14 Years of Fermi Data}

\author{Joshua W. Foster}
\affiliation{Center for Theoretical Physics, Massachusetts Institute of Technology, Cambridge, Massachusetts 02139, U.S.A}

\author{Yujin Park}
\affiliation{Berkeley Center for Theoretical Physics, University of California, Berkeley, CA 94720, U.S.A.}
\affiliation{Theoretical Physics Group, Lawrence Berkeley National Laboratory, Berkeley, CA 94720, U.S.A.}

\author{Benjamin R. Safdi}
\affiliation{Berkeley Center for Theoretical Physics, University of California, Berkeley, CA 94720, U.S.A.}
\affiliation{Theoretical Physics Group, Lawrence Berkeley National Laboratory, Berkeley, CA 94720, U.S.A.}

\author{Yotam Soreq}
\affiliation{Physics Department, Technion—Israel Institute of Technology, Haifa 3200003, Israel}

\author{Weishuang Linda Xu}
\affiliation{Berkeley Center for Theoretical Physics, University of California, Berkeley, CA 94720, U.S.A.}
\affiliation{Theoretical Physics Group, Lawrence Berkeley National Laboratory, Berkeley, CA 94720, U.S.A.}

\date{\today}
\preprint{MIT-CTP/5505}

\begin{abstract}
Dark matter~(DM) in the Milky Way halo may annihilate or decay to photons, producing monochromatic gamma rays. We search for DM-induced spectral lines using 14 years of data from the Large Area Telescope onboard the Fermi Gamma-ray Space Telescope~(\textit{Fermi}-LAT) between $10\,\mathrm{GeV}$ and $2\,\mathrm{TeV}$ in the inner Milky Way leveraging both the spatial and spectral morphology of an expected signal. 
We present new constraints as strong as $\langle \sigma v \rangle \lesssim 6\times 10^{-30}\, \mathrm{cm}^3/\mathrm{s}$ for the two-to-two annihilations and $\tau \gtrsim 10^{30}\,\mathrm{s}$ for one-to-two decays, representing leading sensitivity between $10\,\mathrm{GeV}$ and $\sim$$500\,\mathrm{GeV}$. 
We consider the implications of our line-constraints on the Galactic Center Excess~(GCE), which is a previously-observed excess of continuum $\sim$GeV gamma-rays that may be explained by DM annihilation. 
The Higgs portal and neutralino-like DM scenarios, which have been extensively discussed as possible origins of the GCE, are constrained by our work because of the lack of observed one-loop decays to two photons.
More generally, we interpret our null results in a variety of annihilating and decaying DM models, such as neutralinos, gravitinos, and glueballs, showing that in many cases the line search is more powerful than the continuum, despite the continuum annihilation being at tree level.
\end{abstract}
\maketitle

\section{Introduction}

Monochromatic photons are a smoking-gun signature of dark matter~(DM) annihilation and decay, due to their relative lack of confounding astrophysical backgrounds compared to continuum photon signatures.  
Generically, DM models that produce continuum photons will also produce monochromatic photon lines at energy $E_\gamma = m_\chi$\,($E_\gamma = m_\chi/2$) for DM annihilation\,(decay), with $E_\gamma$ the photon energy and $m_\chi$ the DM mass.  
However, given that DM is known to be electrically neutral (or, at best, millicharged), the photon-line signatures are often loop-suppressed relative to continuum photon contributions.  
Nonetheless, the sensitivity to photon-line signatures is enhanced relative to the sensitivity to continuum signatures because the line-like signal is concentrated in a narrow energy range, set typically by the energy resolution of the telescope.  

The gamma-ray band is an especially promising energy range to look for line-like signatures of DM due to the weakly interacting massive particle~(WIMP) DM paradigm -- electroweak scale DM with weak-scale interactions can explain the observed DM abundance through thermal freeze-out in the early Universe~\cite{Bertone:2004pz}.  
Specific ultraviolet~(UV) WIMP constructions include neutralino DM in supersymmetric models~\cite{Jungman:1995df}, with effective constructions including {\it e.g.} Higgs portal DM models~\cite{Arcadi:2019lka}.  
Additionally, motivated decaying DM models predict observable signals in the gamma-ray band, including gravitino DM with $R$-parity violation~\cite{Takayama:2000uz} and glueball DM~\cite{Faraggi:2000pv,Boddy:2014yra,Soni:2016gzf}. 
In this work we show, using data from the Large Area Telescope~(LAT) onboard the {\it Fermi} Gamma-ray Space Telescope, that line-searches provide the leading sensitivity over continuum searches for many models.

Continuum gamma-rays from DM annihilation and decay in the mass range we consider arise dominantly from quark, lepton, and heavy gauge boson final states, which may produce gamma-rays of varying energy through the subsequent decay chains of the unstable particles in addition to secondary emission from {\it e.g.} inverse-Compton~(IC) scattering of electrons and positrons off of the interstellar radiation field~\cite{Cirelli:2010xx,Bauer:2020jay}.  
For cuspy galactic DM profiles, such as the  Navarro-Frenk-White~(NFW) profile~\cite{Navarro:1995iw,Navarro:1996gj}, the Galactic Center~(GC) region of the Milky Way produces the most gamma-ray flux as seen on Earth from both DM annihilation and decay (see, {\it e.g.},~\cite{Cohen:2016uyg,Chang:2018bpt}), relative to, for example, other nearby galaxies and galaxy clusters.  
However, searches for DM annihilation in the center of the Galaxy are plagued by high and uncertain backgrounds (see~\cite{Leane:2022bfm} for a recent discussion).   
On the other hand, Milky Way dwarf spheroidal galaxies are less bright in terms of DM-induced gamma-ray flux but significantly lower in background flux, given that the dwarf  galaxies are heavily DM dominated~\cite{Geringer-Sameth:2014yza}.  
Strong constraints on the DM annihilation cross-section have been set by the {\it Fermi}-LAT through searches for excess continuum gamma-ray emission from Milky Way dwarf galaxies~\cite{Fermi-LAT:2010cni,Fermi-LAT:2015att,Calore:2018sdx}.  
These searches exclude DM with velocity-averaged $s$-wave annihilation cross-sections at or above that expected for a generic WIMP -- $\langle \sigma v \rangle \approx 2 \times 10^{-26}$ cm$^3/$s -- for DM masses below roughly 50 to 100\,GeV, depending on the annihilation channel and astrophysical uncertainties related the dwarf DM profiles~\cite{Calore:2018sdx}.  
Competitive upper limits on the annihilation cross-section also arise from {\it Fermi} searches for continuum emission in nearby galaxies, such as M31~\cite{Fermi-LAT:2017ztt,Karwin:2020tjw}, galaxy clusters~\cite{Huang:2011xr,Lisanti:2017qlb,Thorpe-Morgan:2020czg}, and also cosmic ray searches with {\it e.g.} AMS-02~\cite{AMS:2016oqu,Lopez:2015uma,Cuoco:2017rxb,Heisig:2020nse}.

The gamma-ray line search towards the GC presented in this work has the advantage of being able to probe the brightest region of the sky from DM annihilation (or decay) without significant concern for the background mismodeling issues that plague continuum searches in this region, since there are few confounding line-like feature in the tens to hundreds of GeV energy range.  
This is especially important in light of the {\it Fermi} GC Excess (GCE)~\cite{Hooper:2010mq,Fermi-LAT:2015sau,Fermi-LAT:2017opo}, which is an excess of $\sim$GeV continuum gamma-rays observed near the GC that could arise from DM annihilation~\cite{Hooper:2010mq,Goodenough:2009gk,Daylan:2014rsa,Fermi-LAT:2017opo,DiMauro:2021qcf}, though alternate explanations exist in terms of {\it e.g.} pulsar emission or simply Galactic diffuse gamma-ray mismodeling~\cite{Abazajian:2010zy,Abazajian:2014fta,Lee:2015fea,Calore:2014xka,Macias:2016nev,Pohl:2022nnd}.  
The origin of the {\it Fermi} GCE has been heavily debated for over a decade~\cite{Leane:2022bfm}.  

One DM framework that has received significant attention for being able to explain the GCE while being consistent with other constraints on DM, such as direct detection constraints, is that of the Higgs portal~\cite{Goodenough:2009gk,Fraser:2020dpy,Carena:2019pwq}, where a Majorana DM particle with mass $m_\chi \sim 40$\,GeV may annihilate through a Yukawa-type interaction in the s-channel to an off-shell Higgs, which decays predominantly to $b$-quark pairs.  
In this model the annihilation branching ratio to gamma-ray pairs may be simply estimated as the branching ratio of an $\sim\!80\,$GeV Higgs boson to decay to photon pairs, which is $\sim\!10^{-3}$~\cite{LHCHiggsCrossSectionWorkingGroup:2016ypw}.
Given that the {\it Fermi} energy resolution is $\sim$5\,\%, and that the continuum signals are spread over more than an order of magnitude in energy, we naively expect that for background-dominated searches -- where we may approximate the detection significance by $S/\sqrt{B}$, with $S\,(B)$ the number of signal\,(background) counts -- the continuum search to be more sensitive by roughly a factor ${\mathcal O}(10)$ relative to the line search in terms of total annihilation cross-section reach. (See App.~\ref{app:0} for a more careful estimate of the relative sensitivity between the continuum and line-like searches.) 

Given that the continuum signal from the GCE is detected at high significance, while we find no evidence for line-like emission, we show in this work that the Higgs portal explanation of the GCE is constrained by the lack of gamma-ray line emission. 
We also consider neutralino like explanations of the GCE~\cite{Agrawal:2014oha,Achterberg:2015srl}, which we show are disfavored by the lack of line-like counterparts to the GCE.
More generally, depending on the DM model in question the constraints on gamma-ray lines from this work may be the strongest to-date on the theory, as we discuss in the context of annihilating and decaying example UV complete models. 
 
Our work directly builds off of~\cite{Fermi-LAT:2015kyq}, and the older~\cite{Abdo:2010nc,Fermi-LAT:2012ugx,Fermi-LAT:2013thd}, which performed a search for gamma-ray lines from DM annihilation in 5.8 years of {\it Fermi} \texttt{Pass 8} event-level data.  
Ref.~\cite{Fermi-LAT:2015kyq} searched for annihilating DM over the mass range 200\,MeV to 500\,GeV, with the analysis being systematics (statistics) limited below (above) $\sim$6\,GeV.  
Our analysis more than doubles the size of the data set with $\sim$14.0 years of \texttt{Pass 8} data.
We search for annihilating DM in the Galactic halo in the vicinity of the GC over the mass range 10\,GeV to 2\,TeV and for decaying DM from 20\,GeV to 4\,TeV. We chose to start our analysis at 10 GeV to avoid possible confounding systematic uncertainties at lower energies.   
In combination with a more sensitive analysis strategy that targets the inner 30$^\circ$ of the Galaxy, we improve upon the annihilation limits in~\cite{Fermi-LAT:2015kyq} by factors of a few over most of the mass range. 
Our annihilation results are the strongest to-date up to annihilating DM masses $\sim$500\,GeV, where our limits are surpassed by those from the ground-based H.E.S.S. gamma-ray telescope~\cite{HESS:2018cbt}.  
Our DM decay results are the most sensitive to-date over the entire mass range probed.  

The remainder of this article is organized as follows. 
We explain our data selection, signal and background modeling, and analysis procedures in Sec.~\ref{sec:fermi_line_analysis}.
Our results are presented in Sec.~\ref{sec:results}.
In Sec.~\ref{sec:TheoryInterp} we discuss the implications of our findings for annihilating DM models in the context of the GCE, while in Sec.~\ref{sec:other} we interpret our results more broadly for a sample of UV-complete annihilating and decaying DM models.
We conclude in Sec.~\ref{sec:conclusions}.  
Additional results and systematic tests are provided in the Appendices.

\section{{\it Fermi} line analysis}
\label{sec:fermi_line_analysis}

In this section we describe our data selection (Sec.~\ref{sec:data}), signal modeling (Sec.~\ref{sec:signal}), and analysis methods (Sec.~\ref{sec:analysis}).

\subsection{Data Selection}
\label{sec:data}

\begin{figure*}[!htb]
	\begin{center}
		\includegraphics[width=0.975\textwidth]{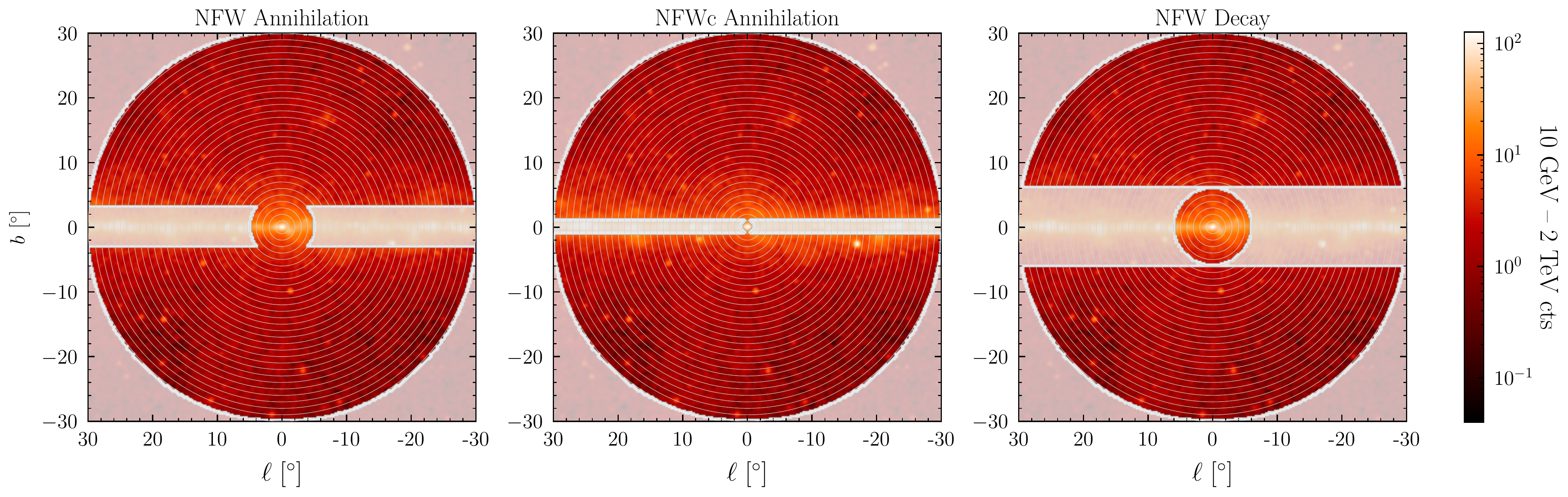}
	\end{center}
	\caption{The stacked photon counts map summed over 10\,GeV - 2\,TeV and the top three EDISP quartiles and smoothed with a Gaussian kernel with standard deviation of $0.5^\circ$ for clarity. Our analysis procedure makes use of the data within the $r < 30^\circ$ with an additional masking of the galactic plane that is independently optimized for each of the spatial morphologies considered in this work. Lightly shaded regions indicate those which are masked in the corresponding analysis. Grey lines indicate the boundaries of the annuli, which are treated independently and then joined in our analysis.}
	\label{fig:Stacked_Image}
\end{figure*}

\begin{figure}[!htb]
	\begin{center}
		\includegraphics[width=0.46\textwidth]{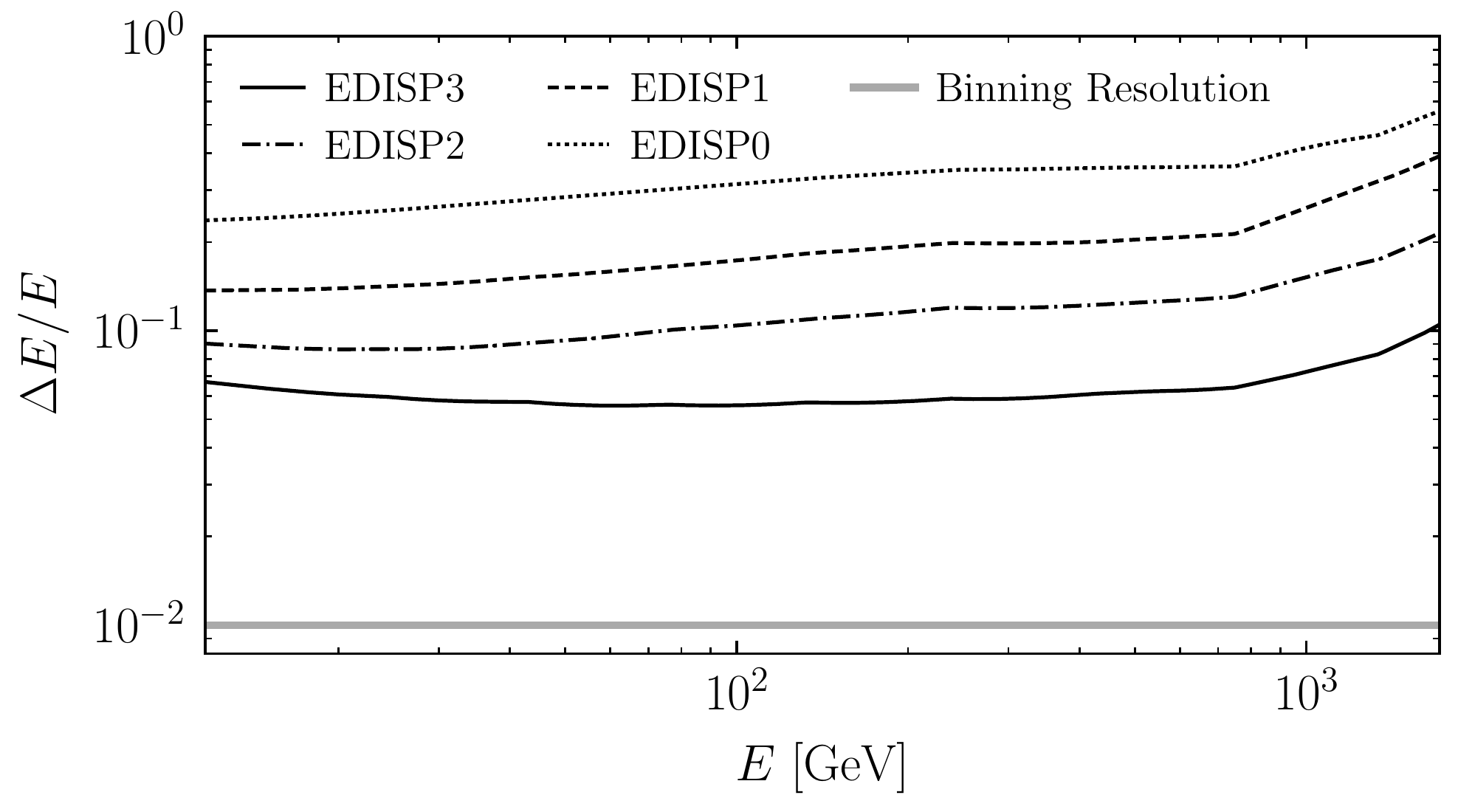}
	\end{center}
	\caption{An illustration of the energy-dependent resolution of \textit{Fermi} as measured by the full width half maximum (FWHM) of a line broadened by the instrumental response relative to its rest energy for each of the four EDISP quartiles. In grey, we compare the resolution at which we bin the data. Our chosen bin widths over-resolve the FWHM of a line by at least a factor of 5 over all energies and all EDISP quartiles. Note that we do not use EDISP quartile 0 in our analysis given its poor energy resolution.}
	\label{fig:Energy_Resolution}
\end{figure}

We make use of \textit{FermiTools}\footnote{\url{https://fermi.gsfc.nasa.gov/ssc/data/analysis/documentation/}} to reduce 729 weeks of Pass 8 \textit{Fermi} gamma-ray data collected between August 4, 2008 and July 25, 2022 restricted to the SOURCE event class photon classification.  
The data is illustrated in Fig.~\ref{fig:Stacked_Image}. 
We apply the recommended quality cuts \texttt{DATAQUAL>0}, \texttt{LATCONFIG==1} and \texttt{zenithangle<90}. 
For each of the four quartiles of data partitioned by the quality of the energy reconstruction, we produce counts and exposure maps binned into \texttt{HEALPIX}~\cite{Gorski:2004by,Zonca2019} maps with $\texttt{nside=256}$. 
The data are binned in energy between $10\,$GeV and $2.0\,$TeV in 531 logarithmically-spaced energy bin, which considerably overresolves the \textit{Fermi} line response at all energies in all energy dispersion (EDISP) quartiles (see Fig.~\ref{fig:Energy_Resolution}).

We generate exposure maps and detector response matrices with \texttt{edisp\_bins=-47}. 
In total, we produce three data sets in our data reduction procedure, which we refer to as $\mathbf{D}_1$, $\mathbf{D}_2$, and $\mathbf{D}_3$ for EDISP quartiles 1 through 3. 
In principle, a fourth data set, $\mathbf{D}_0$ for EDISP quartile 0, could be produced, but its poor energy resolution renders it unsuitable for this work. 
Note that of the data sets considered here, $\mathbf{D}_3$ has the lowest energy dispersion, while the photons in $\mathbf{D}_1$ are the most dispersed. 
Up to analysis choices regarding binning and quality cuts, we directly follow the data reduction procedure described in the \textit{Fermi} analysis guidelines.\footnote{\url{https://fermi.gsfc.nasa.gov/ssc/data/analysis/scitools/LAT_weekly_allsky.html}}

\begin{figure*}[!htb]
	\begin{center}
		\includegraphics[width=0.95\textwidth]{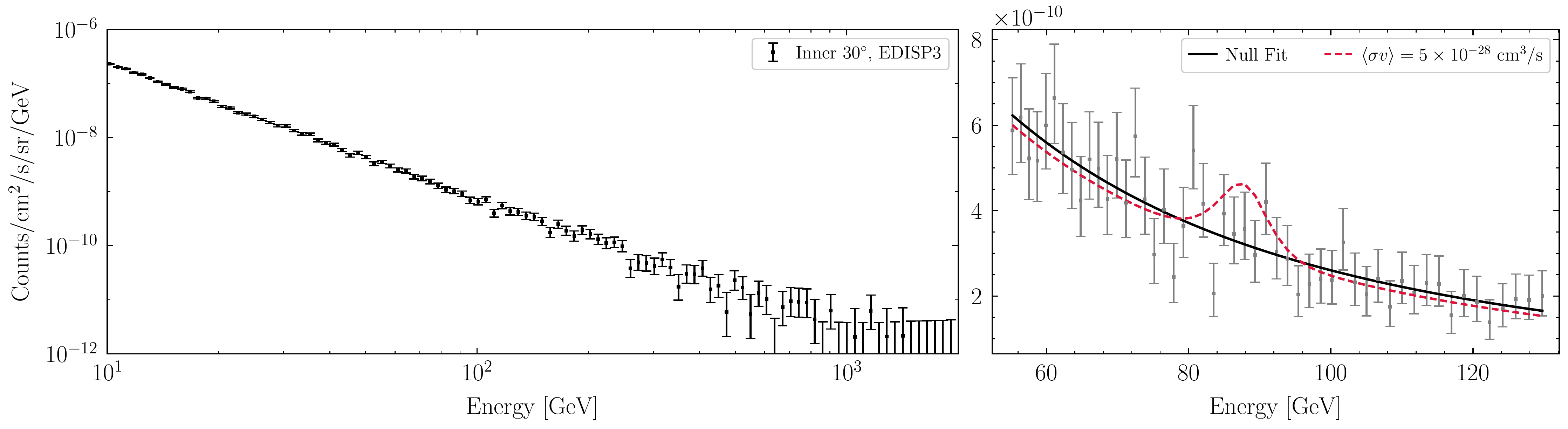}
	\end{center}
	\caption{(\textit{Left}) The total spectrum in the inner 30$^\circ$ of the GC in the EDISP3 quartile.  The data have been down-binned by a factor of five for illustrative purposes only. (\textit{Right}) An example of the analysis for a line search for a $m_\chi = 74$ GeV annihilating DM candidate using data collected within the inner 30$^\circ$ of the GC in the EDISP3 quartile. We show the best-fit null power-law spectrum along with the best-fit continuum spectrum where the annihilation cross-section for $\chi \chi \to \gamma \gamma$ is fixed to $\langle \sigma v \rangle = 5 \times 10^{-28}$ cm$^3$/s. Note that we have stacked the data over all annuli and performed the analysis on the stacked data set for illustrative purposes, whereas our fiducial analysis joins the likelihoods evaluated over the inner 30 annuli and achieves considerably greater constraining power.}
	\label{fig:Spectrum_and_Fit_Example}
\end{figure*}

For our analysis we select pixels within 30$^\circ$ of the GC. 
We then bin the data into 30 concentric annuli of radius $1^\circ$, with the first annulus going from $0^\circ$ to $1^\circ$ away from the GC. 
The annuli are illustrated in Fig.~\ref{fig:Stacked_Image}. 
We stack the data across each annulus. 
In Fig.~\ref{fig:Spectrum_and_Fit_Example} (left panel) we present the stacked data over all 30 annuli within the EDISP3 quartile for illustrative purposes. Note that we analyze the data independently in each quartile and annulus.

\subsection{Signal Modelling}
\label{sec:signal}

The expected number of photons with energy $E = m_\chi$ for DM $\chi$ with mass $m_\chi$ annihilating to two photons ($\chi \chi \to \gamma\gamma$) from a location $\Omega$ on the sky, with differential angular size $d\Omega$, incident upon the $\textit{Fermi}$-LAT
is given by (see, {\it e.g.},~\cite{Gaskins:2016cha} for a review)
\begin{equation}
   {d N_a(\Omega |  m_\chi, \langle \sigma v \rangle) \over d \Omega} 
=   \frac{\langle \sigma v \rangle}{4 \pi m_\chi^2} 
    \times \mathcal{E}(E) \times \mathcal{J}(\Omega) \,,
\end{equation}
where $\langle \sigma v \rangle$ is the velocity averaged annihilation cross-section, 
and $\mathcal{E}$ is the exposure ({\it i.e.}, effective area times exposure time) at $\Omega$ and the true energy $E$. The $\mathcal{J}$-factor is defined by
\begin{equation}
    \mathcal{J}(\Omega) = \int_0^\infty ds \rho_\chi(s, \Omega)^2 \,,
\end{equation}
where $\rho_\chi(s, \Omega)$ is the DM density along the line-of-sight parameterized by the distance from Earth $s$. 
Similarly, the expected number of incident photons with energy $E = m_\chi / 2$ from DM that decays to two photons ($\chi \to \gamma\gamma$) is given by 
\begin{equation}
  {d  N_d(\Omega | m_\chi, \tau) \over d \Omega} = \frac{1}{2 \pi m_\chi \tau} \times  \mathcal{E}(E) \times \mathcal{D}(\Omega) \
\end{equation} 
where $\tau$ is the DM lifetime, and the $\mathcal{D}$-factor is calculated through
\begin{equation}
    \mathcal{D}(\Omega) = \int_0^\infty ds \rho_\chi(s, \Omega).
\end{equation}

We consider two possible DM density profiles in this work. 
Our fiducial results are presented for the NFW DM profile with scale radius $r_s = 20\, \mathrm{kpc}$ normalized to a local DM density of $0.4\,\mathrm{GeV/cm}^3$ at Earth's distance from the GC of $8.5\,\mathrm{kpc}$.  
This profile matches that used in~\cite{Fermi-LAT:2015kyq}, which allows for a direct comparison to the previous {\it Fermi} constraints.
For interpretation of our line constraints within the context of the GCE, however, we additionally consider a contracted NFW profile with identical scale factor and normalization but with an index of $\gamma = 1.25$, as in {\it e.g.}~\cite{Daylan:2014rsa,Fermi-LAT:2017opo}, as the GCE has been found to favor such a contracted NFW profile. 
Note that within the inner regions of the Milky Way the energy density is dominated by baryons and not DM, and thus one expects a deviation from the pure NFW DM profile that is found in DM-only simulations. 
Modern $N$-body simulations find, for example, that Milky Way-like galaxies may have contracted DM density profiles in the inner few degrees (see, {\it e.g.}, the recent FIRE-2 simulations~\cite{Hopkins:2017ycn,2022MNRAS.513...55M}).

The gamma-ray signals are spread spatially and spectrally by the {\it Fermi}-LAT instrumental response.
We use the \texttt{gtsrcmaps} and \texttt{gtmodel} functionality of \textit{FermiTools} to account for these effects by convolving the incident photon counts map $N(\Omega)$ consisting of a monochromatic signal, in the energy bin with energy $E_j$, with the \textit{Fermi} point spread function~(PSF) and detector response matrix by  
\begin{equation}
\begin{split}
    {d N_i^\mathrm{sig}(\Omega | m_\chi, A) \over d \Omega} = \mathrm{DRM}_{ij} &\int d\Omega' {d N(\Omega', m_\chi, A) \over d \Omega} \\
    & \times \mathrm{PSF}_j(\Omega, \Omega') \,.
    \label{eq:sig_pred}
\end{split}
\end{equation}
Here $d N^\mathrm{sig}_i(\Omega) / d\Omega$ is the differential expected number of signal photons in energy bin $E_i$ reconstructed at on-sky location $\Omega$, $\mathrm{PSF}_j$ is the PSF at energy $E_j$, and $\mathrm{DRM}_{ij}$ is the detector response matrix which maps incident photon counts in energy bin $E_j$ into observed photon counts in output energy bin $E_i$. 
Note that $N^{\rm sig}$ is chosen to be either $N_a$ or $N_d$ depending on whether we search for annihilation or decay.  
Additionally, we denote the signal strength parameter by $A$, which is either $\langle \sigma v \rangle$ or $1/\tau$ for annihilation or decay, respectively.  
The PSF is normalized such that $\int d \Omega' \, {\rm PSF}_j(\Omega, \Omega') = 1$. 

\subsection{Analysis Methodology }
\label{sec:analysis}

In this section, we describe the likelihood analyses used to search for evidence of a narrow DM signal on top of continuum background contributions.
Recall that we divide the inner 30$^\circ$ of the Galaxy into concentric annuli centered at the GC with a width of $1^\circ$ in angular radius from the GC. 
We label these such that Annulus 1 spans between $0^\circ$ and $1^\circ$ from the GC, Annulus 2 spans between $1^\circ$ and $2^\circ$ from the GC, and so forth.

We apply an additional plane mask for each of the three spatial morphologies considered in this work: annihilation following an NFW profile, annihilation following a contracted NFW profile (cNFW), and decay following an NFW profile. 
For each spatial morphology, we determine the mask via a data-driven optimization procedure maximizing the expected signal-to-noise ratio.
The Galactic plane is a bright source of gamma-rays; when the plane is completely unmasked, it has the effect of adding background counts to our annuli and diluting the signal-to-noise ratio. 
On the other hand, if the plane mask is too large then we cut down on the signal contributed by the rising DM profile in the inner annuli. 
Accounting for these competing effects, we determine that the following masks provide near optimal sensitivity for our analysis framework. 
(Note that we do not consider more complicated masking structures, where, {\it e.g.}, the plane mask  is adjusted with longitude.)
For annihilation with an NFW profile, we mask regions with $|b| <  3^\circ$ when $r > 5^\circ$. 
For annihilation with a cNFW profile, we mask regions with $|b| > 1^\circ$ when $r > 0.6^\circ$.
For decay with an NFW profile, we mask regions with $|b| > 6^\circ$  when $r > 6^\circ$. Note that our optimization procedure floats the mask width and also a disk size around the GC where the plane mask does not apply, as shown in Fig.~\ref{fig:Stacked_Image}. 
Our masking is incorporated at the level of both the data reduction and the production of the expected signal map; \textit{i.e.}, we modify the integration domain in~\eqref{eq:sig_pred}. 
The subsequent results presented here are constructed with these masking choices.  
We present results without any plane masking in App.~\ref{app:Unmasked_Analysis}.

To search for a DM spectral line in the $j^\mathrm{th}$ annulus of data set $\mathbf{D}_i$, with $i$ referring to the EDISP quartile, that would appear at energy $E$ within energy bin $E_k$,
we first construct the counts data vector by summing over all the pixels within the annulus. 
We denote the annulus-summed counts in $\mathbf{D}_i$ within Annulus $j$ in energy bin $k$ by $\mathbf{d}_{ijk}$. 
Similarly, by averaging over the annulus, we construct the annulus-averaged exposure $\mathcal{E}_{ijk}$ (units of cm$^2$ s).
The binned signal model prediction in this data set, annulus, and energy bin is computed through~\eqref{eq:sig_pred} and denoted by $\mathbf{s}_{ijk}(A | m_\chi)$, with $A$ the signal strength parameter.
As in~\cite{Fermi-LAT:2015kyq}, we model the continuum background model using a
power-law such that the predicted background is given by
\begin{equation}
    \mathbf{B}_{ijk} = \mathbf{a}_{ij} \mathcal{E}_{ijk} E_k^{\mathbf{b}_{ij}} \,,
\end{equation}
with $\mathbf{a}_{ij}$ and $\mathbf{b}_{ij}$
nuisance parameter vectors that control the background amplitudes and spectral indices, respectively, in each data set and annulus independently.
In total, the predicted number of counts in each data set, annulus, and energy bin is given by 
\begin{equation}
    \bm{\mu}_{ijk} = \mathbf{s}_{ijk}(A | m_\chi) + \mathbf{B}_{ijk} \,.
\end{equation}

Our energy binning is chosen so that the 68\% containment interval for a line-like signal is over-resolved by at least a factor of five over the full energy range considered in this work, as illustrated for an example line signal in the right panel of Fig.~\ref{fig:Spectrum_and_Fit_Example}. 
At fixed $m_\chi$ we denote the energy bin that contains our line signal ({\it e.g.}, the energy bin that contains $m_\chi$ for an annihilation signal) by $k = 0$.  
We restrict our analysis to include $k_{\rm max}$ energy bins above and below our signal bin. 
In our fiducial analysis we use $k_{\rm max} = 25$ (see the right panel of Fig.~\ref{fig:Spectrum_and_Fit_Example} for an example), though alternate choices are discussed in App.~\ref{app:energy_range}. 
Note that if $k_{\rm max}$ is too small then our signal model becomes degenerate with our background model, which means that we lose constraining power to our putative signal. 
(Note that we allow the nuisance parameter ${\bm a_{ij}}$ to have either sign.)  
On the other hand, if $k_{\rm max}$ is too large then our analysis becomes more susceptible to mismodeling and systematic differences between the true background shape and our assumed power-law background model.  
Our choice of $k_{\rm max} = 25$ is similar to the energy range chosen in the {\it Fermi} line analysis~\cite{Fermi-LAT:2015kyq}.  
This corresponds to an energy range, roughly, $\Delta E / E \approx 0.64$, where $\Delta E$ is the energy range window and $E$ is the central energy.  

At fixed $m_\chi$ we construct a joint Poisson likelihood over all data sets and annuli by
\begin{equation}
    p({\mathbf{D}} | \bm{\theta}) = \prod_{i=0}^{3}\prod_{j=1}^{30}\prod_{k=-k_{\rm max}}^{k_{\rm max}} \frac{{\bm{\mu}}_{ijk}(\mathbf{\bm\theta})^{{\mathbf{d}}_{ijk}} e^{-{\bm{\mu}}_{ijk}(\mathbf{\bm\theta})}}{{\mathbf{d}}_{ijk}!} \,.
\end{equation}
with the product taken over the four data sets, the 30 annuli, and $2 k_{\rm max} + 1 = 51$ energy bins centered on the bin containing the line energy. 
The model parameter vector is defined by $\bm \theta = \{A, {\bm \theta}_{\rm nuis.}\}$, with ${\bm \theta}_{\rm nuis.} = \{ \bm{a}_{ij}, \bm{b}_{ij}\}$ being the vector of nuisance parameters.
We define the frequentist test statistic~(TS) for discovery by 
\begin{equation}\label{eq:discovery}
    t = 2 \log \bigg[\frac{\mathrm{max}_{\bm{\theta}} p(\mathbf{d} | \bm{\theta})}{\mathrm{max}_{\bm{\theta}_\mathrm{nuis.}}  p(\mathbf{d}|A = 0, \bm{\theta}_\mathrm{nuis.})} \bigg] \,,
\end{equation}
which is asymptotically $\chi^2$-distributed with one degree of freedom under the null hypothesis. Similarly, we construct the TS for upper limits on $A$ by
\begin{equation}\label{eq:limit}
    q(A) = 2 \log \bigg[\frac{\mathrm{max}_{\bm{\theta}} p(\mathbf{d} | \bm{\theta})}{\mathrm{max}_{\bm{\theta}_\mathrm{nuis.}}  p(\mathbf{d}|A, \bm{\theta}_\mathrm{nuis.})} \bigg] \,,
\end{equation}
from which we may compute the 
95\% 
one-sided upper limit $A^{95}$, assuming Wilks' theorem, by $q(A^{95}) \approx 2.71$. 
We also determine the expected 
95\% 
limit and one- and two-sigma containment intervals following~\cite{Cowan:2010js}. 
We power-constrain our limits at the 16$^\mathrm{th}$ percentile expected value~\cite{Cowan:2011an}. 
Tests of our analysis framework's robustness in detecting a signal on simulated data are presented in App.~\ref{app:injection_test}.  

\section{Annihilation and Decay Constraints}
\label{sec:results}

\begin{figure*}[!htb]
	\begin{center}
		\includegraphics[width=0.99\textwidth]{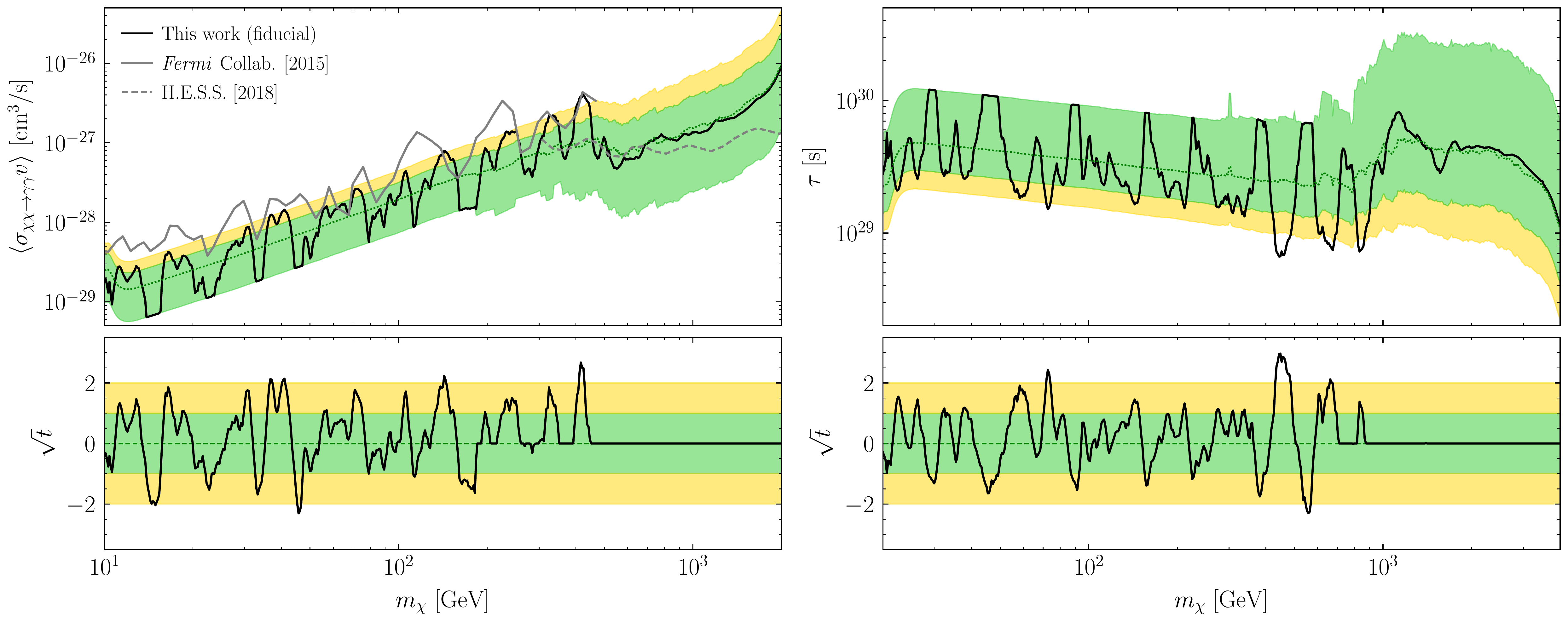}
	\end{center}
	\caption{(\textit{Top, left}) The $95^\mathrm{th}$ percentile power-constrained upper limits from this work on DM annihilation to $\gamma \gamma$ as a function of the DM mass. Expected one- and two-sigma containment intervals for the limit are depicted in green and gold, respectively, with the median expected $95^\mathrm{th}$ percentile limit in dotted green. This result improves upon the prior line constraints set by the \textit{Fermi} collaboration using 5.8 years of data (dotted grey) and is superseded by line constraints set with H.E.S.S. for $m_\chi \gtrsim 500\,\mathrm{GeV}$ \cite{Fermi-LAT:2015kyq, HESS:2018cbt}. (\textit{Bottom, left}). The associated sign-weighted significance for detection of an annihilation line. Green and gold indicate the expected one- and two-sigma containment intervals for the significance under the null hypothesis.  We find no evidence for DM annihilating to gamma-ray lines. We note that $\sqrt{t} \approx 0$ at large $m_\chi$ simply because in this regime there are typically few to no photons within the energy window and ROI. (\textit{Top, right}) As in the top left panel, but for DM that decays  to two photons with partial lifetime $\tau$. (\textit{Bottom, right}) As in the bottom left panel, but for the decay search.}
	\label{fig:Results_NFW_Limits}
\end{figure*}

We apply the analysis framework described in Sec.~\ref{sec:analysis} to the {\it Fermi} data and find the results presented in Fig.~\ref{fig:Results_NFW_Limits} for DM annihilation\,(left) and decay\,(right).  
Note that we over-resolve the energy bins by a factor of at least four in our line search to account for the possibility that a line could {\it e.g.} appear at the edge of an energy bin; that is, for the annihilation search we consider 531 logarithmically-spaced masses between 10\,GeV and 2\,TeV.  
The top panels in Fig.~\ref{fig:Results_NFW_Limits} show our power-constrained 95\% upper limits, with the green (gold) bands indicating the $1\,\sigma(2\,\sigma$) expected containment regions for the limits. 
The bottom panels show the square-root of the discovery TS ($\sqrt{t}$), multiplied by the sign of the best-fit signal parameter $A$.  
In the Wilks' limit, $\sqrt{t}$ may be interpreted as the discovery significance for the two-sided test relative the null hypothesis.
We allow $A$ to be both positive and negative, even though only positive $A$ are physical, to make sure that our upper limits are set with respect to the point of maximum likelihood, which is necessary for employing Wilks' theorem. 
No mass point surpasses our predetermined $5\,\sigma$ threshold for a discovery.   

In Fig.~\ref{fig:Results_NFW_Limits} we compare our upper limits for DM annihilation to those from the previous {\it Fermi} Collaboration analysis in~\cite{Fermi-LAT:2015kyq}, which was statistics limited in the mass range shown. 
Our upper limits improve upon the previous {\it Fermi} work by a factor of a few across most the mass range. 
Above $\sim$500\,GeV our upper limits are surpassed by those of the H.E.S.S. Collaboration~\cite{HESS:2018cbt} using their ground-based Cherenkov telescope, which are shown re-scaled to our fiducial NFW DM profile.  
Note, however, that the H.E.S.S. analysis concentrated on regions much closer to the GC than ours (between 0.3$^\circ$ and $0.9^\circ$ of the GC), where the DM density profile is more uncertain because of baryonic feedback.   

\begin{figure}[!htb]
	\begin{center}
		\includegraphics[width=0.49\textwidth]{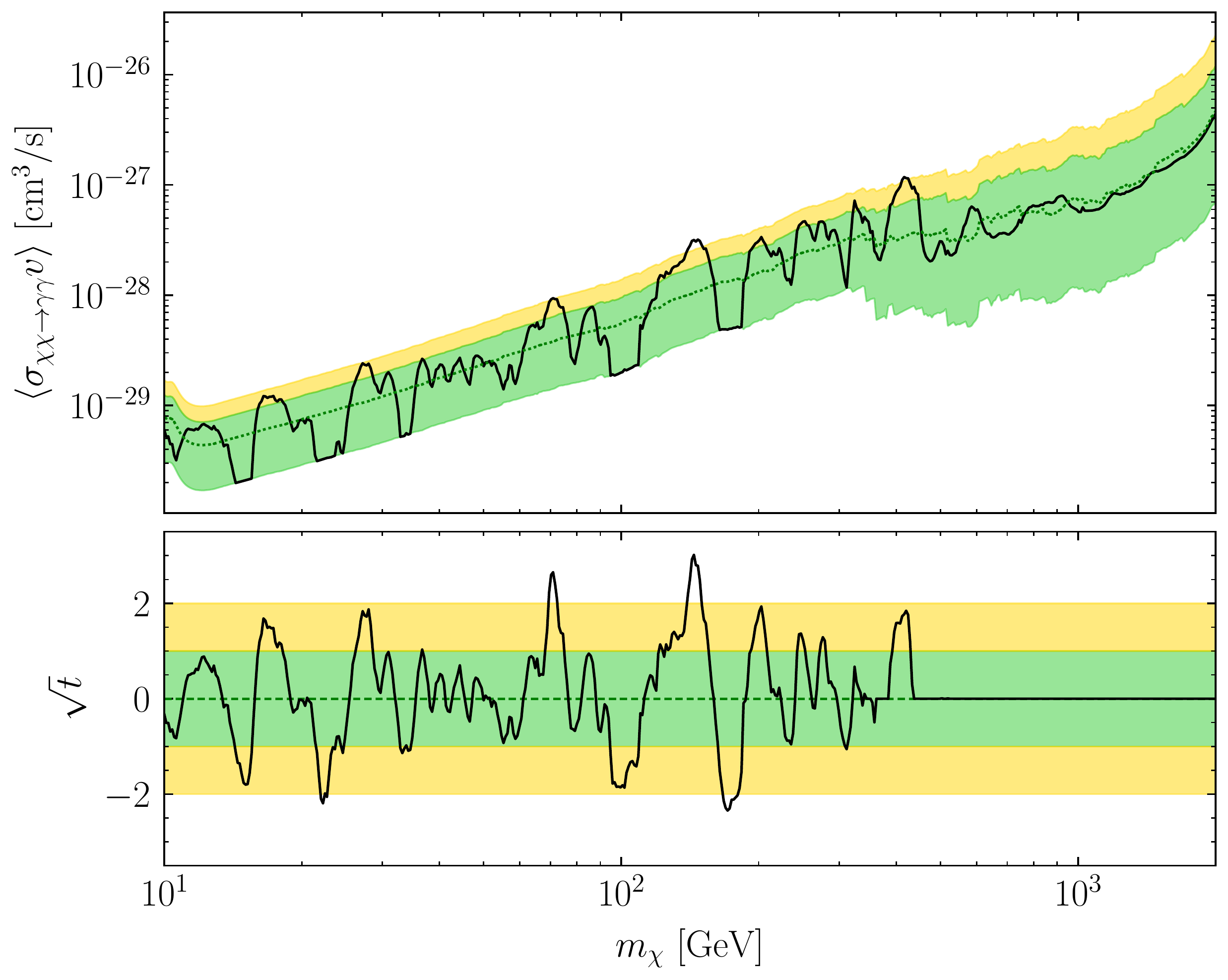}
	\end{center}
	\caption{As in the left panel of Fig.~\ref{fig:Results_NFW_Limits} for DM annihilation, but assuming a contracted NFW profile with index $\gamma = 1.25$.  This upper limit is used in our studies of the {\it Fermi} GCE.}
	\label{fig:NFWc_Annihilation}
\end{figure}

When discussing the GCE we make use of a contracted NFW DM profile with index $\gamma = 1.25$ (see, {\it e.g.},~\cite{Daylan:2014rsa,Fermi-LAT:2017opo}), since this is the profile favored by the GCE morphology under the DM interpretation. 
Using this DM profile leads to the annihilation limits illustrated in Fig.~\ref{fig:NFWc_Annihilation}.

\begin{figure}[!htb]
	\begin{center}
		\includegraphics[width=0.45\textwidth]{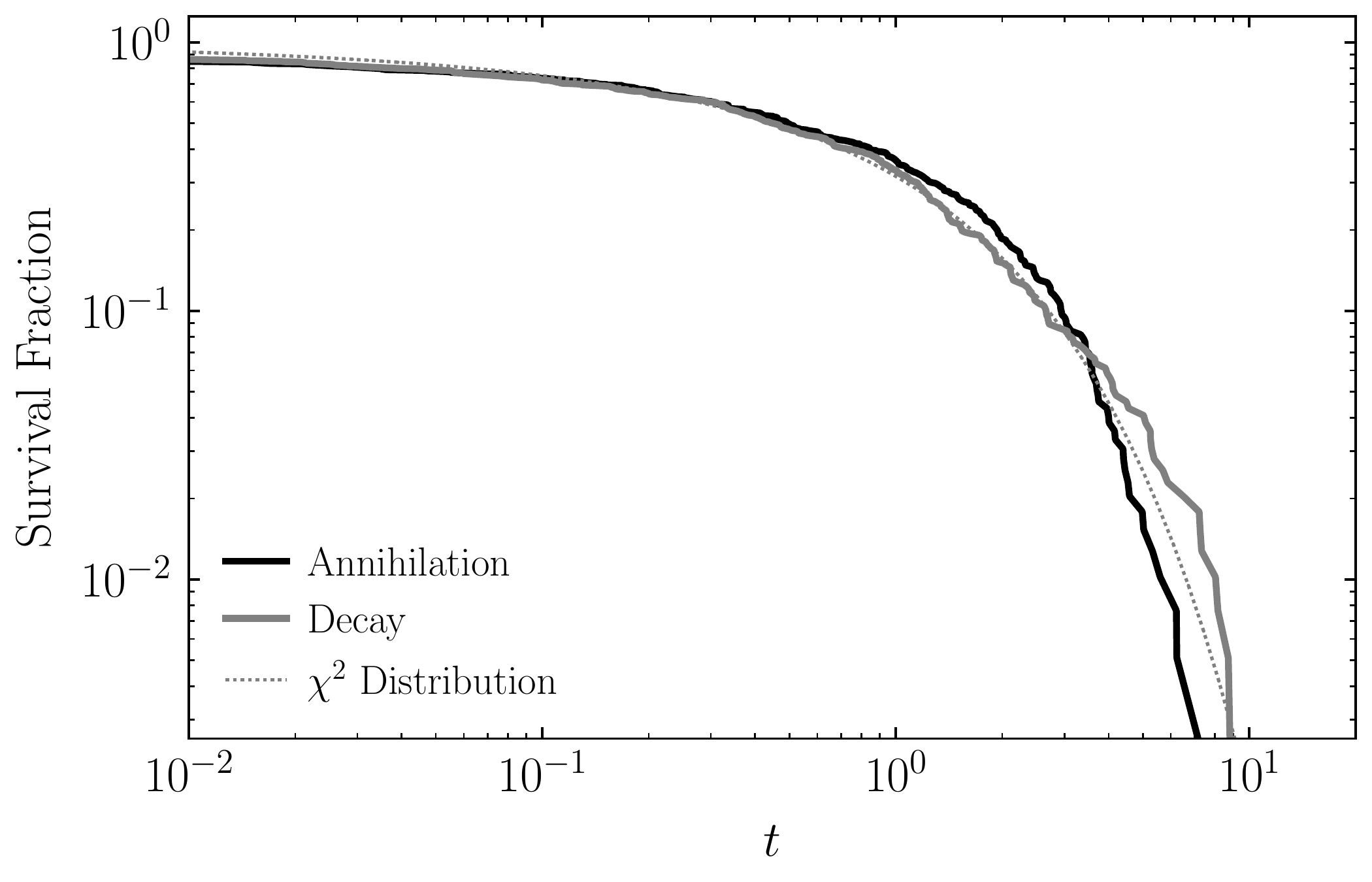}
	\end{center}
	\caption{The survival fraction of the discovery TS in the analysis assuming DM annihilation (decay) following an NFW spatial morphology in black (grey) corresponding to photon lines with a rest energy $E \leq 500$ GeV. Asymptotically, the TS $t$ is expected to be $\chi^2$ distributed with one degree of freedom, giving the result in dotted grey.
	Both the annihilation and decay survival fractions are consistent with the null hypothesis, and, in particular, no high significance excesses are observed.
	}
	\label{fig:Results_NFW_Survival}
\end{figure}

It is instructive to study the distribution of two-sided discovery TSs $t$ as a tool for investigating mismodeling and systematic effects.  
In Fig.~\ref{fig:Results_NFW_Survival} we illustrate the survival fraction for the distribution of $t$ values for the annihilation and decay searches.  
Note that the survival fraction shows the fraction of test points with a $t$ value at or above that indicated on the $x$-axis.  
Under the null hypothesis, and assuming we are in the asymptotic Wilks' limit, we expect the survival fractions to follow that of the $\chi^2$ distribution, which is also indicated in Fig.~\ref{fig:Results_NFW_Survival}.  
Indeed, in neither the annihilation nor the decay scenarios do we see significant departures from the $\chi^2$ distribution, which suggests that both we do not see evidence for DM signals and also that we are limited by statistical uncertainties and not systematic uncertainties.  
We restrict this figure to test points with mass less than $500$\,GeV ($1$\,TeV) for annihilation (decay), since the higher-mass points probe a low-photon-count regime that is likely outside of the asymptotic Wilks' limit.  
Moreover, as seen in {\it e.g.} Fig.~\ref{fig:Results_NFW_Limits}, there is clearly no evidence for line-like emission above 500\,GeV.  
Many annuli have zero or few counts above 500\,GeV, which leads to more test points with $t \approx 0$ than expected under the null hypothesis in the large photon count regime (though this is as expected in the few photon count regime). 

\section{Implications for the Galactic Center Excess}
\label{sec:TheoryInterp}

In this section we use our results, as illustrated in Fig.~\ref{fig:NFWc_Annihilation}, to constrain specific annihilating DM scenarios in the context of the {\it Fermi} GCE.  
In this next section we discuss the implications of our results more generally in the context of decaying and annihilating DM models.  

\subsection{Higgs Portal Dark Matter and the GCE}

A compelling explanation of the GCE arises from the scenario of Majorana DM $\chi$ coupling to the SM through a Higgs portal, with relevant Lagrangian terms
\begin{equation}
    \mathcal{L} \supset - \frac{m_\chi}{2} \bar \chi\chi + i \frac{y_{\chi h}}{\sqrt{2}} h \bar \chi \gamma_5 \chi \,,
\end{equation}
where we fix a pseudoscalar coupling between the DM and Higgs field $h$ with Yukawa coupling constant $y_{\chi h}$.
This results in vanishing  spin-independent DM-SM elastic scattering at tree level, while annihilation rates may be kept at the cross section needed to produce the correct relic abundance from thermal freeze-out by adjusting $y_{\chi h}$ at a given $m_\chi$.

This model has been considered extensively as a potential explanation for the GCE, such as in~\cite{Goodenough:2009gk,Fraser:2020dpy,Carena:2019pwq}, due to the simplicity of the effective theory, the fact that the shape of the GCE is well-fit by a $\bar b b$ spectrum, and the direct detection cross-sections are naturally suppressed below current constraints. 
A straightforward UV-completion may be found in singlet-doublet mixing scenarios~\cite{Fraser:2020dpy,Carena:2019pwq}. 

In this scenario, the bulk of the annihilation proceeds through tree-level $\chi\chi \to \bar f f$ diagrams, for SM fermions $f$, with a cross section given by
\begin{equation}
\langle \sigma v \rangle_{\rm ann} = \sum_f  \frac{N_{c, f} y_{\chi h}^2 y^2_{f h} (m_\chi^2 - m_f^2)^{3/2} }{8 \pi m_\chi (m_h^2 - 4m_\chi^2)^2 },   
\end{equation} where $y_{fh} = m_f/\sqrt{2}v_{EW}$ are the fermion Yukawas, with the SM Higgs Vacuum Expectation Value (VEV) $v_{EW} \approx 178$ GeV, and $N_{c,f} = 3 (1)$ for quarks (leptons). 
Keeping in mind that the GCE is best fit to DM masses $\lesssim 50$ GeV~\cite{Goodenough:2009gk}, on-shell annihilations to $WW$ and other heavy states are kinematically shut off.  
The resultant photons from final-state showers, hadronization, and decay then make up the continuum flux that by assumption constitutes the observed GCE. 

\begin{figure}
\centering
\begin{tikzpicture}
  \begin{feynman}
    \vertex (a){$\chi$};
    \vertex [right=1cm of a] (b);
    \vertex [below=1cm of b] (c);
    \vertex [below=2cm of a] (d){$\chi$};
    \vertex [right=1cm of c] (e);
    \vertex [right=3cm of a] (f){$\;f \; (b, \tau, c...)$};
    \vertex [right=3cm of d] (g){$f$};
    \diagram* {
      (a) -- [fermion] (c),
      (c) -- [fermion] (d),
      (c) -- [scalar, edge label =\(h\)] (e),      
      (e) -- [fermion] (f),
      (g) -- [fermion] (e),
    };
  \end{feynman}
\end{tikzpicture}
\begin{tikzpicture}
  \begin{feynman}
    \vertex (a){$\chi$};
    \vertex [right=1cm of a] (b);
    \vertex [below=1cm of b] (c);
    \vertex [below=2cm of a] (d){$\chi$};
    \vertex [right=1cm of c] (e);
    \vertex [right=1cm of e] (f);
    \vertex [right=4cm of a] (h){$\gamma$};
    \vertex [right=4cm of d] (i){$\gamma$};
    \diagram* {
      (a) -- [fermion] (c),
      (c) -- [fermion] (d),
      (c) -- [scalar, edge label =\(h\)] (e),      
      (e) -- [photon, half left, edge label =\(W\)] (f),
      (e) -- [photon, half right] (f),
      (f) -- [photon] (h),
      (f) -- [photon] (i)
    };
  \end{feynman}
\end{tikzpicture}

\begin{tikzpicture}
  \begin{feynman}
    \vertex (a){$\chi$};
    \vertex [right=1cm of a] (b);
    \vertex [below=1cm of b] (c);
    \vertex [below=2cm of a] (d){$\chi$};
    \vertex [right=0.8cm of c] (e);
    \vertex [right=2.655cm of a] (f);
    \vertex [right=2.655cm of d] (g);
    \vertex [right=1cm of f] (h){$\gamma$};
    \vertex [right=1cm of g] (i){$\gamma$};
    \diagram* {
      (a) -- [fermion] (c),
      (c) -- [fermion] (d),
      (c) -- [scalar, edge label =\(h\)] (e),      
      (e) -- [photon] (f),
      (g) -- [photon, edge label =\(W\)] (e),
      (f) -- [photon] (g),
      (f) -- [photon] (h),
      (g) -- [photon] (i)
    };
  \end{feynman}
\end{tikzpicture}
\begin{tikzpicture}
  \begin{feynman}
    \vertex (a){$\chi$};
    \vertex [right=1cm of a] (b);
    \vertex [below=1cm of b] (c);
    \vertex [below=2cm of a] (d){$\chi$};
    \vertex [right=0.8cm of c] (e);
    \vertex [right=2.655cm of a] (f);
    \vertex [right=2.655cm of d] (g);
    \vertex [right=1cm of f] (h){$\gamma$};
    \vertex [right=1cm of g] (i){$\gamma$};
    \diagram* {
      (a) -- [fermion] (c),
      (c) -- [fermion] (d),
      (c) -- [scalar, edge label =\(h\)] (e),      
      (e) -- [fermion] (f),
      (g) -- [fermion, edge label =\(t\)] (e),
      (f) -- [fermion] (g),
      (f) -- [photon] (h),
      (g) -- [photon] (i)
    };
  \end{feynman}
\end{tikzpicture}
    \caption{The leading contributions to annihilation of higgs portal DM in general (top left), and to the $\chi\chi \to \gamma\gamma$ channel in specific (remaining three). Two additional diagrams contributing to $\gamma\gamma$ have been omitted for brevity, from exchanging the final state photons in the lower diagrams.}
    \label{fig:feyn_higgs_portal}
\end{figure}
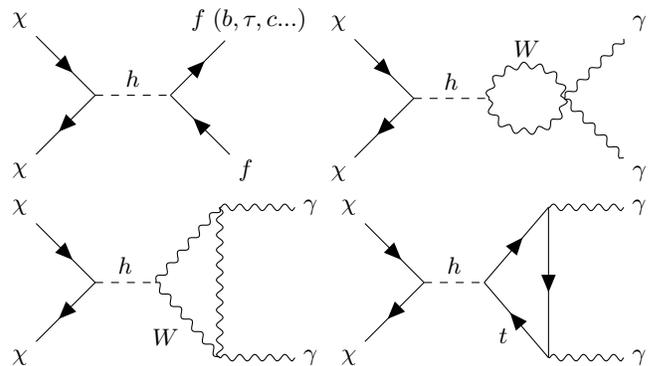

Associated to the continuum gamma-rays, the annihilation to $\gamma\gamma$ final states proceeds dominantly through diagrams with $W$ or top loops. 
These channels, kinematically forbidden to contribute towards bulk annihilation but dominating the monochromatic signal, allow a relatively enhanced production of photon lines for fixed total annihilation cross section. The diagrams associated with both the bulk (tree-level dominated) and $\gamma\gamma$ (1-loop leading) annihilations are shown in Fig.~\ref{fig:feyn_higgs_portal}, though note that two additional diagrams from exchanging external photon legs have been omitted. 

Furthermore, the ratio between these two rates, the bulk annihilation that produces continuum flux and the $\gamma\gamma$ channel that produces the photon line, is dependent only on known SM quantities and the DM mass:
\begin{equation}
    \left. \frac{\langle \sigma_{\rm \chi\chi \to \gamma\gamma} v \rangle}{\langle \sigma v \rangle_{\rm ann}} =  {\rm BR}_{h \to \gamma \gamma} \right|_{m_h = 2 m_\chi} \,,
    \label{eq:line_HP}
\end{equation}
which is $\sim 10^{-3}$ for $\sim 50$ GeV DM masses~\cite{LHCHiggsCrossSectionWorkingGroup:2016ypw}. 

By fitting the predicted continuum flux from DM annihilations to the GCE spectrum, we may obtain a benchmark mass and cross section for the Higgs-portal model. 
The results of this fit inform the mass and cross-sections of interest for the monochromatic photon signal.  
We take the spectral information of the GCE from~\cite{DiMauro:2021qcf}.
While the normalization of the excess spectrum changes significantly as data selection and astrophysical models are varied, we are only interested in obtaining a benchmark value from this portion of the analysis.  With that in mind, we use the fiducial result from~\cite{DiMauro:2021qcf} along with a rough accounting of systematic uncertainties, as described shortly. 
 
 \begin{figure}
    \centering
    \includegraphics[width = 0.9\linewidth]{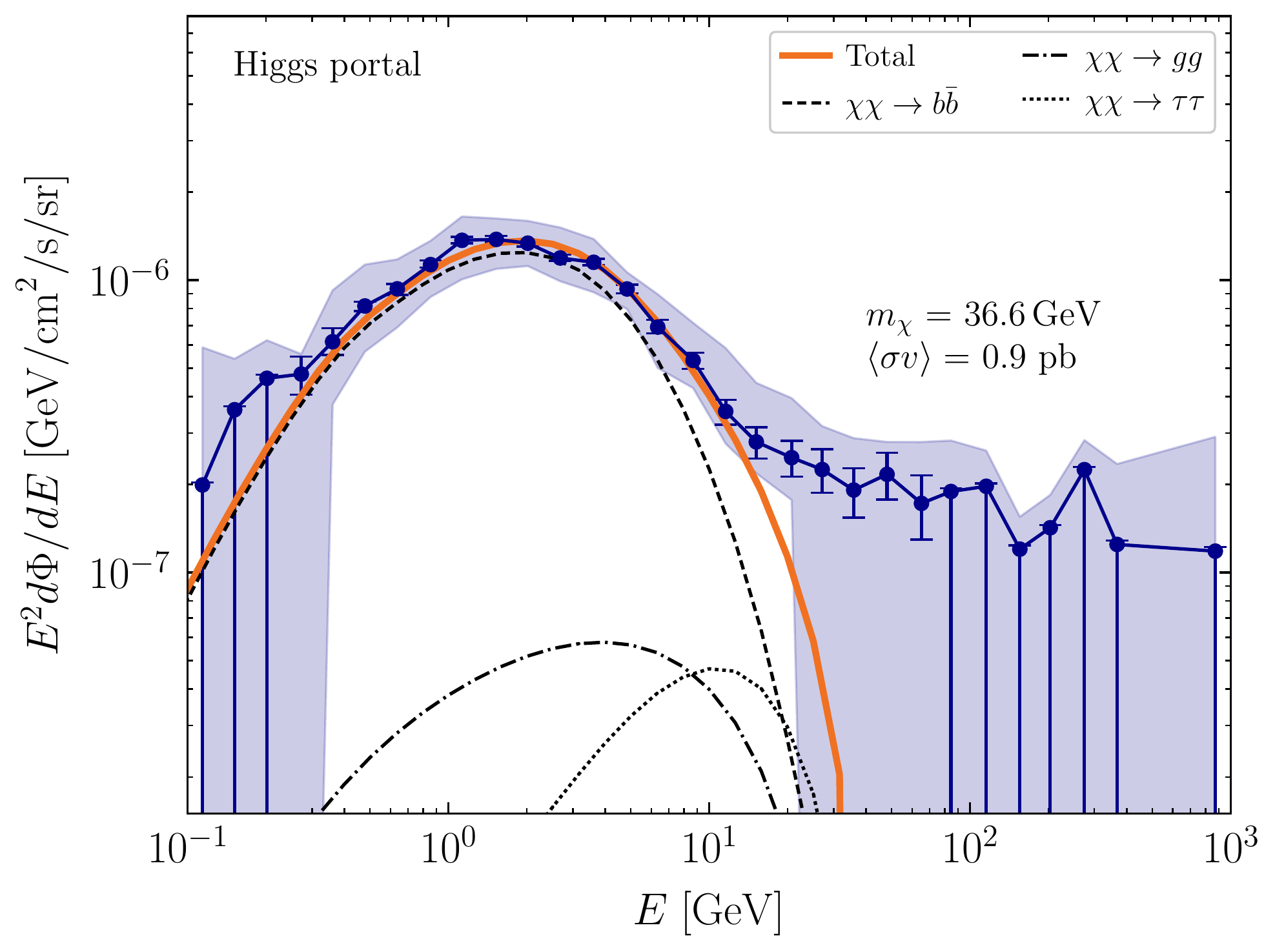}
    \caption{The best-fit model for annihilating Higgs portal DM to the GCE measured with 11 years of Fermi data, reproduced from Ref.~\cite{DiMauro:2021qcf}.  The ROI is a $40^\circ \times 40^\circ$ region centered at the GC.  The error bars are statistical for the fiducial analysis choice in~\cite{DiMauro:2021qcf}, with the blue band illustrating their estimate of systematic uncertainties. For the Higgs portal model with indicated mass and cross-section, assuming a contracted NFW DM profile with $\gamma = 1.25$, the overall continuum spectrum is illustrated in addition to its breakdown into its most dominant three channels. The GCE is dominantly measured at $\sim$GeV energies, which favores a $\sim$40 GeV Higgs portal model. The DM spectral shape is largely set by the $\bar b b$ flux, though the contribution from the much harder $\tau\tau$ spectrum is also important. }
    \label{fig:GCE_HP}
\end{figure}

In Fig.~\ref{fig:GCE_HP} we reproduce the GCE spectrum from~\cite{DiMauro:2021qcf}, which is normalized to a $40^\circ \times 40^\circ$ ROI around the GC.  The error bars are statistical for the fiducial analysis, with the blue bands expressing systematic uncertainties inferred in~\cite{DiMauro:2021qcf} from {\it e.g.} background mismodeling. 
 
The predicted continuum flux of annihilating DM is a sum of its {\it prompt} and {\it secondary} gamma rays. 
The prompt signal is generated directly as final-state annihilation products, while the secondary signal arises from stable non-photon final states that propagate through the Galactic medium and only later produce gamma-ray signals.  For the final states that we are interested in, secondary production is subdominant (see, {\it e.g.},~\cite{DiMauro:2021qcf}), and so we ignore it in this analysis.  
The prompt spectrum, in units of cts/cm$^2$/s/GeV/sr, is computed by
\es{eq:prompt_continuum}{
 \frac{{\rm d}\Phi}{{\rm d} E {\rm d}\Omega} = \frac{ \mathcal{J}}{4 \pi m_\chi^2} \sum_{X} \langle \sigma_{\chi\chi \to XX } v \rangle \frac{{\rm d} N_{X \to \gamma }}{{\rm d} E} \,,
}
with $X$ denoting the final states ({\it e.g}, $b$ quark pairs) and ${{\rm d} N_{X \to \gamma }} / {{\rm d} E}$ the decay spectrum of gamma-rays produced through $X$ decay.  Note that in this section we assume a contracted NFW profile, with $\gamma = 1.25$, when computing the ${\mathcal J}$ factor in order to match onto previous results for the GCE.
 
\begin{figure}
    \centering
    \includegraphics[width = 0.9\linewidth]{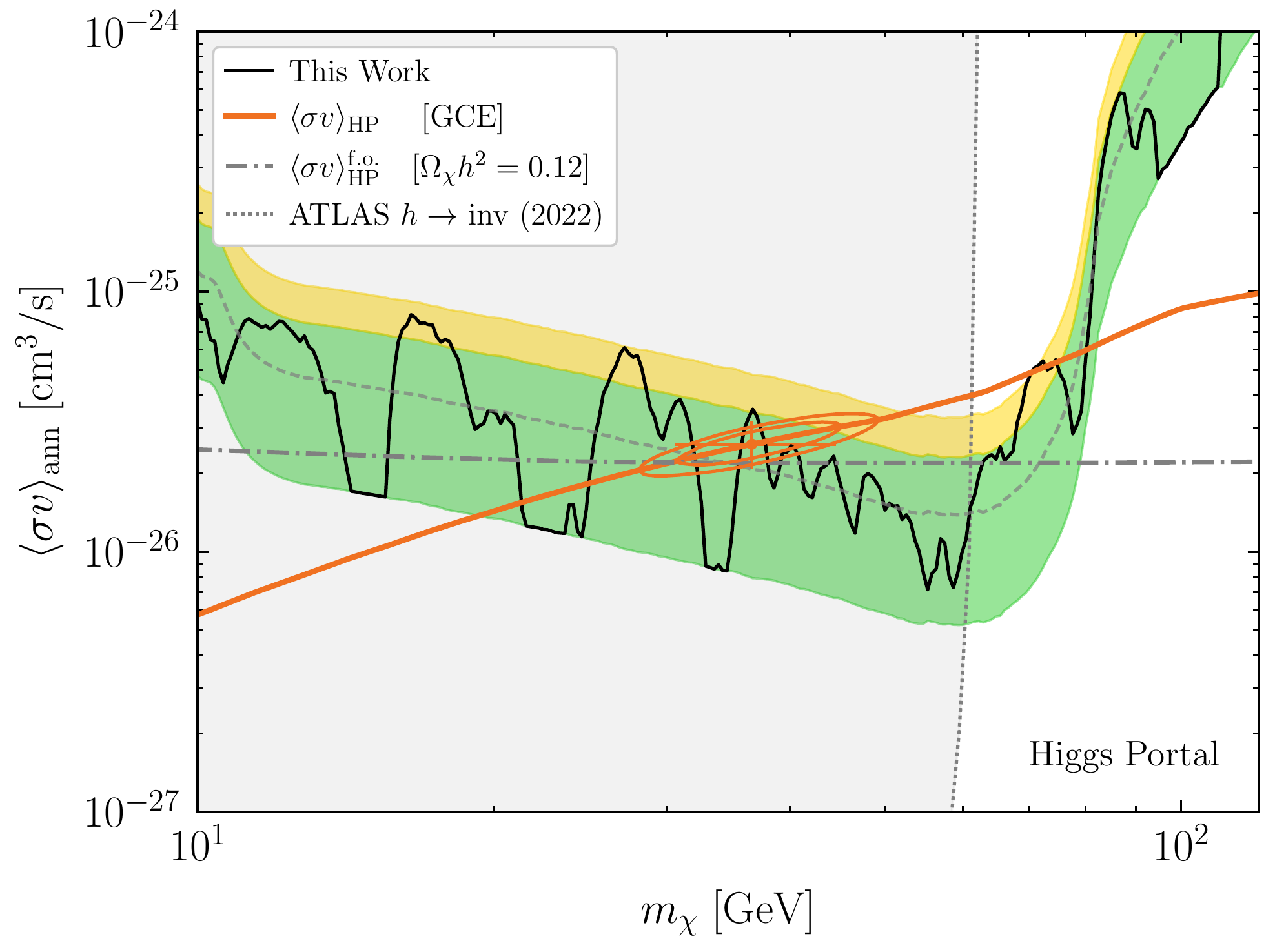}
    \caption{The $95^{\rm th}$ percentile limits of our line search, reinterpreted to constrain the total annihilation cross section of Higgs portal DM, as a function of DM mass (black). The best-fit cross section to the GCE spectrum~\cite{DiMauro:2021qcf} as a function of mass, as well as the globally favored 68\% and 95\% CL ellipse regions, are shown in orange. Also shown is the expected cross section from the thermal relic abundance (dashed grey), and the region of parameter space constrained at 95\% confidence by Higgs to invisible decays (shaded, dotted grey)~\cite{ATLAS:2022yvh,Wang:2841498}, though these constraints can be evaded by invoking more dynamical degrees of freedom.}
    \label{fig:HP}
\end{figure}

The continuum annihilation flux is shown in Fig.~\ref{fig:GCE_HP} for the Higgs portal scenario, as well as its breakdown into dominant channel contributions, for the best-fit mass and cross-section when the model is fit to the GCE. 
 We determine the best-fit model parameters for the Higgs portal model to be
\begin{align}
    m_{\chi}^{\rm HP} 
=&   36.7_{-6}^{+8} \,{\rm GeV} \, , \nonumber\\
    \langle \sigma v \rangle_{\rm ann}^{\rm HP} 
=&  2.7_{-0.5}^{+0.6} \times 10^{-26} \, {\rm cm^3/s} \,,
\end{align}
where in constructing the likelihood we enlarge the error bars on the GCE spectrum, illustrated in Fig.~\ref{fig:GCE_HP}, to be the quadratic sum of the statistical uncertainties and the systematic uncertainties from~\cite{DiMauro:2021qcf} (illustrated as the shaded region). 
Note that this is a rough accounting of systematics, which are correlated bin-to-bin, but the purpose of this analysis is to get a general sense for the parameter space needed to explain the GCE and not to rigorously fit the Higgs portal parameter space to the {\it Fermi} continuum data.  
The best-fit parameter space, at 1$\sigma$ (inner ring) and 2$\sigma$ (outer ring) significance, is illustrated accounting for the correlation between $m_\chi$ and $\langle \sigma v \rangle_{\rm ann}$ in Fig.~\ref{fig:HP}.  Intriguingly, the best-fit parameter space is consistent with the thermal annihilation cross-section, illustrated by the horizontal grey curve, needed to obtain the correct DM abundance.  The solid, orange curve in Fig.~\ref{fig:HP} shows the best-fit cross-section at each, fixed $m_\chi$.

Through~\eqref{eq:line_HP} we translate our 95\% constraints on the line-like annihilation signal, with cross-section  $\langle \sigma_{\chi \chi \to \gamma\gamma} v \rangle$, in Fig.~\ref{fig:NFWc_Annihilation}  to the total annihilation cross-section $\langle \sigma v \rangle_{\rm ann}$ for the Higgs portal model, illustrated in Fig.~\ref{fig:HP}.  Our line search significantly constrains the parameter space where the Higgs portal model may explain the GCE.
Note that there is a small ($\sim$1$\sigma$) upward fluctuation in our line upper limit at $\sim$35 GeV, which prevents us from ruling out the best-fit point.
Our limits on gamma-ray lines 
significantly narrow the preferred parameter space for the Higgs portal model to explain the GCE.

As the preferred DM mass is less than $m_h/2$, with $m_h$ the SM Higgs mass, and generally sizable DM-Higgs couplings are required to achieve thermal cross sections, it is pertinent to consider the impact of bounds on invisible Higgs decays on this parameter space. The same coupling $y_{\chi h}$ that facilitates the DM annihilation induce a contribution to the SM Higgs decay width given by
\begin{equation}
    \Gamma_{h \to \bar \chi \chi} = \frac{y_{\chi h}^2  m_h}{16 \pi} \left( 1- \frac{4 m_\chi^2}{m_h^2}\right)^{1/2},
\end{equation} 
which is constrained at 95\% confidence to be no more than 11\% of the total Higgs width~\cite{ATLAS:2022yvh,Wang:2841498}.

In Fig.~\ref{fig:HP} we also show the constraint on the annihilation cross-section from invisible Higgs decay. A Higgs portal explanation of the GCE with mass $\lesssim m_h/2$ is disfavored by invisible Higgs decays; our gamma-ray limits extend this constraint to masses below $\sim$80 GeV.   However, the invisible Higgs decay limits may be evaded if the Higgs portal realization is non-minimal (see, {\it e.g.},~\cite{Mondal:2014goi,Bell:2017rgi,Ipek:2014gua,Yang:2018fje,Cuoco:2016jqt}), while avoiding the gamma-ray line limits may be more difficult. For scenarios that augment the dark sector, for instance with a singlet scalar mixing with the SM Higgs~\cite{Mondal:2014goi}, our constraints may apply as presented while Higgs decay bounds are relaxed.  For more involved scenarios such as 2HDM constructions~\cite{Yang:2018fje,Cuoco:2016jqt}, our bounds will be somewhat modified due to the increased amount of free parameters available in the theory. In all such cases, our photon line search provides independent and complementary constraints to collider-based probes on Higgs portal dark sectors. 

\subsection{Neutralino description of the GCE}
\label{sec:neutralino}

Since the GCE peaks at energies $\sim$GeV, as illustrated in Fig.~\ref{fig:GCE_HP}, the data generally prefer relatively low DM masses, $m_\chi \lesssim 50$ GeV, annihilating to light fermions. However, as the extraction of the GCE is subject to significant systematic uncertainties (see, {\it e.g.},~\cite{Murgia:2020dzu}), heavier DM candidates with $m_\chi \sim 100$ GeV have also been considered in the literature, specifically in the context of supersymmetric models where the DM is a neutralino~\cite{Agrawal:2014oha,Achterberg:2015srl,Cao:2015loa,Butter:2016tjc,Achterberg:2017emt,Murgia:2020dzu}. 
These models are compelling in part because of the additional motivation for supersymmetry near the electroweak scale due to the electroweak hierarchy problem. In this section we consider how neutralino explanations of the GCE are constrained by the search for associated gamma-ray lines.

For this discussion, for simplicity, we adopt a Split-SUSY~\cite{Wells:2003tf,Giudice:2004tc,Arkani-Hamed:2004ymt,Arvanitaki:2012ps,Arkani-Hamed:2012fhg} scenario that effectively suppresses the sfermion-mediated annihilation channels and simplifies the parameter space under consideration. The remaining, dominant annihilation channels, after decoupling the sfermions, are then to the electroweak gauge bosons. As an ansatz, we model the dark sector to be approximately wino, consisting of a Majorana DM $\chi$ and a nearly-degenerate chargino counterpart $\chi^\pm$.  Small variations to this picture, where the neutralino is {\it e.g.} a more significant admixture of bino and Higgsino, lead to similar results, so long as the there is a sizeable annihilation $WW$ final state, which is typical for {\it e.g.} the well-tempered neutralinos that can naturally make up the observed DM abundance in the mass range of interest~\cite{Arkani-Hamed:2006wnf}.  Note that pure winos with masses $\sim$100 GeV are only expected to be a sub-fraction of DM unless the DM is produced non-thermally; the GCE may arise from annihilation of a DM sub-fraction, or the neutralino could be a more significant fraction bino and make up all of the DM. While any particular neutralino model may differ in detail from the pure wino case, the wino phenomenology is sufficient to illustrate the constraining power of our gamma-ray line limits for neutralino explanations of the GCE. The relevant interaction with the SM is given by
\begin{equation}
    \mathcal{L} \supset -g W^{\mp}_\mu \bar \chi \gamma^\mu \chi^{\pm}  - e A_\mu \bar \chi^\pm \gamma^\mu \chi^\pm.
\end{equation}

In the pure wino case, the difference between the neutral and charged component masses is radiatively set, $\Delta m_+ \sim 150 - 160$ MeV, depending on the wino mass~\cite{Ibe:2012sx}.
However, light charginos $m^\pm_\chi \lesssim 270$ GeV with a small mass gap $\Delta m_+ \lesssim 220$ MeV are ruled out by collider searches~\cite{ATLAS:2013ikw}, and indeed charginos with $m^\pm_\chi \lesssim 95$ GeV are disfavored by LEP altogether~\cite{DELPHI:2003uqw,ALEPH:2002gap} (see~\cite{Agrawal:2014oha} for a discussion).  In reality, because of these stringent collider constraints on light gauginos, the vast majority of viable neutralino explanations of the GCE will likely be significantly mixed, may annihilate into $WW$, $ZZ$, and $hh$ at various branching ratios, and require scans of the full parameter space to identify. 

\begin{figure}
\begin{tikzpicture}
  \begin{feynman}
    \vertex (a){$\chi$};
    \vertex [right=1.5cm of a] (b);
    \vertex [below=2cm of b] (c);
    \vertex [below=2cm of a] (d){$\chi$};
    \vertex [right=1.2cm of b] (e){$W$};
    \vertex [right=1.2cm of c] (f){$W$};
    \diagram* {
      (a) -- [fermion] (b),
      (b) -- [fermion, edge label =\(\chi_\pm \)] (c),
      (c) -- [fermion] (d),
      (b) -- [photon] (e),
      (c) -- [photon] (f)
    };
  \end{feynman}
\end{tikzpicture}
\begin{tikzpicture}
  \begin{feynman}
    \vertex (a){$\chi$};
    \vertex [right=1.5cm of a] (b);
    \vertex [below=2cm of b] (c);
    \vertex [below=2cm of a] (d){$\chi$};
    \vertex [right=1.1cm of b] (e);
    \vertex [right=1.1cm of c] (f);
    \vertex [right=1cm of e] (g){$\gamma$};
    \vertex [right=1cm of f] (h){$\gamma$};
    \diagram* {
      (a) -- [fermion] (b),
      (b) -- [fermion, edge label =\(\chi_\pm\)] (c),
      (c) -- [fermion] (d),
      (b) -- [photon, edge label =\(W\)] (e),
      (c) -- [photon] (f),
      (e) -- [photon, edge label =\(W\)] (f),
      (e) -- [photon] (g),
      (f) -- [photon] (h),      
    };
  \end{feynman}
\end{tikzpicture}

\begin{tikzpicture}
  \begin{feynman}
    \vertex (a){$\chi$};
    \vertex [right=1.5cm of a] (b);
    \vertex [below=2cm of b] (c);
    \vertex [below=2cm of a] (d){$\chi$};
    \vertex [right=1.1cm of b] (e);
    \vertex [right=1.1cm of c] (f);
    \vertex [right=1cm of e] (g){$\gamma$};
    \vertex [right=1cm of f] (h){$\gamma$};
    \diagram* {
      (a) -- [fermion] (b),
      (b) -- [photon, edge label =\(W\)] (c),
      (c) -- [fermion] (d),
      (b) -- [fermion, edge label =\(\chi_\pm\)] (e),
      (f) -- [fermion] (c),
      (e) -- [fermion, edge label =\(\chi_\pm\)] (f),
      (e) -- [photon] (g),
      (f) -- [photon] (h),      
    };
  \end{feynman}
\end{tikzpicture}
\begin{tikzpicture}
  \begin{feynman}
    \vertex (a){$\chi$};
    \vertex [right=1.5cm of a] (b);
    \vertex [below=2cm of b] (c);
    \vertex [below=2cm of a] (d){$\chi$};
    \vertex [right=1cm of b] (e);
    \vertex [below=1cm of e] (f);
    \vertex [right=3.6cm of a] (g){$\gamma$};
    \vertex [right=3.6cm of d] (h){$\gamma$};
    \diagram* {
      (a) -- [fermion] (b),
      (b) -- [fermion, edge label =\(\chi_\pm\)] (c),
      (c) -- [fermion] (d),
      (b) -- [photon] (f),
      (f) -- [photon, edge label =\(W\)] (c),
      (f) -- [photon] (h),
      (f) -- [photon] (g),      
    };
  \end{feynman}
\end{tikzpicture}
    \caption{As in Fig.~\ref{fig:feyn_higgs_portal}, but for wino-like DM. Extra diagrams from the exchange of final-state bosons have been omitted for brevity.}
    \label{fig:feyn_neutralino}
\end{figure}
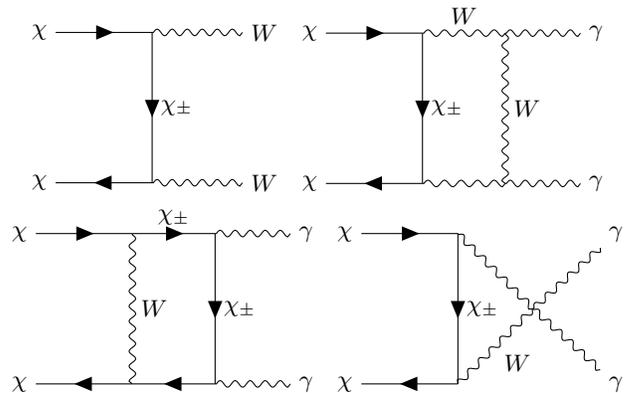

We restrict our discussion to the wino case for the remainder of this section, except for relaxing the mass gap $\Delta m_+$ to ensure $m^\pm_\chi \gtrsim 100$ GeV (we may assume this comes from incorporating a small bino admixture without significantly altering the phenomenology).  The DM annihilates at tree-level to $WW$ and to $\gamma \gamma$ via $W - \chi^\pm$ loops; the relevant diagrams governing both the continuum and photon-line annihilation are shown in Fig.~\ref{fig:feyn_neutralino}. The annihilation cross section in this case is given explicitly by
\begin{equation}
    \langle \sigma_{\chi\chi\to WW} v \rangle = \frac{g^4 (m_\chi^2 - m_W^2)^{3/2} }{2 \pi m_\chi (2 m_\chi^2 - m_W^2)^2 } +  \mathcal{O}(v^2) \,,
\end{equation}
where $v$ is the relative DM velocity.
For the DM masses under consideration, electroweak corrections and Sommerfeld enhancement effects are negligible.  The tree-level cross-section evaluates to $\langle \sigma v \rangle \approx 4 \times 10^{-24} \, \rm{cm}^3/\rm{s}$ for $m_\chi \sim 100 \,\rm{GeV}$. This is far larger than the cross section needed to achieve the observed relic density, and if the wino experiences a thermal history it will freeze-out to only a small fraction of the DM abundance, $f_\chi \approx \left( \langle \sigma v \rangle / \langle \sigma v \rangle_{\rm f.o.} \right)^{-1} \sim 10^{-2}$. Correspondingly, the observed annihilation signal is suppressed by a factor of $f_\chi^2$. Alternatively, the wino may be populated non-thermally and make up any fraction or all of the DM abundance.  We consider both scenarios, but note that while this changes the \emph{theoretical} prediction of the wino annihilation signal, it does not change the relationship between the \emph{observed} continuum annihilation products from the GCE and its corresponding photon line signal, as the same amount of DM is producing both.

For near-threshold masses $m_\chi \sim m_W$, which is the case preferred by the data, thermal corrections to the s-wave annihilation may become relevant and even dominant. In the regime where $m_\chi = m_W (1 + \delta)$, with $\delta \ll 1$, the leading order contributions to the $WW$ annihilation are given by  
\begin{equation}
    \langle \sigma v \rangle  \approx \frac{g^4}{2 \pi m_W^2} \left( 5 v^3  +  7 \delta v + \mathcal{O}(v^5, \delta v^3) \right).
\end{equation}
For masses below threshold, the dominant annihilations are to $\bar qq$ and $\bar \ell\ell$ at 1-loop, but we simply truncate our analysis at $m_\chi = m_W$.  The cross section to monochromatic photons, illustrated in Fig.~\ref{fig:feyn_neutralino},
evaluates to $\langle \sigma v \rangle_{\gamma\gamma} \sim 10^{-2} \langle \sigma v \rangle_{WW}$ for $m_\chi \sim 100$ GeV. 

\begin{figure}
    \centering
    \includegraphics[width = 0.9\linewidth]{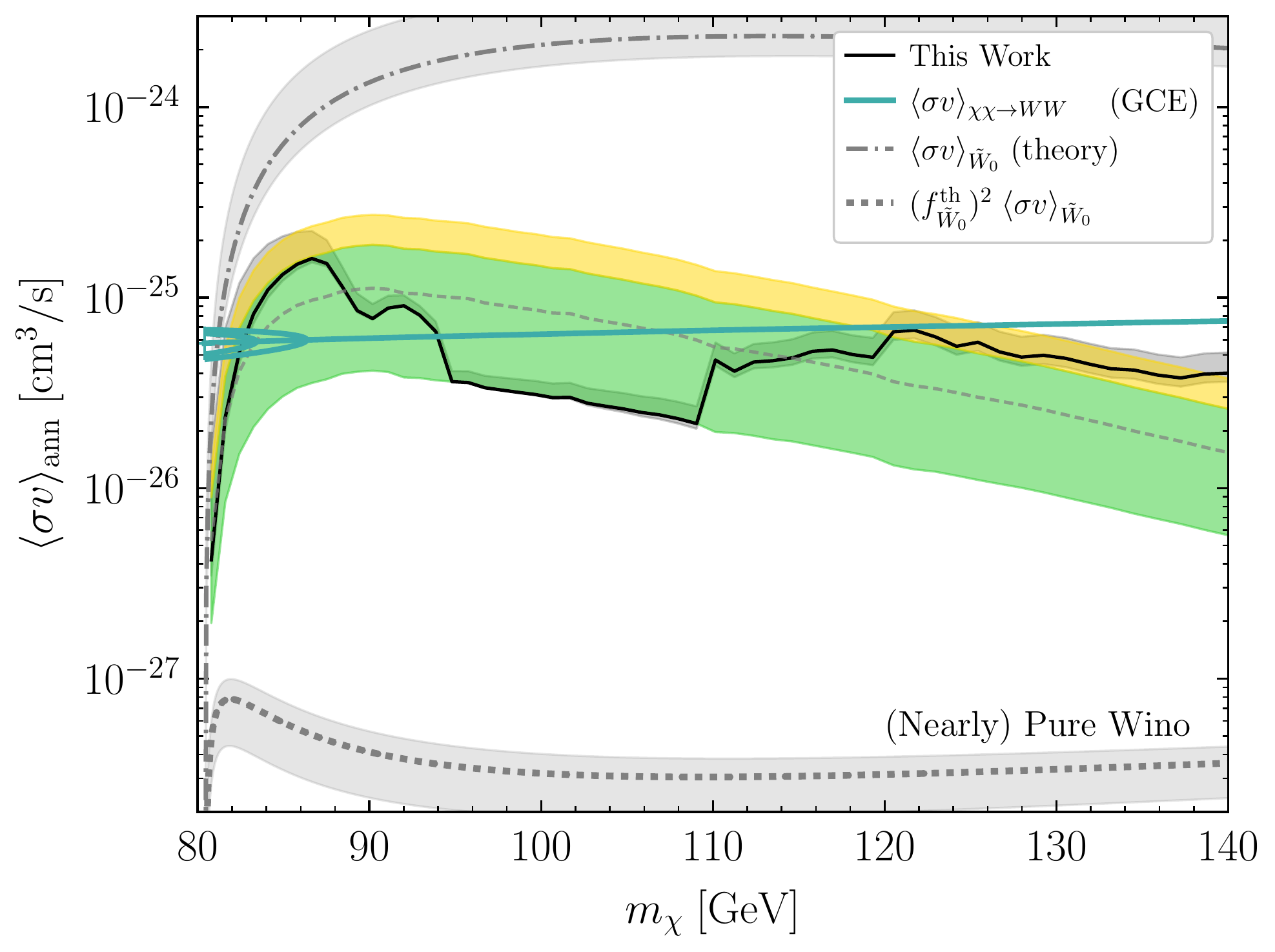}
    \caption{The 95\% upper limits of our line search, applied to constrain wino-like explanations of the GCE, as a function of the DM mass (black, and shaded green and yellow containment bands), assuming that at each mass the wino is 100\% of the DM. The best-fit cross section to the GCE at each fixed mass is shown (blue line), as well as the 68\% and 95\% containment ellipses to illustrate the best-fit parameter space when the mass and cross-section are both treated as free parameters. The theoretical wino annihilation cross sections are illustrated assuming it constitutes all (dot dashed) or a thermal fraction (dotted) of the DM, for the case where the neutralino-chargino mass gap is set at $\Delta m_+ = 0.2 m_\chi$. The gray shaded regions represent varying across $\Delta m_+  \in [0, 0.3] m_\chi$, though note that charginos with mass $\lesssim 100$ GeV are experimentally disfavored~\cite{ATLAS:2013ikw,DELPHI:2003uqw,ALEPH:2002gap,Agrawal:2014oha}.  In the case where the wino is a DM sub-fraction we illustrate the annihilation cross-section multiplied by the sub-fraction squared.}
    \label{fig:Wino}
\end{figure}

The blue line in Fig.~\ref{fig:Wino} shows  the best-fit cross section to explain the GCE for fixed wino DM mass, assuming the wino is all of the DM. The half-ellipses delineate the 68\% and 95\% CL containment regions preferred by the data, allowing both the mass and cross-section to vary. The fit of the model to the GCE data is conducted as in the Higgs portal case.
As expected, the data ultimately prefers a near-threshold wino (see, {\it e.g.},~\cite{Agrawal:2014oha,Achterberg:2015srl}).
Superimposed is the effective theoretical cross section of an annihilating wino, both as a fraction and all of the DM, for a range of chargino masses, $\Delta m_+ /m_\chi \in [0, 0.3]$. We take the scenario of $\Delta m_+ = 0.2 m_\chi$ as fiducial, which guarantees $m_\chi^\pm > 95$ GeV for the entire considered range.  Last, we show the limits of our line search on this parameter space.
The parameter space that is maximally preferred by the GCE is tightly constrained by our search for photon lines.  Only a small mass range around 85 GeV is allowed for the wino at 95\% confidence  as an explanation of the GCE, given our null results for an associated gamma-ray line. 

\subsection{EFT description of the GCE}

Let us suppose that the GCE originates from a DM model with mediator mass much larger than $m_\chi$, so that we may use an EFT framework. EFTs for the GCE have been extensively studied (see, {\it e.g.},~\cite{Alves:2014yha,Liem:2016xpm,Karwin:2016tsw,Roszkowski:2017nbc,GAMBIT:2021rlp}).  
For example, let us consider pseudo-scalar operators of the form, for Dirac DM $\chi$,
\es{eq:EFT}{
{\mathcal L} =  {m_f \over \Lambda_f^3} \bar f \gamma_5 f \bar \chi \gamma_5 \chi\,,
}
where $f$ is a SM fermion with mass $m_f$. 
The pseudo-scalar form of the interactions induce velocity suppression on the DM-SM elastic scattering and help the DM candidate be compatible with direct detection constraints~\cite{Alves:2014yha}. 
In this EFT the DM annihilates at tree-level to SM fermion pairs $\bar f f$ with cross-section~\cite{Karwin:2016tsw}
\es{}{
\langle \sigma_{\chi \bar \chi \to \bar f f} v\rangle = {N_c m_f^2 m_\chi^2 \over \pi \Lambda_f^6} \sqrt{1 - {m_f^2 \over m_\chi^2}} \,,
}
to leading order in the small DM velocity and where $N_c = 3$ ($N_c  = 1$) for quark (lepton) final states.
For $f = b$ quarks, we need $\langle \sigma_{\chi \bar \chi \to \bar b b} v\rangle \sim 1.5 \times 10^{-26}$ cm$^3$/s for $m_\chi \approx 40$ GeV to explain the GCE, while using inverse Compton emission of final state electrons off of the interstellar radiation field allows us to explain the GCE for the $\mu^+ \mu^-$ final state with $m_\chi \approx 56$ GeV and $\langle \sigma_{\chi \bar \chi \to \mu^+ \mu^-} v\rangle \sim 4 \times 10^{-26}$ cm$^3$/s~\cite{DiMauro:2021qcf}. 

By closing the fermion loop, the DM $\chi$ acquires a one-loop annihilation channel to $\bar \chi \chi \to \gamma\gamma$. 
This leads to the result 
\es{}{
\frac{\langle \sigma_{\bar \chi\chi\to \gamma\gamma}  v \rangle}{\langle \sigma_{\bar \chi\chi\to \bar ff} v \rangle }  \approx \frac{ \alpha^2 q_f^4 x^2  \left| \log \left( -x^2 \right) \right|^4 }{4 \pi^2},  \qquad x \equiv \frac{m_f}{2 m_\chi} \,,
}
with $\alpha$ the fine structure constant. This loop-induced line cross-section is below our sensitivity for both $\bar b b$ and $\mu^+ \mu^-$ final states.  Note that in principle there is also the $\gamma Z$ final state, though this is not kinematically accessible for the parameter space to explain the GCE.

On the other hand, a UV complete model will likely give rise to multiple, correlated terms in the DM EFT, and so it is likely not a good approximation to only consider the Lagrangian term in~\eqref{eq:EFT} for a single fermion in isolation. In App.~\ref{app:EFT} we consider, at the opposite extreme, the effective Lagrangian terms that involve DM couplings to $W$ and $Z$ bosons, which directly give rise to photon lines. 

\section{Implications for motivated DM models}
\label{sec:other}

Independent of the GCE, there are many other compelling DM candidates that might be competitively constrained with gamma-ray line searches.  We discuss the implications of our results for a number of such models here, including 
Higgsino DM (annihilating),  gravitino DM (decaying), and glueball DM (decaying). 

\subsection{Higgsino DM}

Arguably the most theoretically compelling of the experimentally viable DM models at present is the nearly-pure thermal Higgsino. Like the other MSSM neutralinos, it has strong theoretical motivation (see, {\it e.g.},~\cite{Co:2021ion}), but unlike the bino~\cite{ATLAS:2014jxt,CMS:2015flg} and wino~\cite{Cohen:2013ama,Fan:2013faa} it is not currently disfavored by data. 

\begin{figure}
    \centering
    \includegraphics[width = 0.9\linewidth]{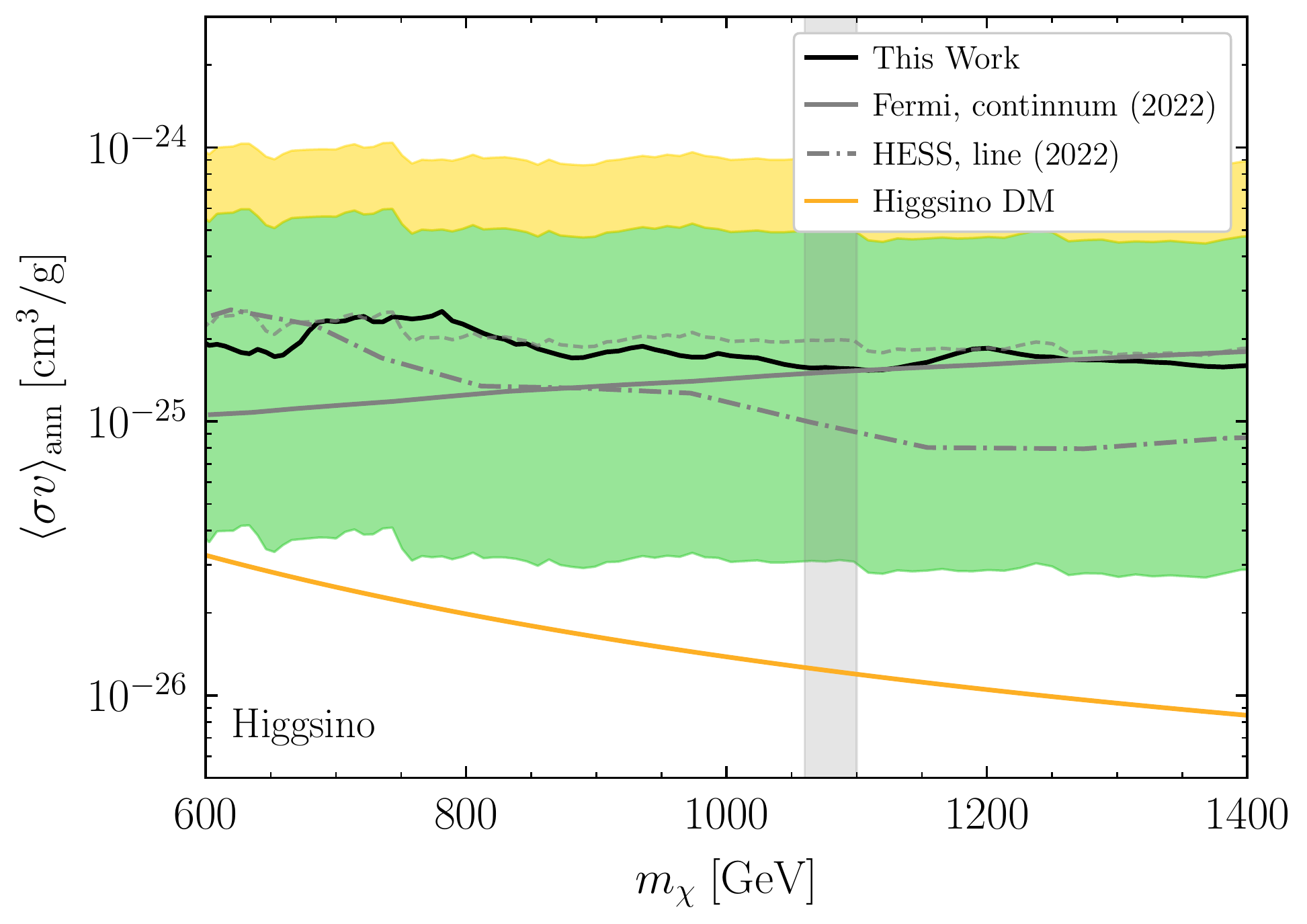}
    \caption{Our line search reinterpreted as a constraint for Higgsino DM, assuming that it constitutes all of the observed DM. The orange line denotes the expected annihilation cross section for the MSSM Higgsino, with the vertical shaded region denoting the mass that yields the correct DM abundance under a thermal cosmology, $m_\chi \approx 1.1$ TeV. Also shown are limits from continuum searches with Fermi data~\cite{Dessert:2022evk} (solid grey)  and line searches with HESS~\cite{HESS:2018cbt} (dot-dashed grey).}
    \label{fig:higgsino}
\end{figure}

Much like the wino scenario discussed in Sec.~\ref{sec:neutralino}, the Higgsino is extremely predictive as a model, and the mass required to achieve the full thermal relic abundance is narrowly fixed at 1.1 TeV~\cite{Bottaro:2022one}. The dark sector consists of the DM $\chi$ and quasi-degenerate neutralino $\tilde \chi$ and chargino $\chi^\pm$ counterparts. 
The relevant interaction terms are given by
\begin{equation}
    \mathcal{L} \supset - \frac{g}{2} W^{\mp} \bar \chi \gamma^\mu \chi^\pm - \frac{g}{4 \cos \theta_W} Z_\mu \bar \chi \gamma^\mu \tilde \chi  \,.
\end{equation}
 As in the case of the wino, the Higgsino acquires a one-loop decay to two photons, which, depending on the mass, receives a Sommerfeld enhancement~\cite{Kowalska:2018toh,Rinchiuso:2020skh,Krall:2017xij,Dessert:2022evk}. 
Our line search may thus be interpreted as a constraint on the total Higgsino annihilation cross-section $\langle \sigma v \rangle_{\mathrm ann}$, as illustrated in Fig.~\ref{fig:higgsino}, assuming that the Higgsino is 100\% of the DM at each mass $m_\chi$. Our constraint is surpassed by the Fermi continuum limits~\cite{Dessert:2022evk} and the HESS line limits~\cite{HESS:2018cbt} at low and high masses, respectively, though note that the line search is more robust than the continuum one, since the line morphology has less confounding astrophysical backgrounds. 

\subsection{Gravitino DM}

We now consider models that invoke a finite DM lifetime, beginning with the case of gravitino DM destabilized by bilinear R-parity violation. 
The R-parity violation can be contained in the following soft-SUSY breaking terms~\cite{Ibarra:2007wg,Ishiwata:2008cu,Takayama:2000uz},
\begin{equation}
\mathcal{L} \supset  B_i \tilde L_i H_u + m^2_{H L_i} \tilde L_i H_d^* ,
\end{equation}
where $\tilde L_i$ are the left-handed slepton doublets and $H_{u (d)}$ are the up-(down-) type Higgs doublets. The sneutrinos acquire a VEV proportional to the size of R-parity violation,
\begin{equation}
\langle \tilde \nu_i \rangle = v_{EW} \frac{B_i \sin \beta + m^2_{H L_i} \cos \beta}{m_{\tilde \nu_i}^2},
\end{equation}
where $\tan\beta = \langle H_u^0 \rangle / \langle H_d^0 \rangle$. In the case where the NLSP is a neutralino, the gravitino decay proceeds through its coupling to gauge-boson gauginos, resulting in decay channels to a lepton and a gauge boson via sneutrino insertion: $\ell W, \nu Z, \nu h, \nu \gamma$. In scenarios where $m_{3/2} \lesssim 80$ GeV, however, the remaining channels are kinematically forbidden and the gravitino is forced to decay via  $\nu \gamma$, meaning that in this case the model has a 100\% branching ratio to a final state that gives a monochromatic photon line.

\begin{figure}
    \centering
    \includegraphics[width =0.9\linewidth]{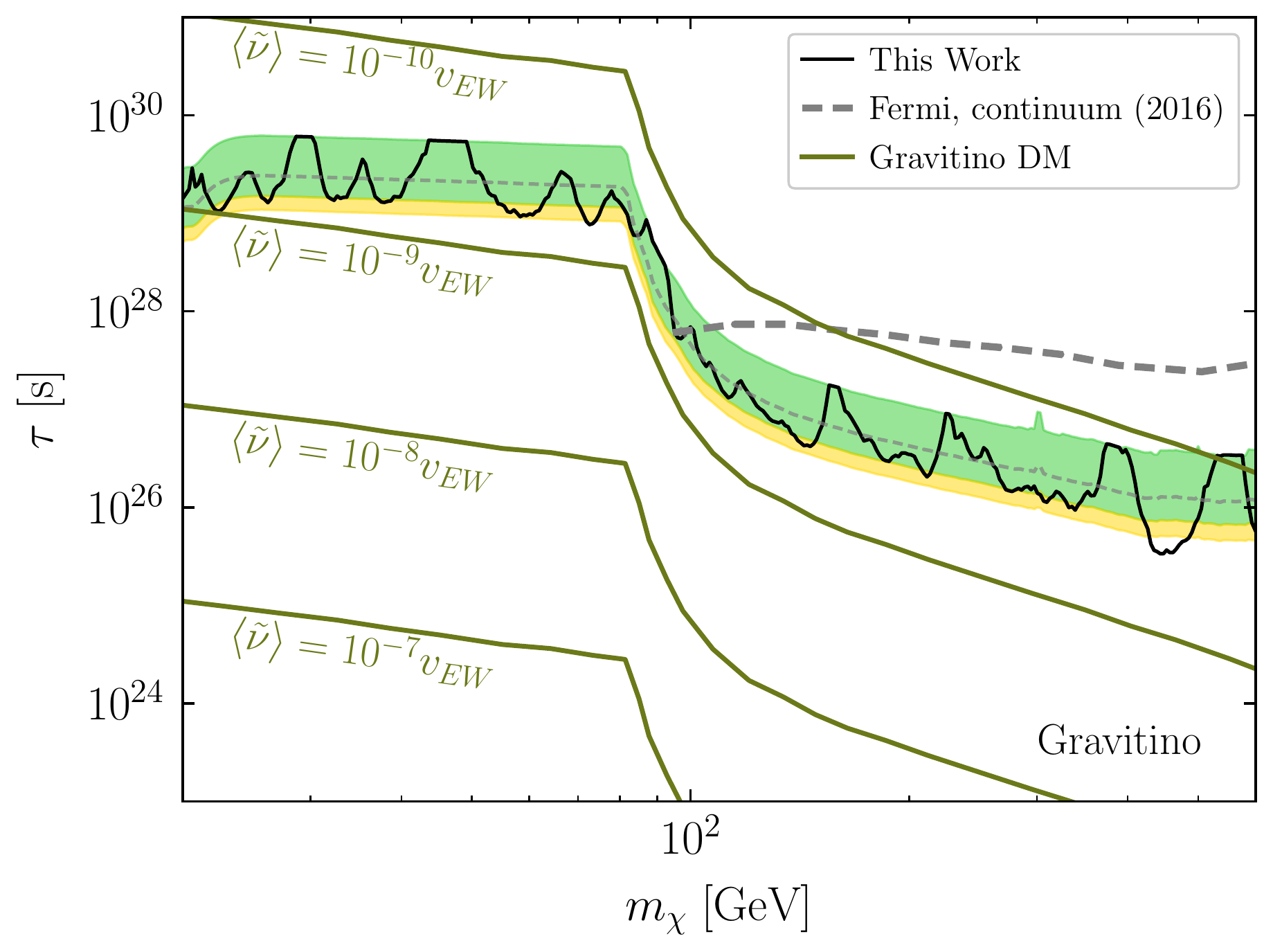}
    \caption{Our line search reinterpreted as constraints for decaying gravitino DM. The dark green lines are the expected lifetime of the gravitino, assuming model parameters from~\cite{Ishiwata:2008cu} and taking $\langle \tilde \nu \rangle =10^{[-7, -10]} v$, which is the parameter that carries R-parity violation. Superimposed are the limits from continuum emission from gravitino decay to $\ell W, \nu Z, \nu h$~\cite{Cohen:2016uyg}, which subsume our limits at high masses but have no sensitivity to low mass scenarios where $m_{3/2} \lesssim m_W$.}
    \label{fig:gravitino}
\end{figure}

The decay width of the gravitino into $\nu \gamma$ is given by~\cite{Ishiwata:2008cu,Takayama:2000uz}
\begin{equation}
 \Gamma_{\psi \to \nu \gamma} = \frac{g^2}{128 \pi \cos^2 \theta_W}  \frac{\langle \tilde \nu \rangle^2 m_{3/2}^3}{M_{\rm pl}^2} \theta^2_{\tilde \gamma}, 
\end{equation}
where $\theta_{\tilde \gamma}$ depends on the neutralino masses and mixing, and the size of R-parity breaking is encapsulated in the sneutrino vev $\langle \tilde \nu \rangle$.  We take the model parameters from~\cite{Ishiwata:2008cu} as a fiducial scenario, where the NLSP is bino type with $m_{\tilde B} = 1.5 m_{3/2}$, $\tan \beta = 10$, and $m_{\tilde \nu} = 2 m_{3/2}$. Constraints on gravitino lifetime may then be interpreted as limits on the sneutrino VEV,  and we find that our line search is able to disfavor at the 95\% level scenarios with $\langle \tilde \nu \rangle \gtrsim 10^{-9} v_{EW} $. We illustrate these limits in Fig~\ref{fig:gravitino}, noting that our line search, while subdominant to the continuum limits at higher energies~\cite{Cohen:2016uyg}, probe the only observable signature of metastable gravitinos at low masses and are thus leading in this regime. 

\subsection{Glueball DM}

Next, we consider the case of glueball DM. In this scenario, the dark sector confines at scale $\Lambda_D$, and the dark matter is made up of the lightest, $0^{++}$, glueball state with mass $m_\chi \simeq \Lambda_D$.  In this phase the interaction with the SM is facilitated by the operator 
\begin{equation}
\mathcal{L} \supset \lambda_\chi \chi H^\dagger H, 
\end{equation}
which induces both the decay of $\chi \to hh$ and $\chi -h$ mixing.  Note that this operator arises from the dimension six operator $G^2 H^\dagger H$, with $G$ the dark gauge field strength: thus, we expect small $\lambda_\chi \sim \Lambda_D^3 / \Lambda^2$, with $\Lambda$ the UV cut-off of the theory.  The DM inherits all the decay modes of the Higgs, weighted by a mixing angle $\theta$. For $m_\chi \lesssim m_W$, the dominant channel is $b\bar{b}$, while for $m_\chi \gg m_W$, it will decay mostly to $WW$, $ZZ$ and $hh$~\cite{Cohen:2016uyg}.  Explicitly, 
\begin{align}
    \Gamma_{\chi} = \Gamma_{\chi \to hh} + \left. \sin^2\theta \, \Gamma_{h}\right|_{m_h = m_\chi}  ,
\end{align} 
where 
\begin{align}
    \Gamma_{\chi \to hh}
=   \frac{\lambda_\chi^2 }{32\pi m_\chi }\sqrt{1-\frac{4m_h^2}{m_\chi^2}}\left( \cos^3\theta+2\cos\theta\sin^2\theta \right)^2 ,
\end{align}
and $\Gamma_{h}\mid_{m_h = m_\chi}$ is the SM Higgs decay width for a Higgs with mass $m_\chi$~\cite{LHCHiggsCrossSectionWorkingGroup:2016ypw}. The $h-\chi$ mixing angle $\theta$ is set by $\Lambda_D$, $\lambda_\chi$, and the SM Higgs VEV $v_{EW}$,
\begin{equation}
{\rm tan} \theta =  \frac{2 \lambda_\chi v_{EW}}{\xi - \sqrt{4 \lambda_\chi^2 v_{EW}^2 + \xi^2} } ,
\end{equation}
where 
\begin{equation}
   \xi \equiv  m_h^2 - m_\chi^2 - \frac{v_{EW}^2 \lambda_\chi }{2 \Lambda_D}  .
\end{equation}

\begin{figure}
    \centering
    \includegraphics[width = 0.9\linewidth]{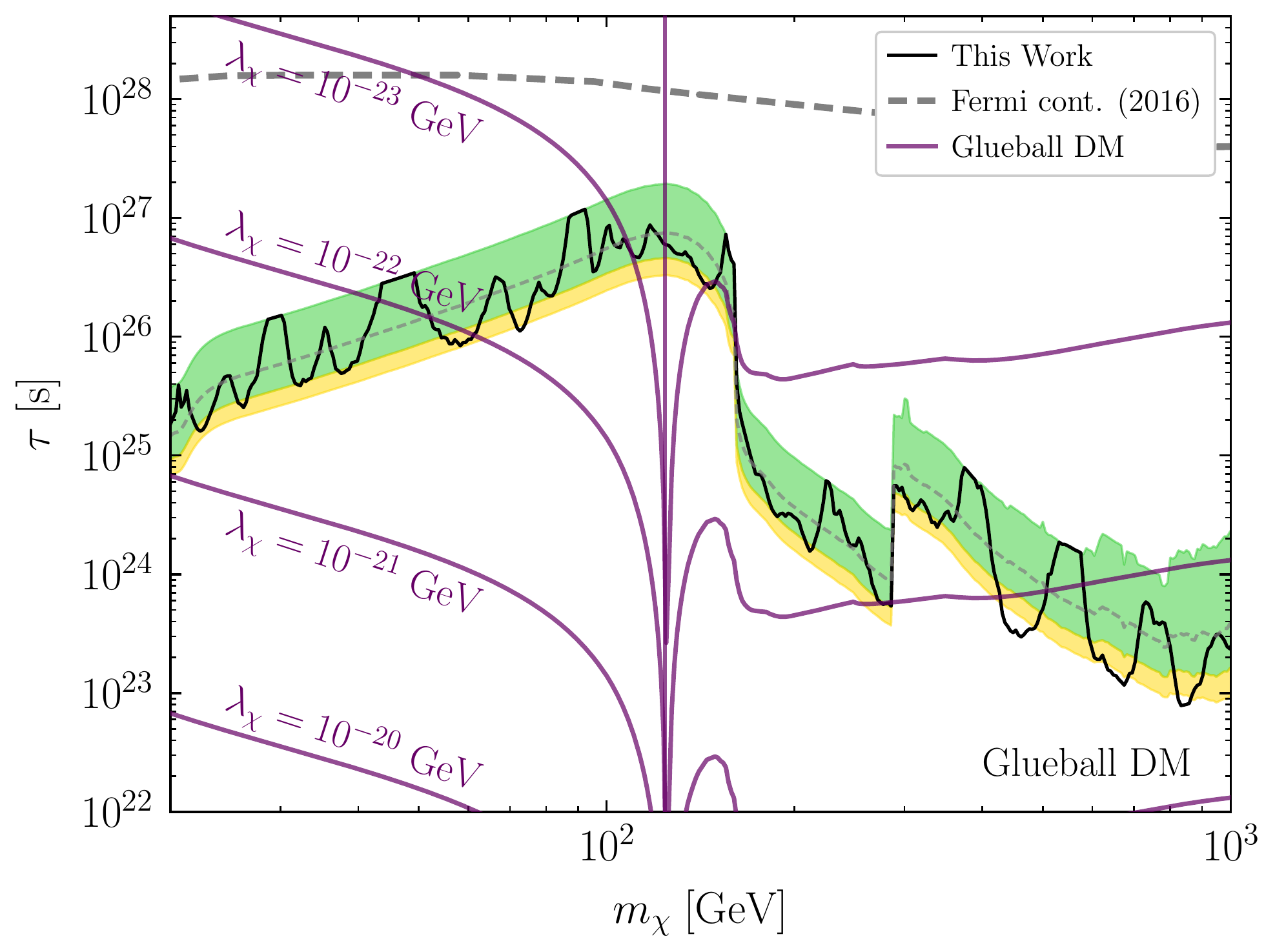}
\caption{Our line search reinterpreted as a constraint on glueball DM. The violet lines indicate the expected lifetime of the glueball, assuming $\Lambda_D \simeq m_\chi$, for various values of $\lambda_\chi$.  Shown also in solid grey are limits from Fermi continuum emission searches~\cite{Cohen:2016uyg}, which are more competitive as they probe the comparatively much more prominent decay channels to $\bar b b, hh, WW,$ and $ZZ$.}
    \label{fig:glueball}
\end{figure}

The contribution to monochromatic photons therefore comes entirely via the Higgs mixing term,  
\begin{align}
    {\rm BR}_{\chi \to \gamma X} = \frac{\sin^2 \theta }{{\Gamma_\chi}} \left. \left( \Gamma_{h \to \gamma\gamma} +  \frac{1}{2}\Gamma_{h \to Z \gamma} \right) \right|_{m_h = m_\chi},
\end{align}
at least in the case of $m_\chi \gg m_Z$. We note, however, that the channel $\chi \to Z \gamma$ produces a monochromatic photon that is offset from the $E^{(\gamma\gamma)}_\gamma = m_\chi$ carried by the $\gamma\gamma$ decay products, with $E^{(Z\gamma)}_\gamma = m_\chi \left( 1 - m_Z^2/m_\chi^2 \right)$.  For the energy resolutions shown in in Fig.~\ref{fig:Energy_Resolution}, the $Z \gamma $ and $\gamma\gamma$ channels can be considered to contribute to the same signal when $m_Z^2/m_\chi^2 \lesssim 0.1$. The results of our line search interpreted in the context of glueball decay are given in Fig.~\ref{fig:glueball}. Our results disfavor scenarios where $\lambda_\chi \gtrsim 10^{-22}$ GeV,  though we find that they are subsumed by the {\it Fermi} continuum limits~\cite{Cohen:2016uyg}.  Note, for example, that for a confinement scale $\Lambda_D = 100$ GeV, the theory achieves $\lambda_\chi = 10^{-22}$ GeV for a UV-completion scale of $\Lambda \approx 10^{14}$ GeV, which is near where one may expect the UV completion to be if the dark gauge group unifies or interacts non-trivially with the SM near the scale of Grand Unification. 

\section{Discussion}
\label{sec:conclusions}

In this work we present the most sensitive search to date for annihilating and decaying DM in the gamma-ray band between 10 GeV and $\sim$500 GeV, for annihilation, and 2 TeV for decay.  Our annihilation sensitivity is surpassed by that of the H.E.S.S. experiment above $\sim$500 GeV. In principle H.E.S.S. would also be sensitive to decaying DM in this mass range, though such an analysis has not been performed to-date.  We find no evidence for annihilating or decaying DM and thus set leading constraints on the DM annihilation cross-section and decay rate.  

In most DM models the annihilation channel to $\gamma\gamma$ final states is loop-suppressed relative to the tree-level annihilation to unstable final states that produce continuum gamma-rays during their decays.  Naively, in these cases the continuum search is more sensitive, since it corresponds to a tree-level process versus a loop process for the monochromatic signal. However, in this work we show that in reality the interplay between continuum and line searches is more nuanced. In some DM models, such a neutralino models and Higgs portal models, the line searches are competitive in sensitivity relative to the continuum searches because of relative enhancements of the one-loop monochromatic decays. Moreover, the line searches have lower background rates, since they are confined to narrow energy ranges, and especially important is the fact that they have fewer confounding astrophysical backgrounds.  This latter point is especially important in the context of the {\it Fermi} GCE. Models such as neutralino and Higgs portal models that could explain the {\it Fermi} GCE predict associated line signatures that could be in reach of our search. However, since we find no evidence for gamma-ray lines, our analyses constrain the allowable parameter spaces for these models to explain the GCE. 

Given that our search is statistics limited and uses all available {\it Fermi} gamma-ray data to-date, it is unlikely that additional analyses of {\it Fermi} data will provide substantially increased sensitivity relative to our work.  With that said, a slight increase in sensitivity may be gained in future work by using an analysis strategy that incorporates spatial modeling at the pixel level, in addition to spectral modeling.  Such an approach should be contrasted with ours, which models the background in a given ring spectrally given a phenomenological power-law model.  Preliminary estimates indicate that the improvement in sensitivity from such an approach would be minimal, however, and this approach would also potentially be more susceptible to mismodeling, given {\it e.g.}  known failures of the Galactic diffuse models to accurately reproduce all of the small-scale variations found in the {\it Fermi} data. On the other hand, the upcoming Cherenkov Telescope Array (CTA) will have an increased effective area relative to {\it Fermi} and slightly worse, comparable, or improved energy resolution, depending on whether the gamma-ray energies are low or high~\cite{CTAConsortium:2017dvg}. In particular, CTA will extend down to $\sim$20 GeV, though the effective area and energy resolution degrade sharply at low energies.  Moreover, CTA will be subject to much more significant cosmic-ray backgrounds than {\it Fermi} and will acquire less exposure time over smaller regions of the sky, since it will have a field of view of a few degrees and only operate under {\it e.g.} optimal moonlight conditions, whereas the {\it Fermi}-LAT has a field of view covering approximately 20\% of the sky and takes data continuously.  Still, given the superior effective area of CTA it seems likely that future studies with that instrument for DM annihilation and decay will surpass those in this work in sensitivity for DM mass at least above roughly $100$ GeV. CTA may even provide leading sensitivity at lower masses, but understanding precisely where the CTA versus {\it Fermi}-LAT sensitivity cross-over is reached requires a dedicated study beyond the scope of this work. 

\begin{acknowledgments}
{\it 
We thank N. Rodd, S. Mishra-Sharma, and T. Slatyer for helpful discussions, and we thank N. Rodd for comments on the manuscript. 
We also thank C. Dessert for collaboration at early stages of the work. 
J.F. was supported by a Pappalardo Fellowship. 
B.R.S. and Y.P. were supported in part by the DOE Early Career Grant DESC0019225.  
Y.S. was supported by grants from NSF-BSF (No.~2021800), ISF (No.~482/20) and the Azrieli foundation.  
B.R.S. and Y.S. were supported in part by the BSF grant (No.~2020300). 
W.L.X. is supported by the U.S. Department of Energy under Contract DE-AC02-05CH11231.  
This research used resources from the Lawrencium computational cluster provided by the IT Division at the Lawrence Berkeley National Laboratory, supported by the Director, Office of Science, and Office of Basic Energy Sciences, of the U.S. Department of Energy under Contract No.  DE-AC02-05CH11231.
}

\end{acknowledgments}
\appendix

\section{Monochromatic signals in DM EFTs}
\label{app:EFT}

In this Appendix we consider a few minimal EFT descriptions of scenarios where the DM couples to the SM photon.  Assuming that the coupling enters before electroweak symmetry breaking, a corresponding coupling to other electroweak gauge bosons can be inferred. The operators mediating the DM annihilation can be written as   
\begin{equation}
    \mathcal{L}^{B}_{\rm anni} \subset \frac{\chi^\dagger \chi}{\Lambda^2}   B_{\mu\nu} B^{\mu\nu}, \quad \mathcal{L}^{W}_{\rm anni} \subset \frac{\chi^\dagger \chi}{\Lambda^2}   W^i_{\mu\nu} W_i^{\mu\nu},
\end{equation}
and likewise for decay,
\begin{equation}
    \mathcal{L}^{B}_{\rm decay} \subset \frac{\chi}{\Lambda}   B_{\mu\nu} B^{\mu\nu}, \quad \mathcal{L}^{W}_{\rm decay} \subset \frac{\chi}{\Lambda}  W^i_{\mu\nu} W_i^{\mu\nu},
\end{equation}
where $B_{\mu\nu}$ is the field strength of the hypercharge gauge boson $B_\mu$ and $W^i_{\mu\nu}$ that of the $SU(2)$ gauge bosons $W^i_\mu$. Here, we will assume scalar DM candidates for both annihilation and decay, though this approach easily extends to other cases. For both annihilation and decay, the low-energy phenomenology is simply depletion of DM into $\gamma\gamma$, $\gamma Z$ and $ZZ$ (and $WW$ in the case of $W^i_\mu$-coupling). The branching ratio to monochromatic photons is then given by
\begin{widetext}
\es{}{
    \mathrm{Br}^{B}_{\chi [\chi] \to \gamma\gamma} & = \frac{ c_W^4  + s_W^2 c_W^2 f_{\chi [\chi] \to \gamma Z} \Theta \left(   \sqrt{\frac{\Delta E}{E}} - \frac{m_Z}{[2]m_\chi} \right) }{  c_W^4  + 2 s_W^2 c_W^2 f_{\chi [\chi] \to \gamma Z} +  s_W^4 f_{\chi [\chi] \to ZZ} } \\ 
    \mathrm{Br}^{W}_{\chi [\chi] \to \gamma\gamma}  &= \frac{ s_W^4  + s_W^2 c_W^2 f_{\chi [\chi] \to \gamma Z} \Theta \left( \sqrt{\frac{\Delta E}{E}} - \frac{m_Z}{[2] m_\chi}\right)  }{  s_W^4  + 2 s_W^2 c_W^2 f_{\chi [\chi] \to \gamma Z} +  c_W^4 f_{\chi [\chi] \to ZZ} + 2 f_{\chi [\chi] \to WW} },  \\
    f_{\chi [\chi] \to \gamma Z }  & = \left( 1- \frac{m_Z^2}{[4] m_\chi^2} \right) \Theta \left( [2] m_\chi - m_Z \right)  \\
    f_{\chi [\chi] \to Z Z }  & = \sqrt{ 1- \frac{4 m_Z^2}{[4] m_\chi^2} } \Theta \left( [2] m_\chi - 2 m_Z \right)   \\ 
    f_{\chi [\chi] \to WW }  & = \sqrt{ 1- \frac{4 m_W^2}{[4] m_\chi^2} } \Theta \left( [2] m_\chi - 2 m_W \right) \,,  \\ 
}
\end{widetext}
where $s_W (c_W) = \sin \theta_W (\cos \theta_W)$ is given by the Weinberg angle $\theta_W$. As before, note that this expression takes the $\frac{1}{2} \gamma Z$ channel as contributing to the same photon line as $\gamma\gamma$,  an assumption valid only when the difference in photon energy is smaller than the energy resolution.   

\begin{figure}
    \centering
\includegraphics[width = 0.9\linewidth]{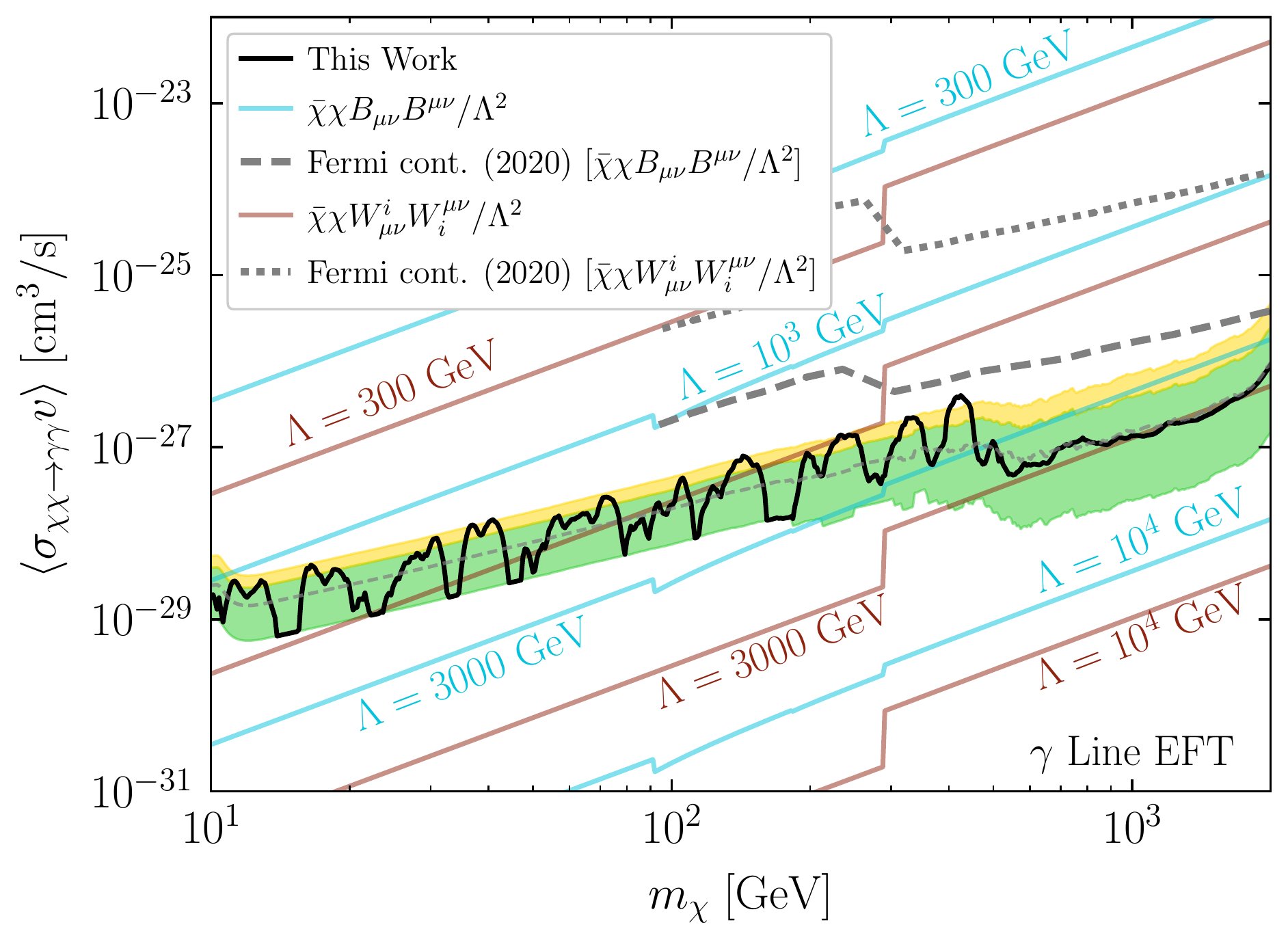}
    \caption{Our annihilation line search results assuming an NFW profile, reinterpreted to constrain interactions obeying the effective operators $\chi^\dagger \chi B_{\mu\nu} B^{\mu\nu}$ and $\chi^\dagger \chi W^i_{\mu\nu} W_i^{\mu\nu}$. These constraints are presented in the parameter space of the photon line cross section, and the {\it Fermi} continuum constraints~\cite{Abazajian:2020tww} (grey dashed and dotted, respectively) have been scaled accordingly. Also shown are the predictions from assuming various benchmark values of $\Lambda$ in both scenarios. }

    \label{fig:minimal_EFT_anni}
\end{figure}

\begin{figure}
    \centering
\includegraphics[width = 0.9\linewidth]{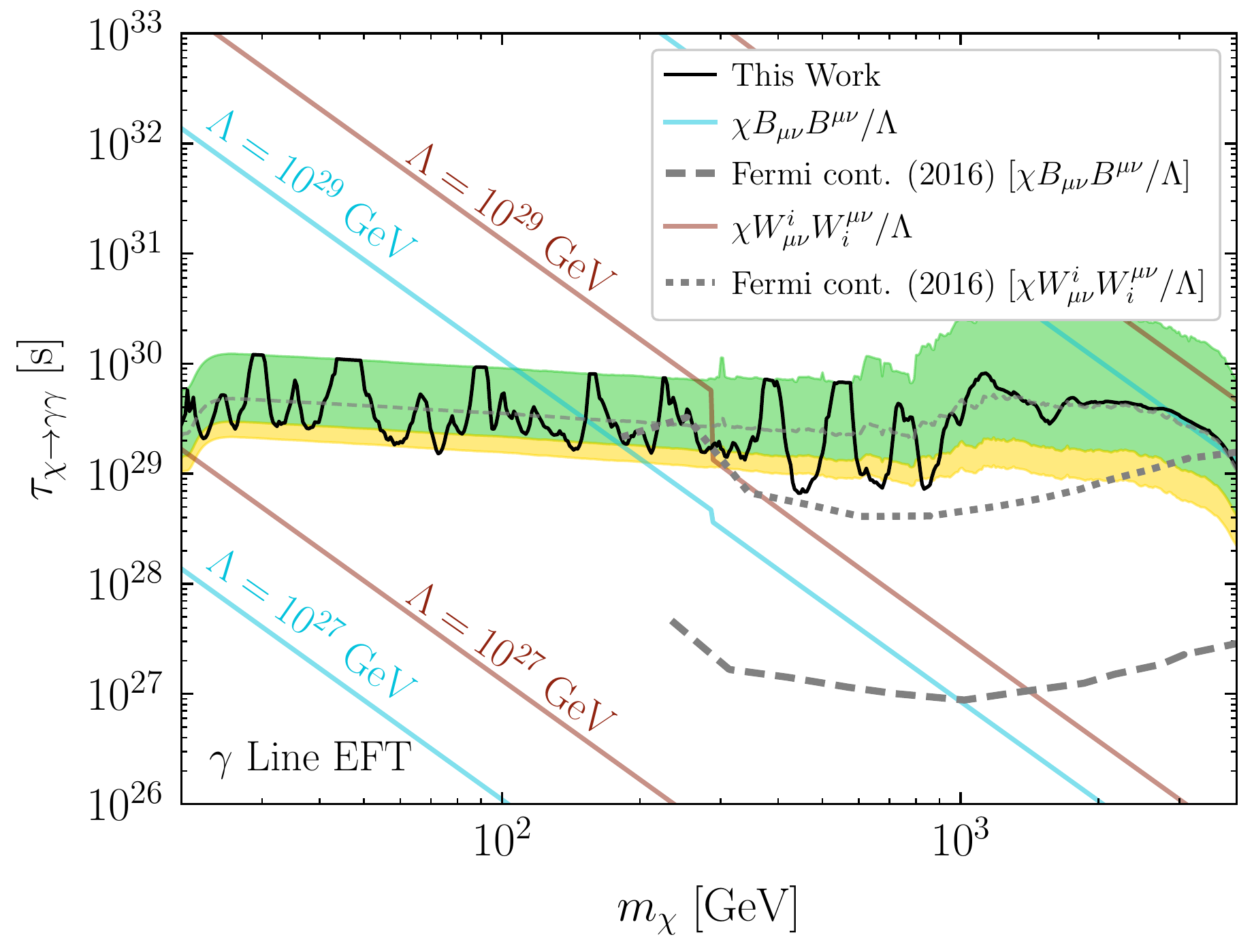}
    \caption{As in Fig.~\ref{fig:minimal_EFT_anni} but for DM decay.}
    \label{fig:minimal_EFT_decay}
\end{figure}

This branching ratio, approximately $10-20\%$ in regimes where $m_\chi \gg m_Z$ and 100\% where $m_\chi < m_W$,  offers a notably larger monochromatic photon yield than the other scenarios discussed in this work.
We demonstrate this relative advantage in Fig.~\ref{fig:minimal_EFT_anni} for annihilation and Fig.~\ref{fig:minimal_EFT_decay} for decay, noting that our results subsume continuum limits in all cases.  We map out the parameter space of the cut-off scale $\Lambda$ that is disfavored by our search.

\section{Direct comparison between continuum and line sensitivity}
\label{app:0}

In this Appendix we directly compare the expected sensitivity between a continuum and line-like signal from DM annihilation, taking as an illustration a DM candidate with mass $m_\chi = 40$ GeV that may decay, at tree-level, to $b \bar b$ (cross-section $\langle \sigma_{\chi \chi \to b \bar b} v \rangle$) or, at loop level, to $\gamma\gamma$ (cross-section $\langle \sigma_{\chi \chi \to \gamma\gamma} v \rangle$).  Under the null hypothesis, where we suppose that no DM signal is present in the data, we ask the question: what is the ratio of 95\% upper limits, $\langle \sigma_{\chi \chi \to \gamma \gamma} v \rangle^{\rm 95} / \langle \sigma_{\chi \chi \to b \bar b} v \rangle^{\rm 95}$, between the line-like and continuum cross-sections? To compute this ratio, we must estimate the sensitivity to the continuum signal under the null hypothesis. Using the same ROI as in the line search, we consider the energy range 1-100 GeV, where we assume that the data is described by the  {\it Fermi} Galactic diffuse model \texttt{gll\_iem\_v07} (\texttt{p8r3}). We generate the model expectation at 442 bins so as to reproduce an identical $\Delta E/E = 0.01$ binning resolution as used in our main results and use the Asimov approach~\cite{Cowan:2010js} to quantify the expected sensitivity to either annihilation directly to photons or annihilation to $b \bar b$.  Note that we use a spectral likelihood, without incorporating spatial information within an annulus, whereas most analyses of the GCE use a spatial likelihood, with nuisance parameters uncorrelated between energy bins (see, {\it e.g.},~\cite{Murgia:2020dzu}).  We adopt the simpler analysis strategy because we are simply interested in roughly estimating $\langle \sigma_{\chi \chi \to b \bar b} v \rangle^{\rm 95}$ under the null hypothesis. 

\begin{figure}[!htb]
	\begin{center}
		\includegraphics[width=0.45\textwidth]{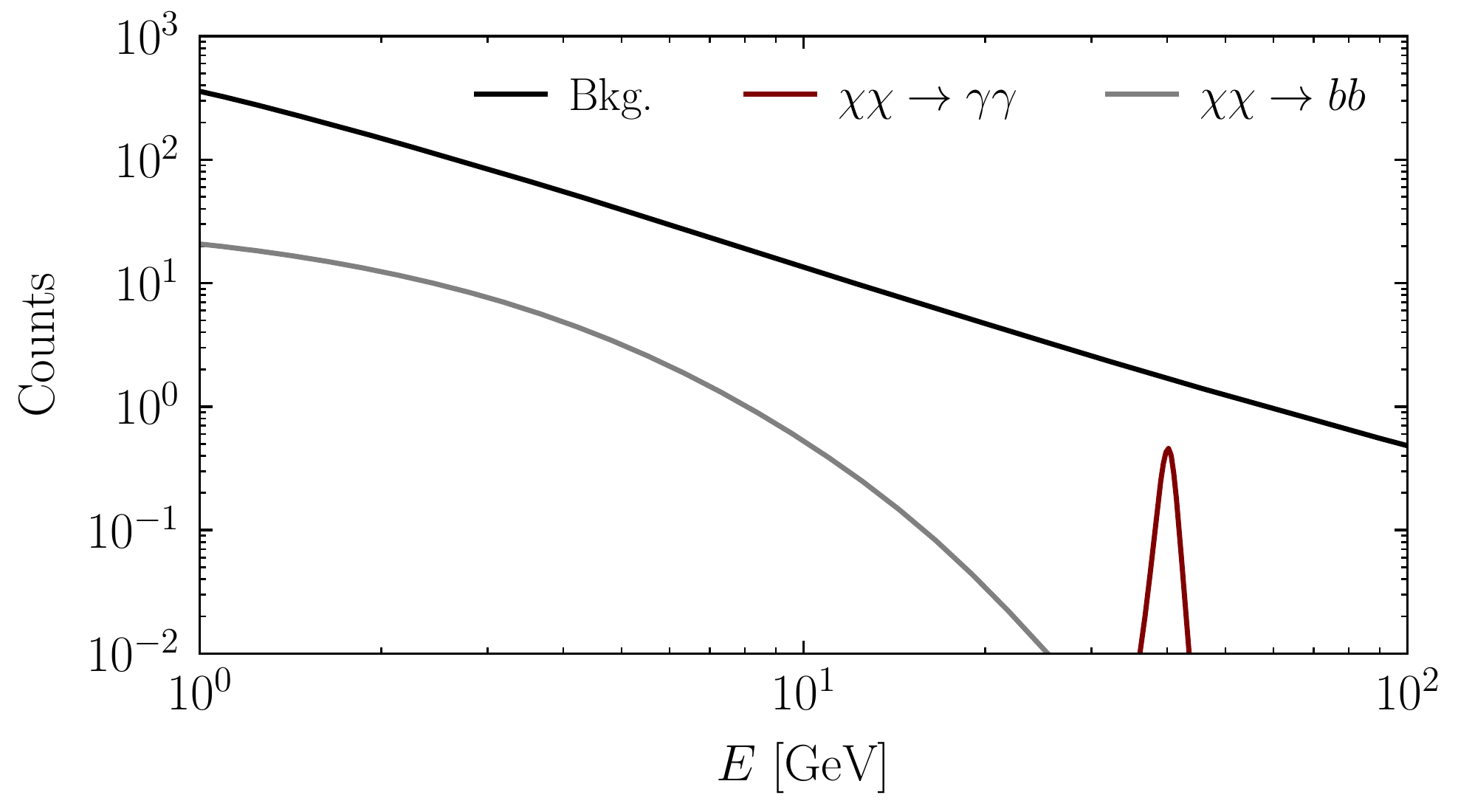}
	\end{center}
	\caption{A comparison of the diffuse background emission in Annulus 1 for EDISP3 data with the line-like signal of annihilation to photons and the               continuum signal of annihilation to $b\bar b$. See text for details.}
	\label{fig:ContinuumComparison}
\end{figure}

In Fig.~\ref{fig:ContinuumComparison}, we illustrate the continuum gamma-ray spectrum for $\langle \sigma_{\chi \chi \to b \bar b} v \rangle = 10^{-26} \, \mathrm{cm}^3/\mathrm{s}$ and the line-like signal convolved with the detector response for $\langle \sigma_{\chi \chi \to \gamma \gamma} v \rangle = 10^{-28} \, \mathrm{cm}^3/\mathrm{s}$. We overlay the expected background emission within this ROI on top of the expected signals. To estimate the sensitivity we perform a spectral fit of the continuum emission to the mock data, also including a continuum background component given precisely by the diffuse emission spectral template but with a free normalization parameter that is treated as a nuisance parameter. Note that this analysis is idealized in that it does not account for the possibility of mismodeling. In the innermost ring, we estimate the ratio $\langle \sigma_{\chi \chi \to \gamma \gamma} v \rangle^{\rm 95} / \langle \sigma_{\chi \chi \to b \bar b} v \rangle^{\rm 95} \approx 4.6 \times 10^{-2}$, with a similar value found in the joint likelihood across all rings ($\langle \sigma_{\chi \chi \to \gamma \gamma} v \rangle^{\rm 95} / \langle \sigma_{\chi \chi \to b \bar b} v \rangle^{\rm 95} \approx 4.2 \times 10^{-2}$).

\section{Unmasked Analysis}
\label{app:Unmasked_Analysis}
\begin{figure*}[!htb]
	\begin{center}
		\includegraphics[width=0.99\textwidth]{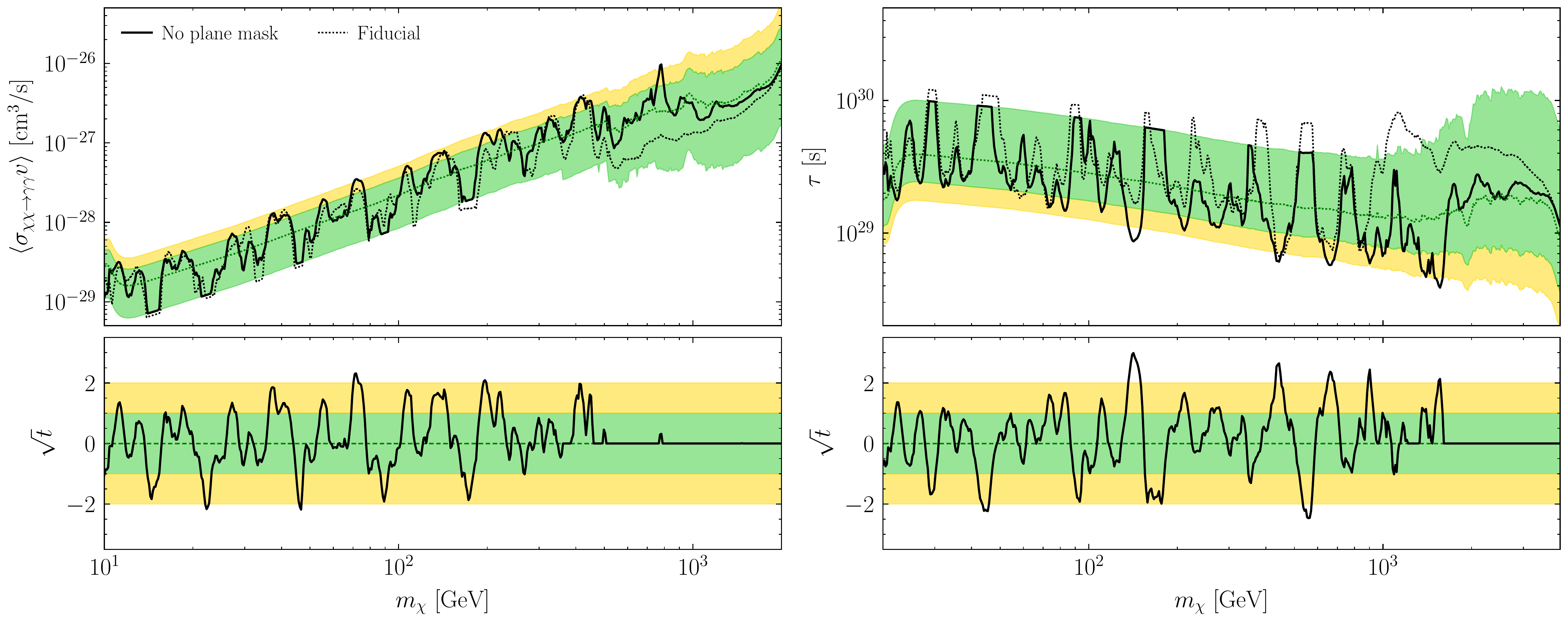}
	\end{center}
	\caption{As in Fig.~\ref{fig:Results_NFW_Limits}, but without plane masking. The 95$^\mathrm{th}$ percentile upper limit obtained using the fiducial masked analysis is indicated with a black dotted line.}
	\label{fig:Unmasked_Results_NFW_Limits}
\end{figure*}

In this Appendix, we consider the impact of our plane-masking procedure by repeating our analysis with no masking applied. The results are presented for the NFW annihilation and decay searches in Fig.~\ref{fig:Unmasked_Results_NFW_Limits}, which achieve generally weaker sensitivity to line-like signals than our fiducial analysis. Of possible interest is the moderate significance detection at $m_\chi \approx 140\, \mathrm{GeV}$, which does not appear in the masked analysis, suggesting that this feature is associated with Galactic plane emission rather than decaying DM.

\section{Analysis Energy Range}
\label{app:energy_range}
\begin{figure*}[!htb]
	\begin{center}
		\includegraphics[width=0.49\textwidth]{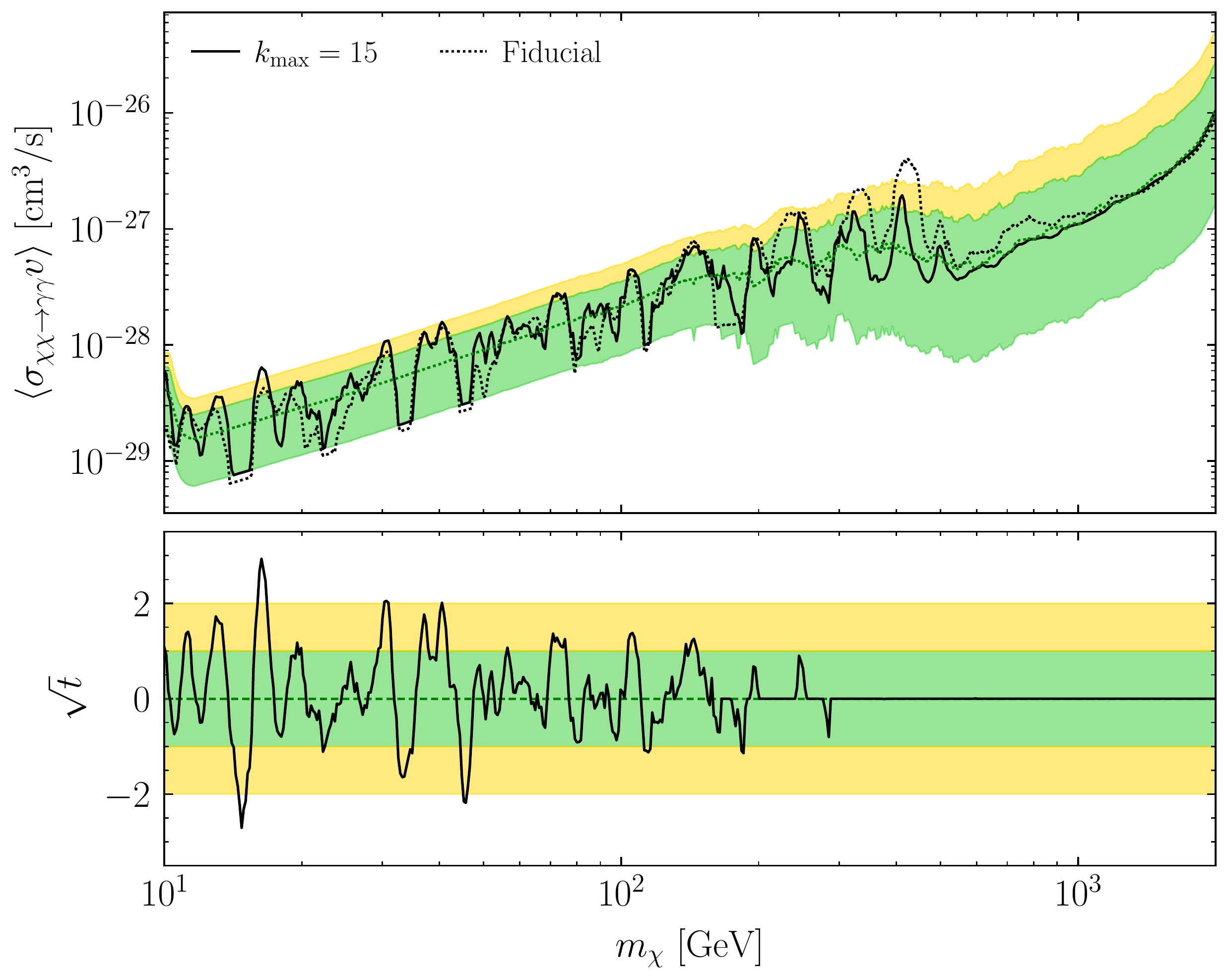}
		\includegraphics[width=0.49\textwidth]{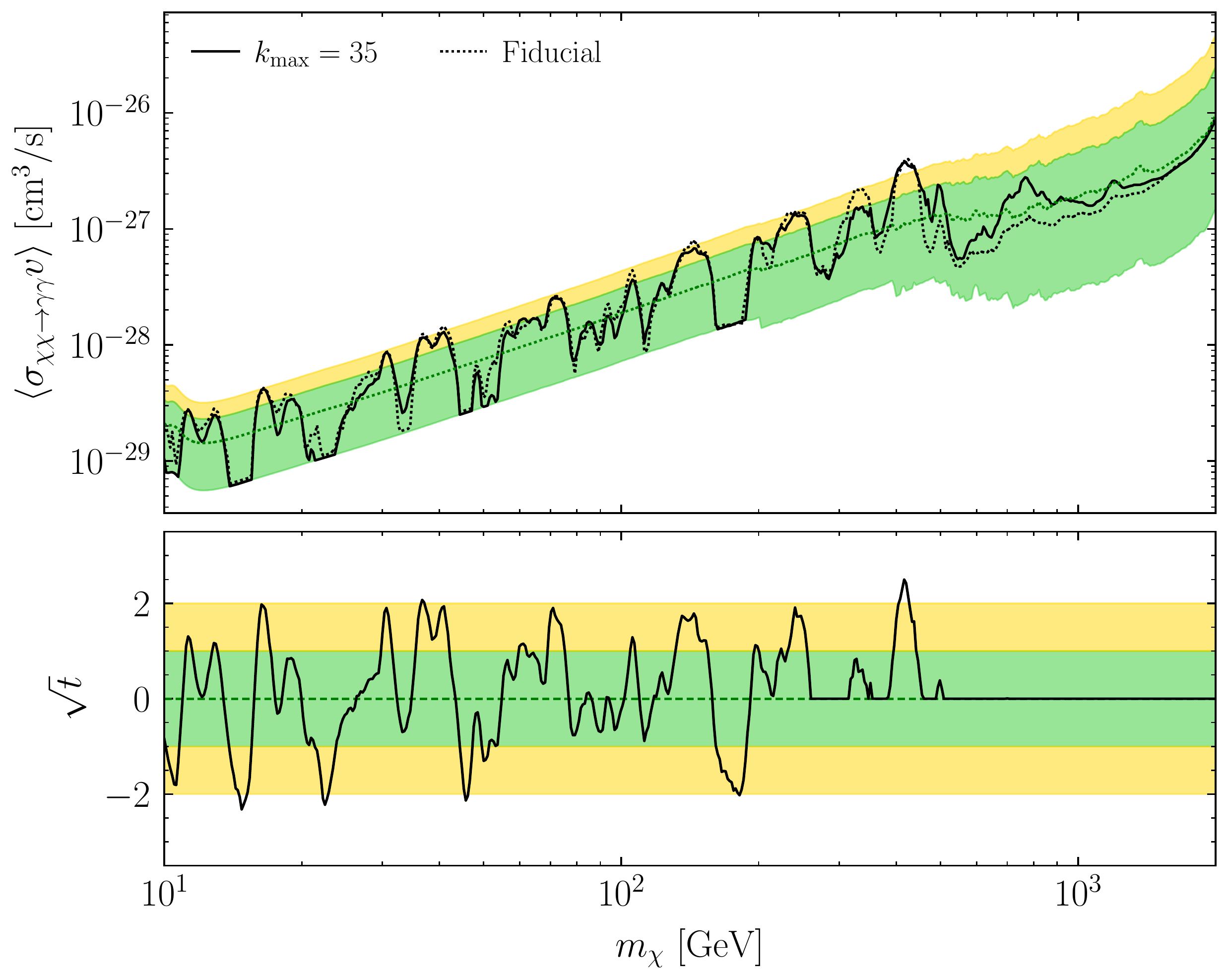}
	\end{center}
	\caption{As in the left panel of Fig.~\ref{fig:Results_NFW_Limits}, but with the analysis window narrowed to $k_{\rm max} = 15$ and widened to $k_{\rm max} = 35$. The limits obtained  with the fiducial window  $k_{\rm max} = 25$ are indicated with a dotted black line.}
	\label{fig:EnergyWindow}
\end{figure*}

Here, we consider the effect of narrowing and enlarging the analysis energy range. In our fiducial analysis we use $k_{\rm max} = 25$ energy bins above and below the bin containing the central line location. In this Appendix, we consider the effect of narrowing the energy range to $k_{\rm max} = 15$ and widening it to $k_{\rm max} = 35$, with results presented in Fig.~\ref{fig:EnergyWindow}. The sensitivities and detection significances are minimally changed by these adjustments to the analysis energy range.

\section{Independent Annulus Results}
\begin{figure*}[!htb]
	\begin{center}
		\includegraphics[width=0.49\textwidth]{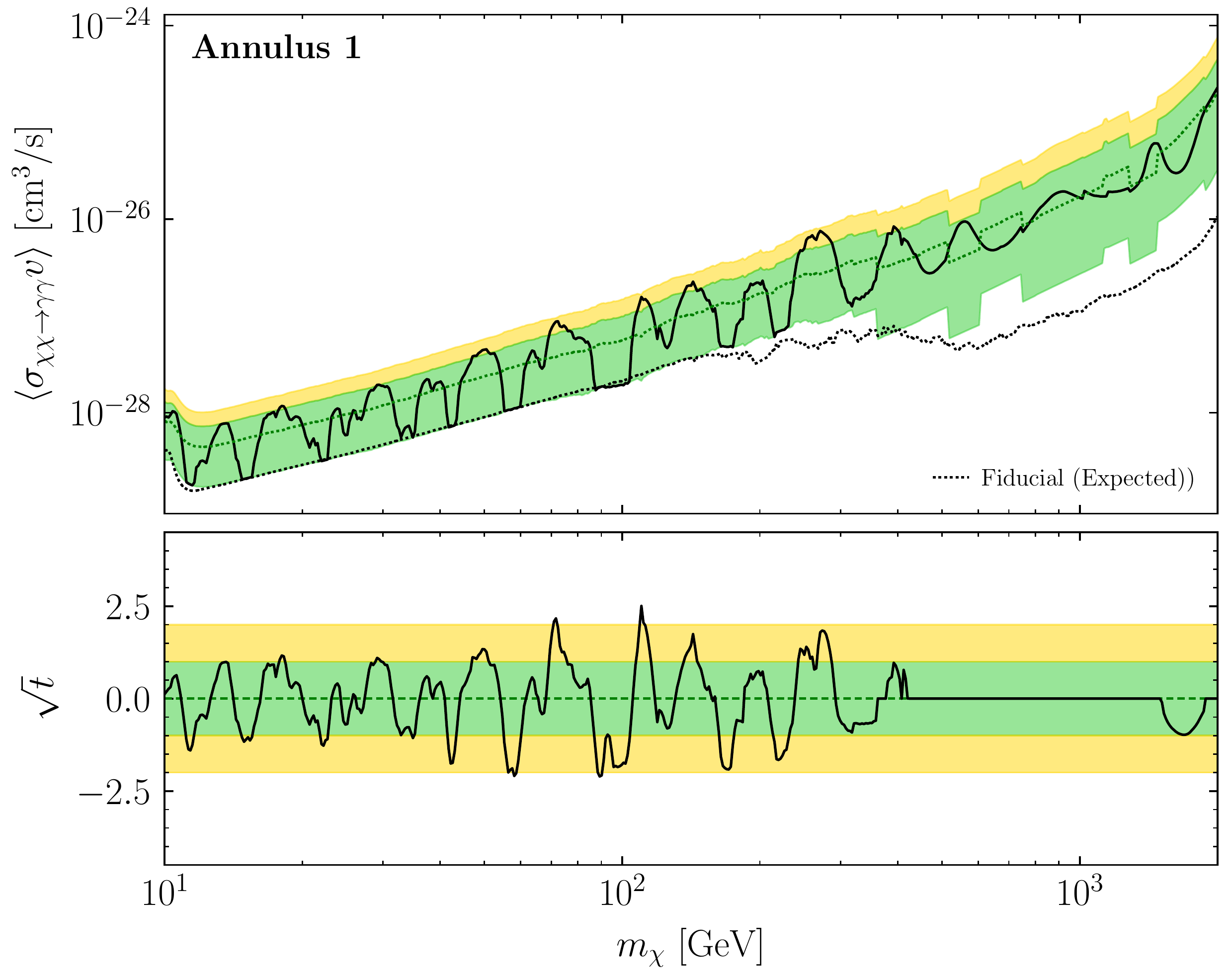}
		\includegraphics[width=0.49\textwidth]{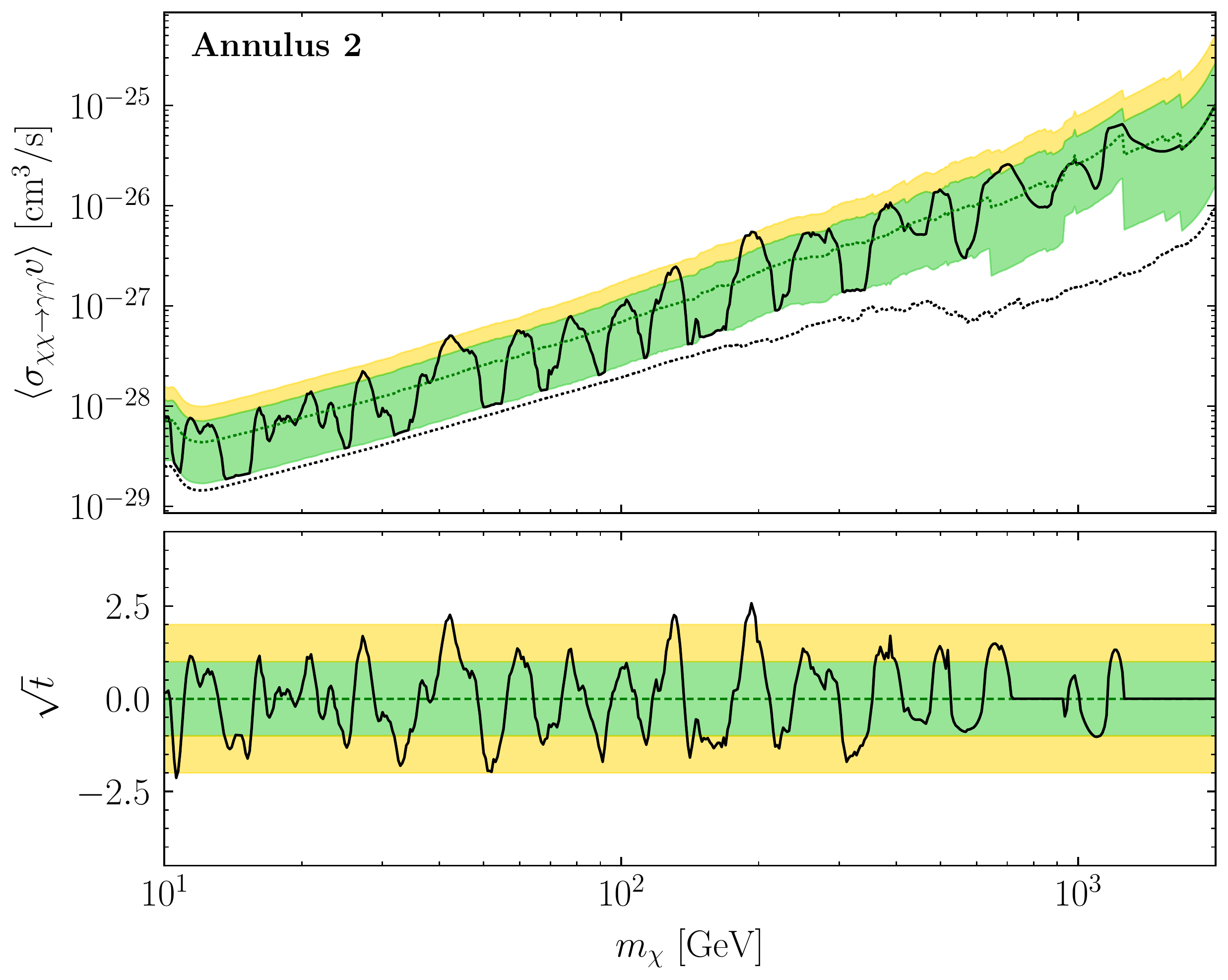}
	\end{center}
	\caption{As in the left panel of Fig.~\ref{fig:Results_NFW_Limits} but for Annulus 1 and Annulus 2. The expected 95$^\mathrm{th}$ percentile limit for the joint analysis over all annuli is indicated by a dotted black line.}
	\label{fig:Annuli12}
\end{figure*}

\begin{figure*}[!htb]
	\begin{center}
		\includegraphics[width=0.49\textwidth]{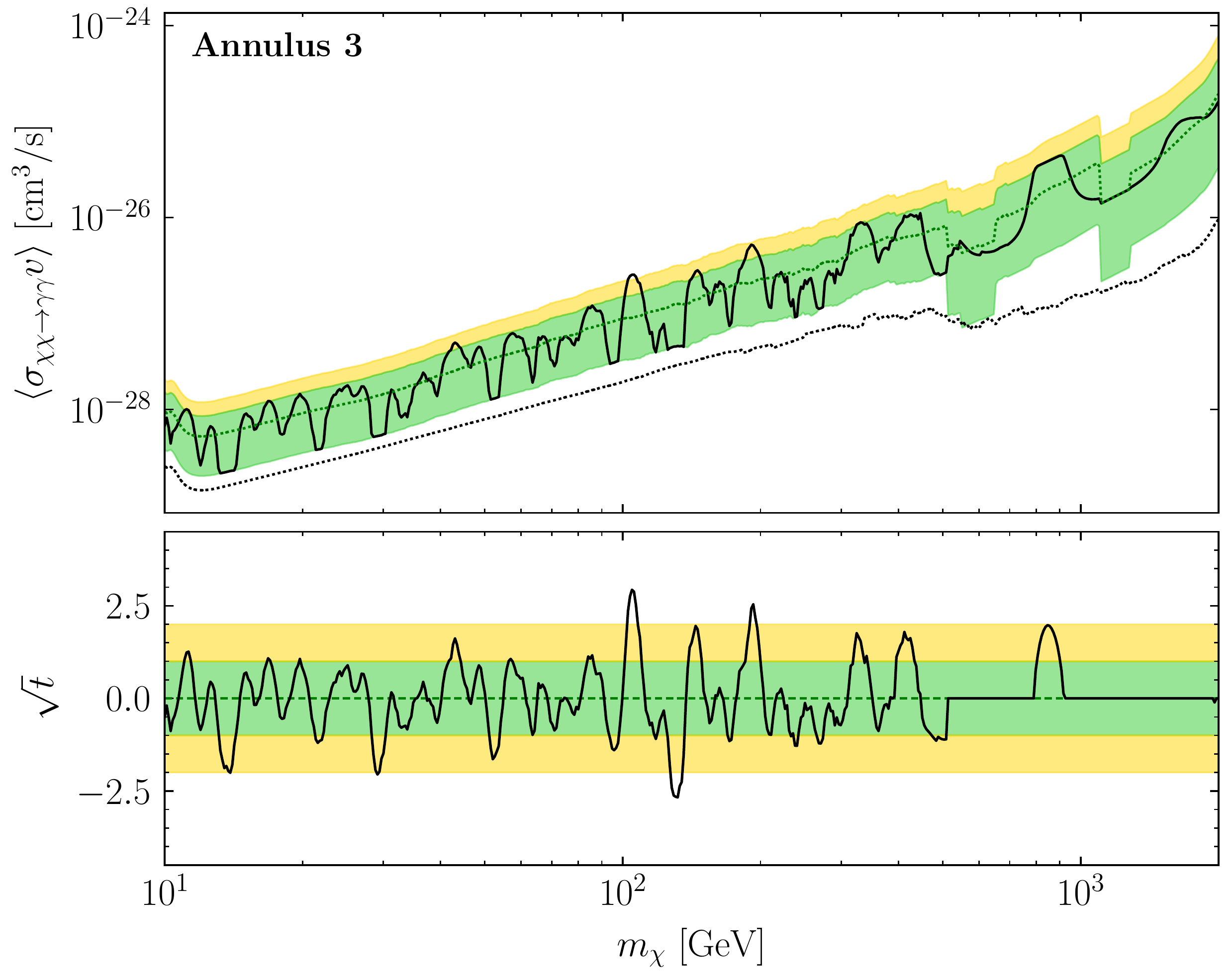}
		\includegraphics[width=0.49\textwidth]{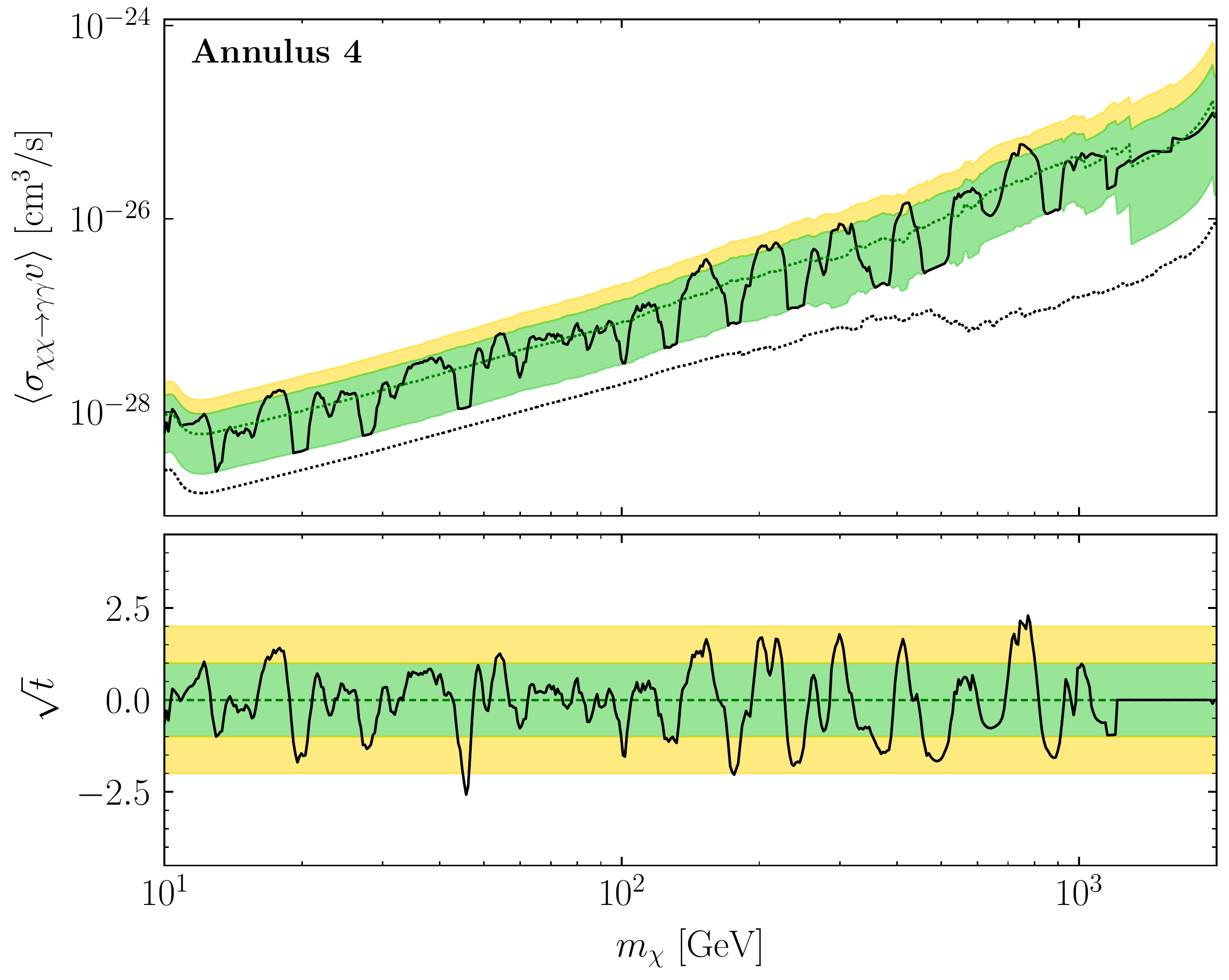}
	\end{center}
	\caption{As in the left panel of Fig.~\ref{fig:Results_NFW_Limits} but for Annulus 3 and Annulus 4.}
	\label{fig:Annuli34}
\end{figure*}

\begin{figure*}[!htb]
	\begin{center}
		\includegraphics[width=0.49\textwidth]{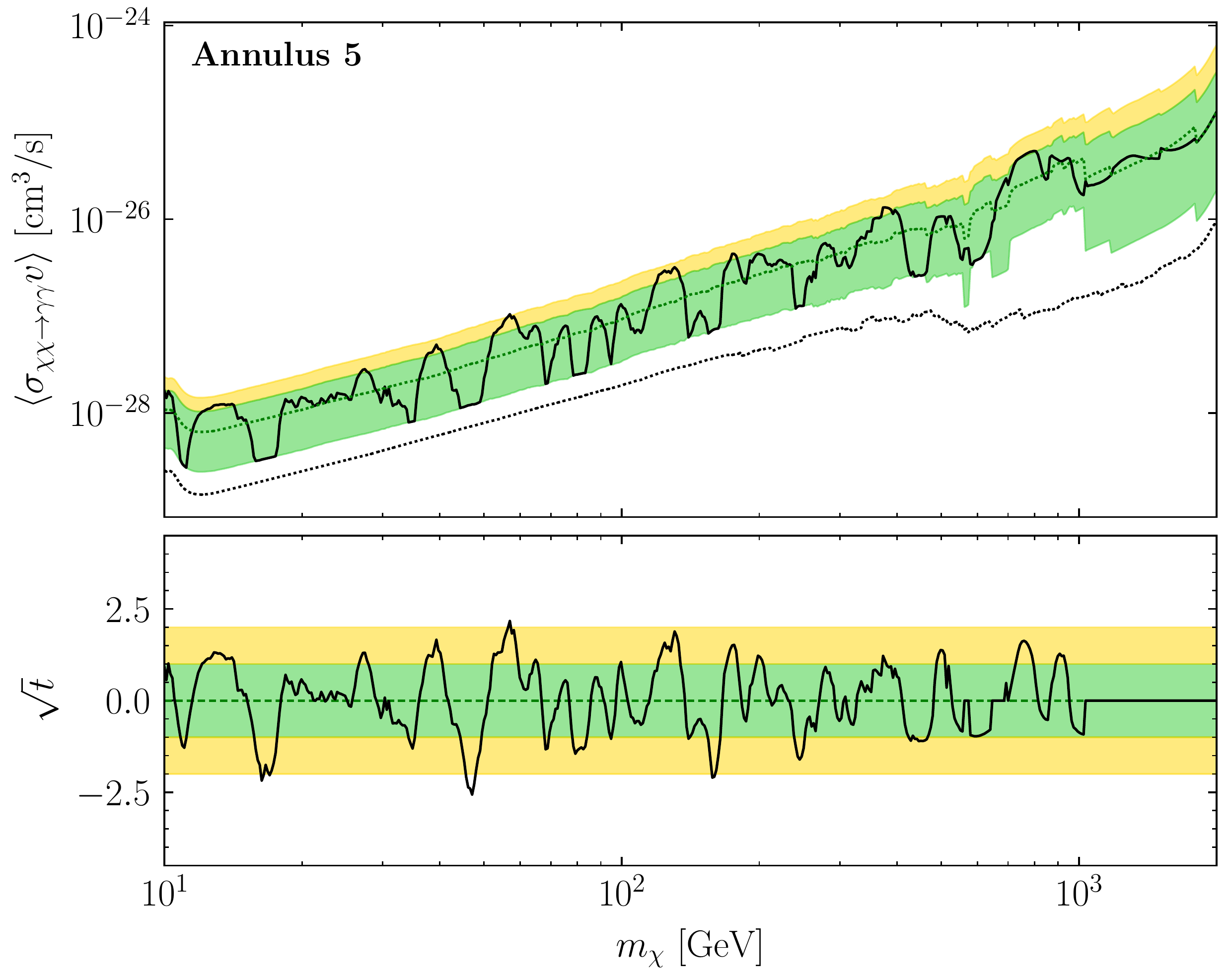}
		\includegraphics[width=0.49\textwidth]{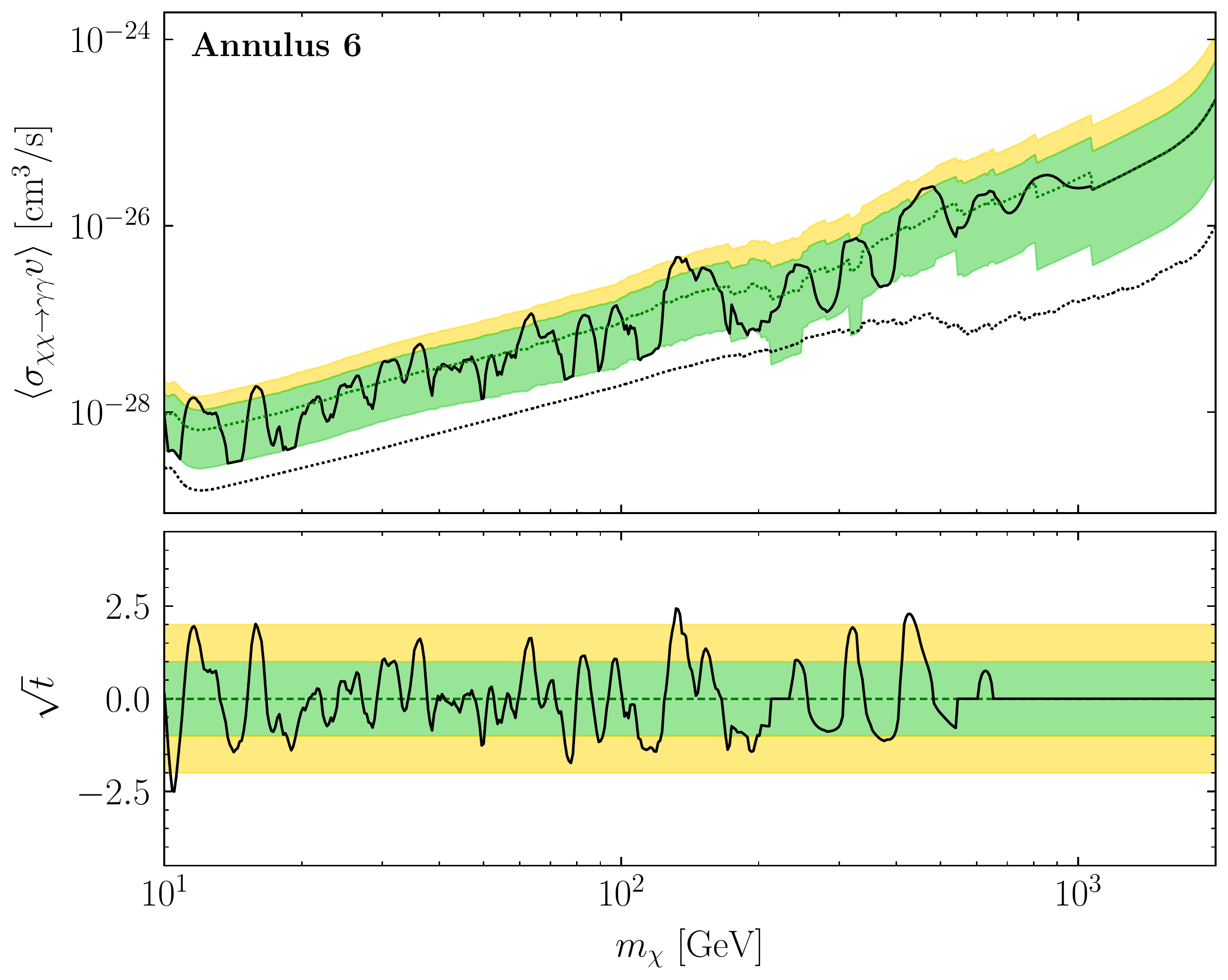}
	\end{center}
	\caption{As in the left panel of Fig.~\ref{fig:Results_NFW_Limits} but for Annulus 5 and Annulus 6.}
	\label{fig:Annuli56}
\end{figure*}

\begin{figure*}[!htb]
	\begin{center}
		\includegraphics[width=0.49\textwidth]{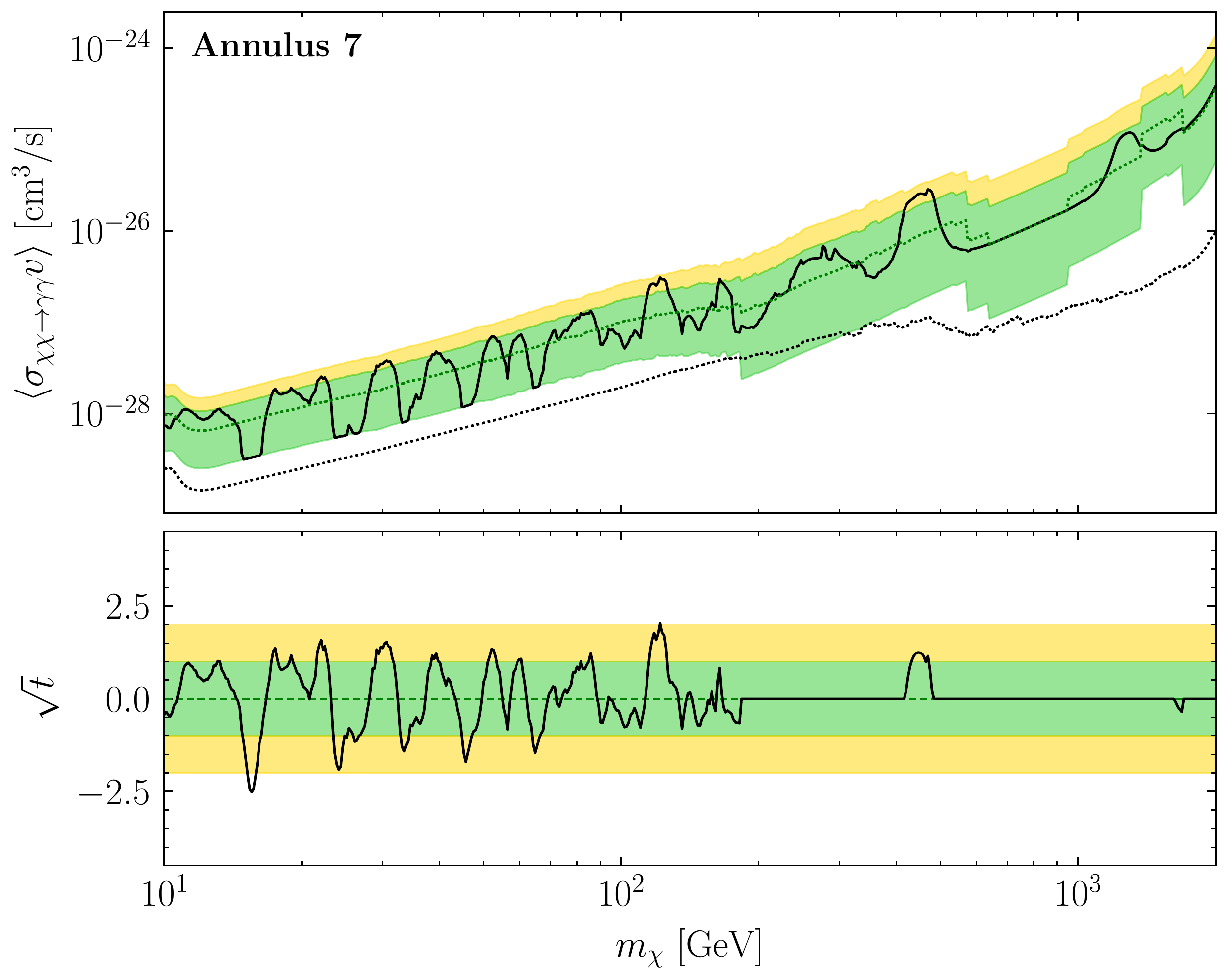}
		\includegraphics[width=0.49\textwidth]{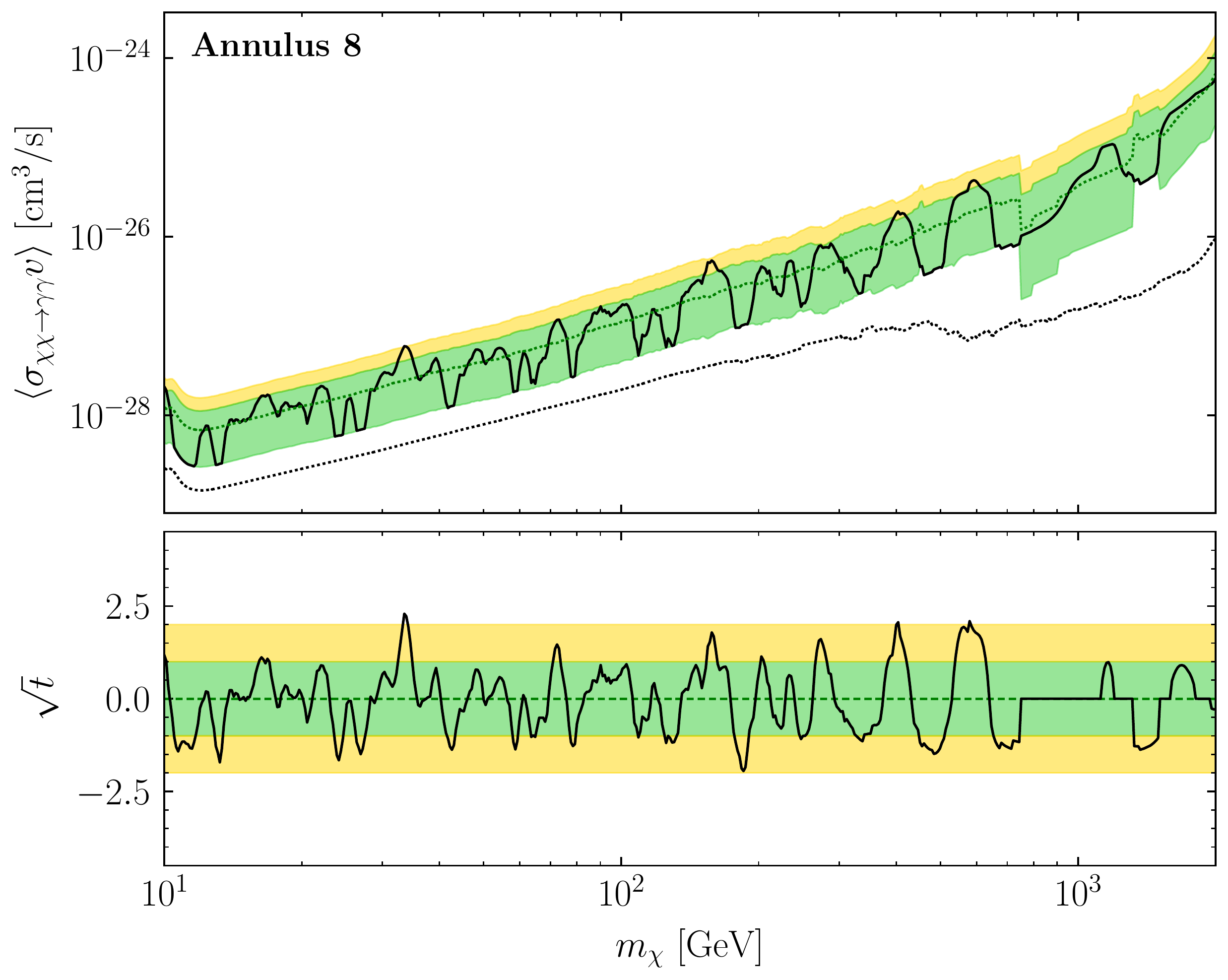}
	\end{center}
	\caption{As in the left panel of Fig.~\ref{fig:Results_NFW_Limits} but for Annulus 7 and Annulus 8.}
	\label{fig:Annuli78}
\end{figure*}

\begin{figure*}[!htb]
	\begin{center}
		\includegraphics[width=0.49\textwidth]{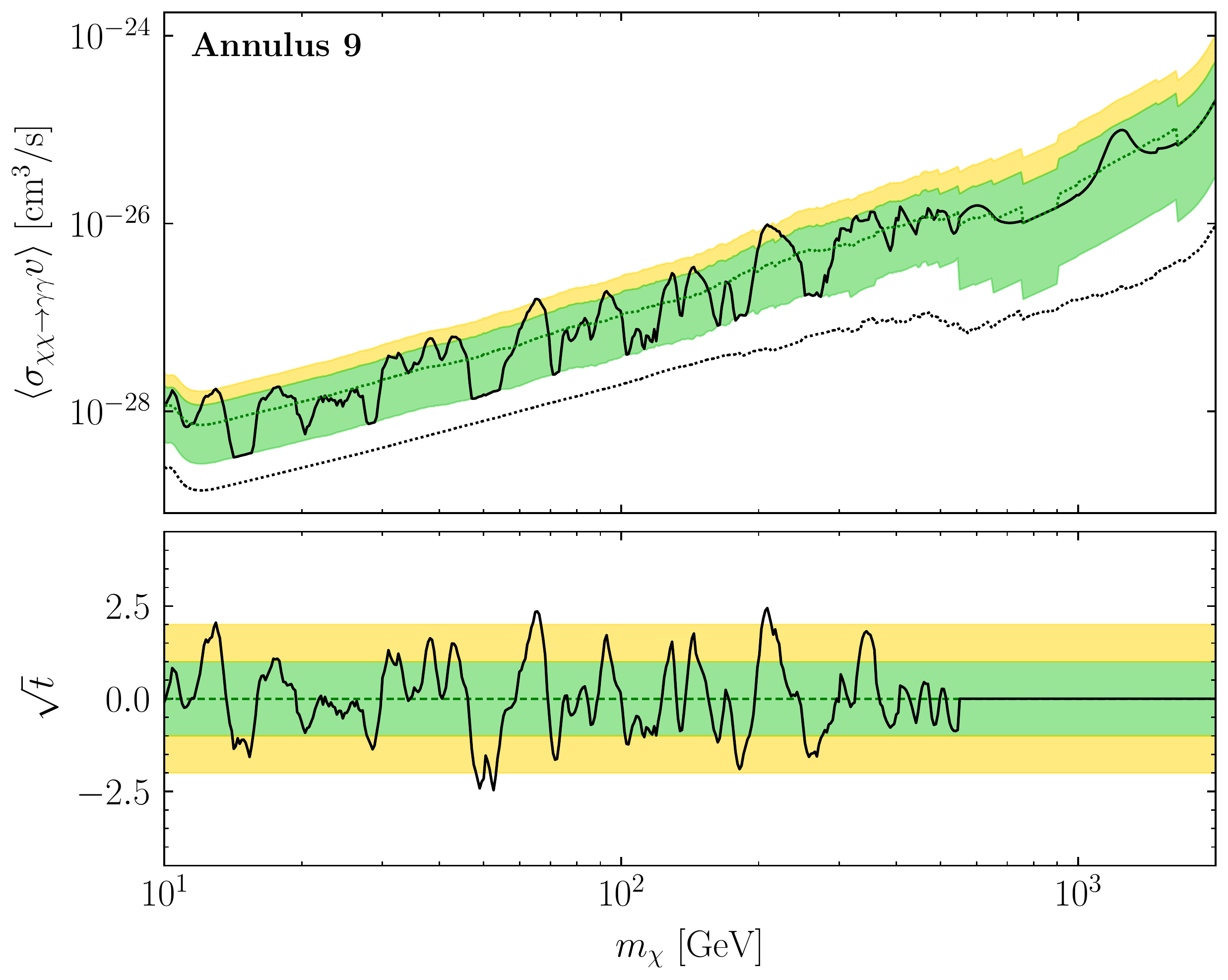}
		\includegraphics[width=0.49\textwidth]{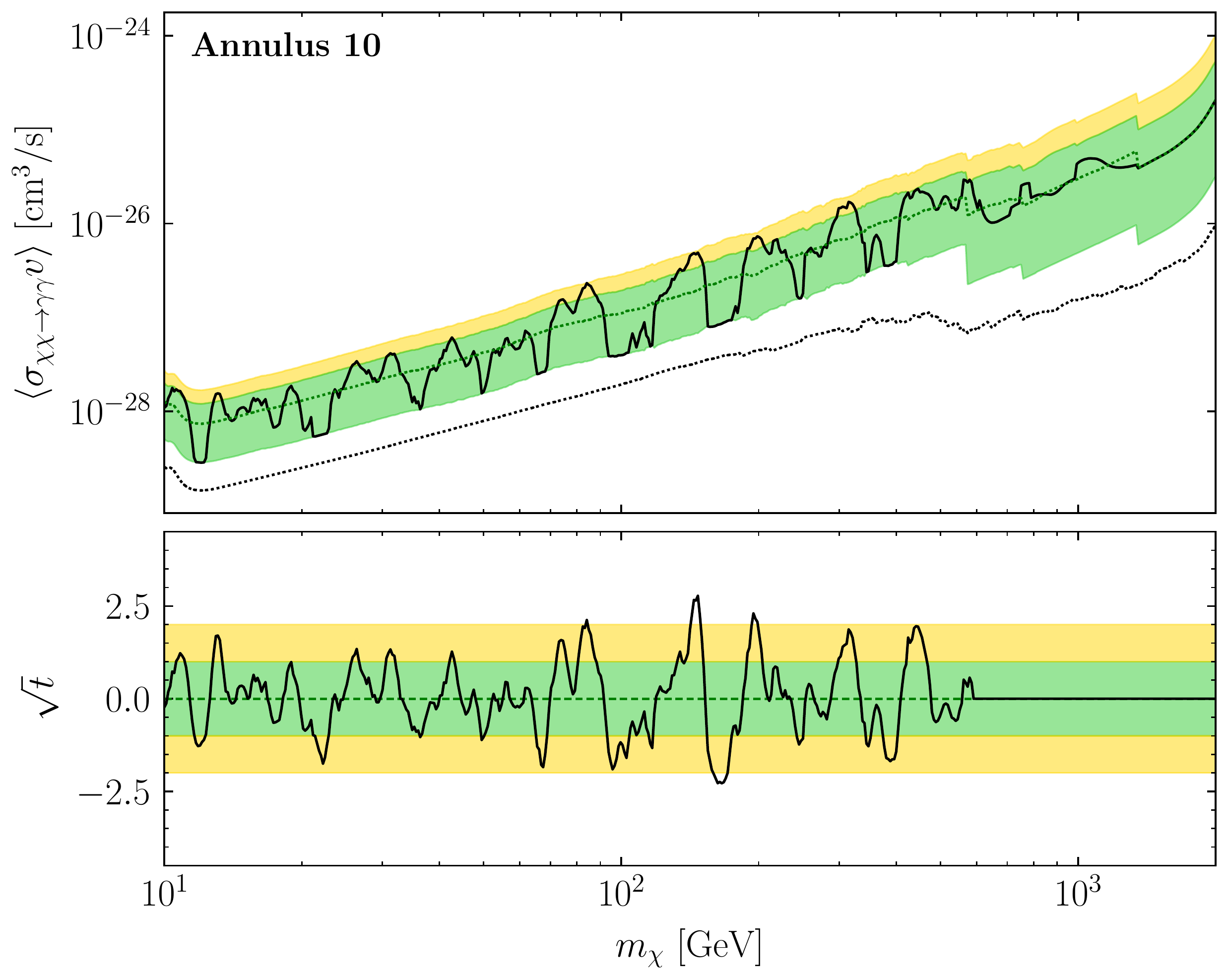}
	\end{center}
	\caption{As in the left panel of Fig.~\ref{fig:Results_NFW_Limits} but for Annulus 9 and Annulus 10.}
	\label{fig:Annuli910}
\end{figure*}

\begin{figure*}[!htb]
	\begin{center}
		\includegraphics[width=0.49\textwidth]{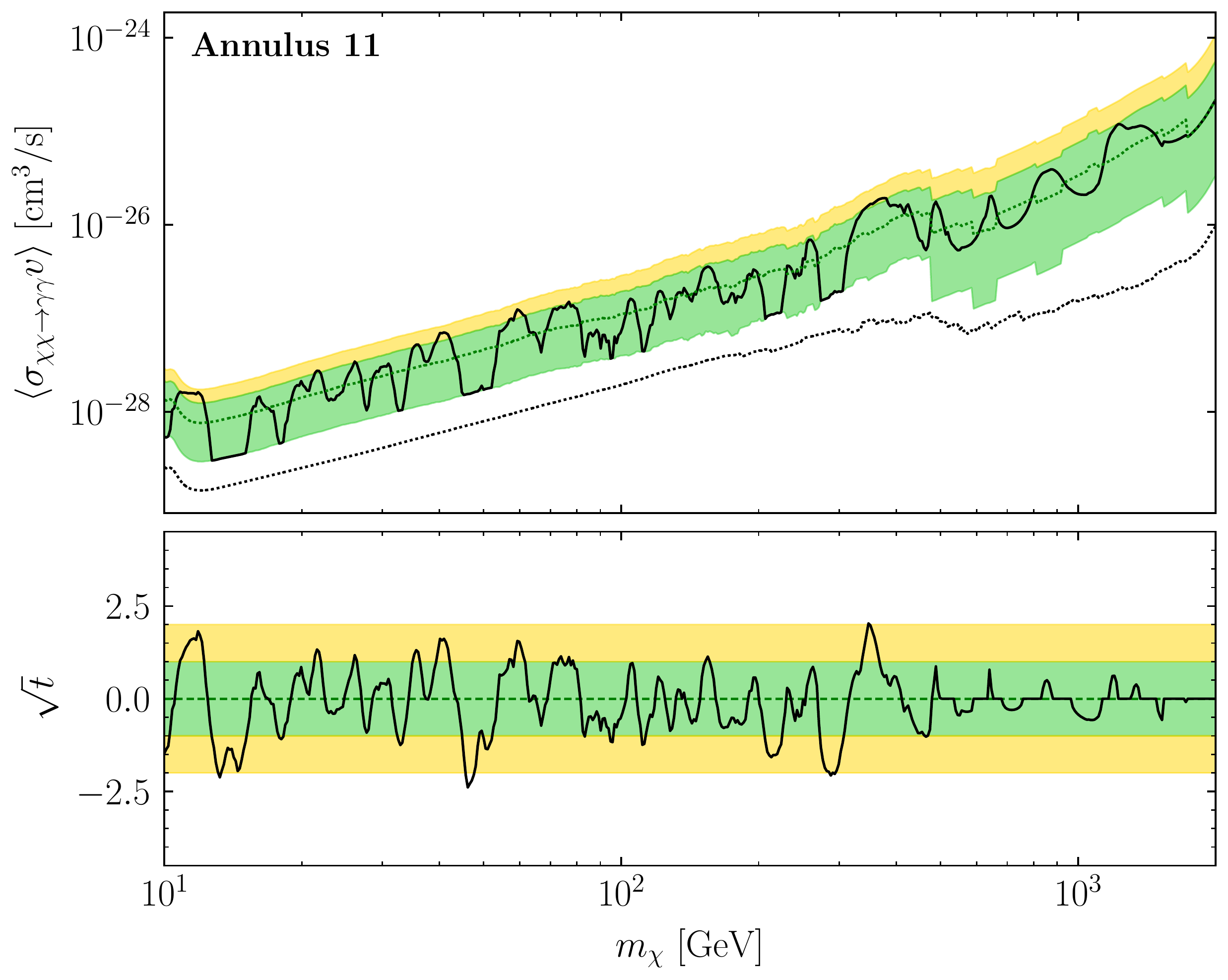}
		\includegraphics[width=0.49\textwidth]{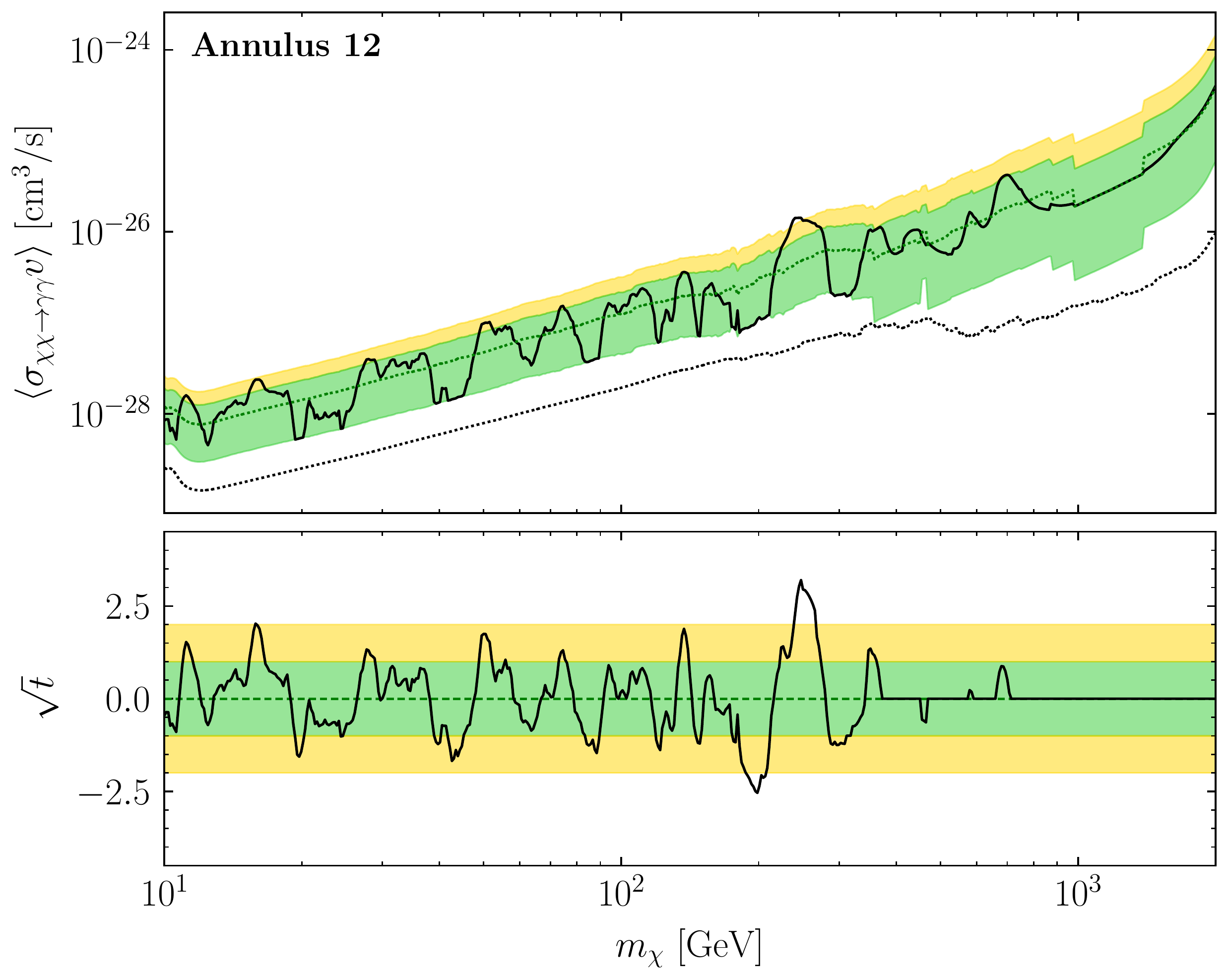}
	\end{center}
	\caption{As in the left panel of Fig.~\ref{fig:Results_NFW_Limits} but for Annulus 11 and Annulus 12.}
	\label{fig:Annuli1112}
\end{figure*}

\begin{figure*}[!htb]
	\begin{center}
		\includegraphics[width=0.49\textwidth]{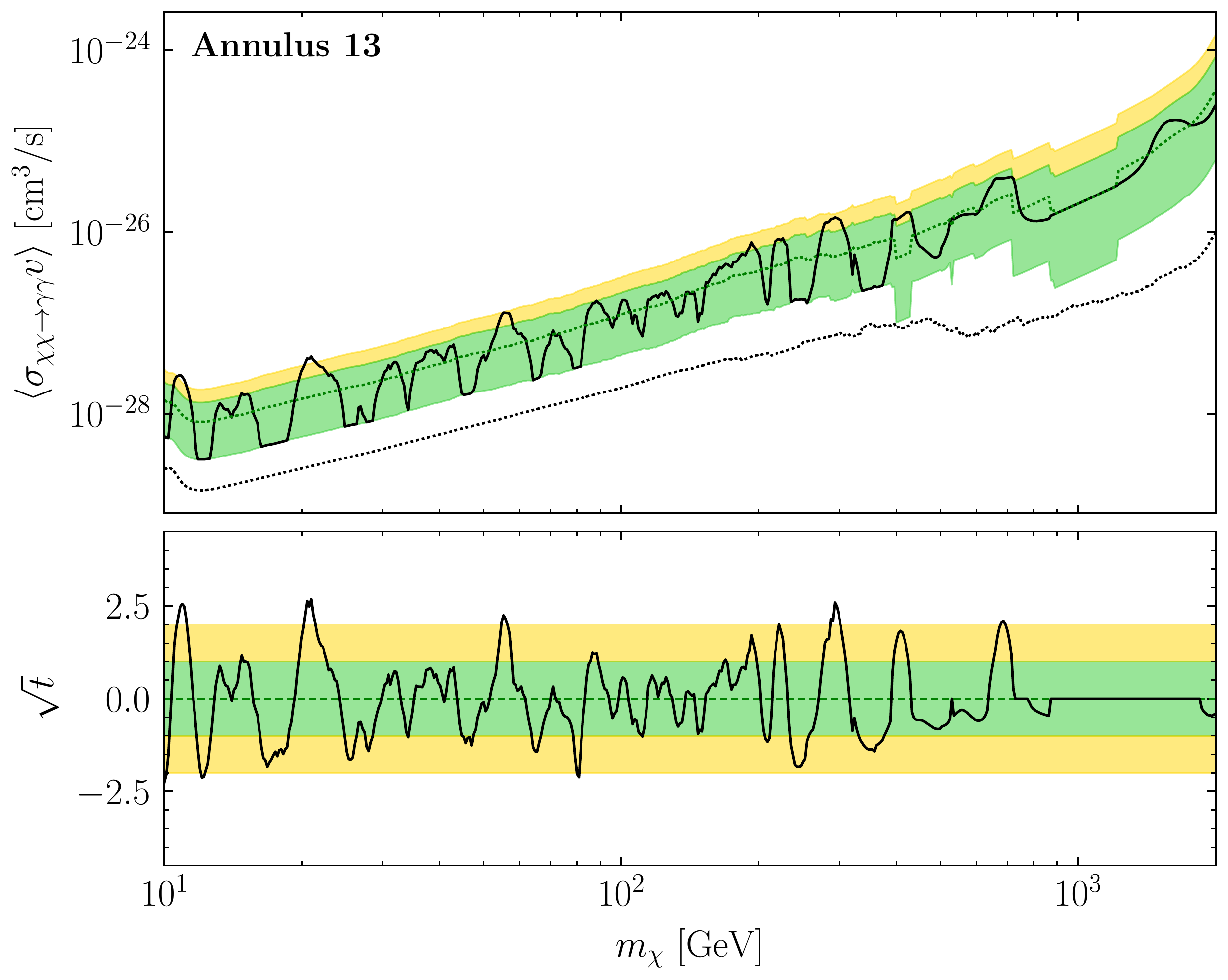}
		\includegraphics[width=0.49\textwidth]{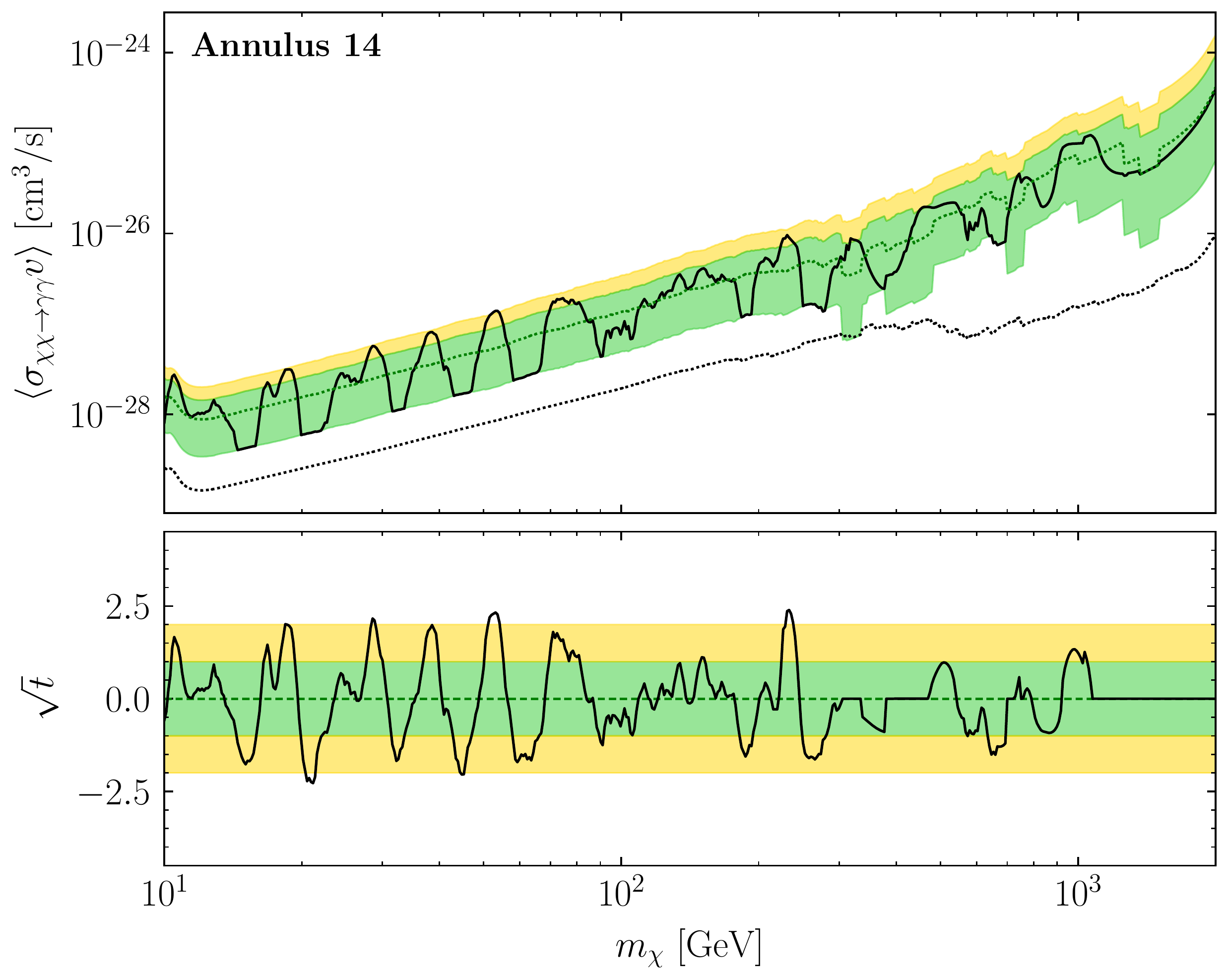}
	\end{center}
	\caption{As in the left panel of Fig.~\ref{fig:Results_NFW_Limits} but for Annulus 13 and Annulus 14.}
	\label{fig:Annuli1314}
\end{figure*}

\begin{figure*}[!htb]
	\begin{center}
		\includegraphics[width=0.49\textwidth]{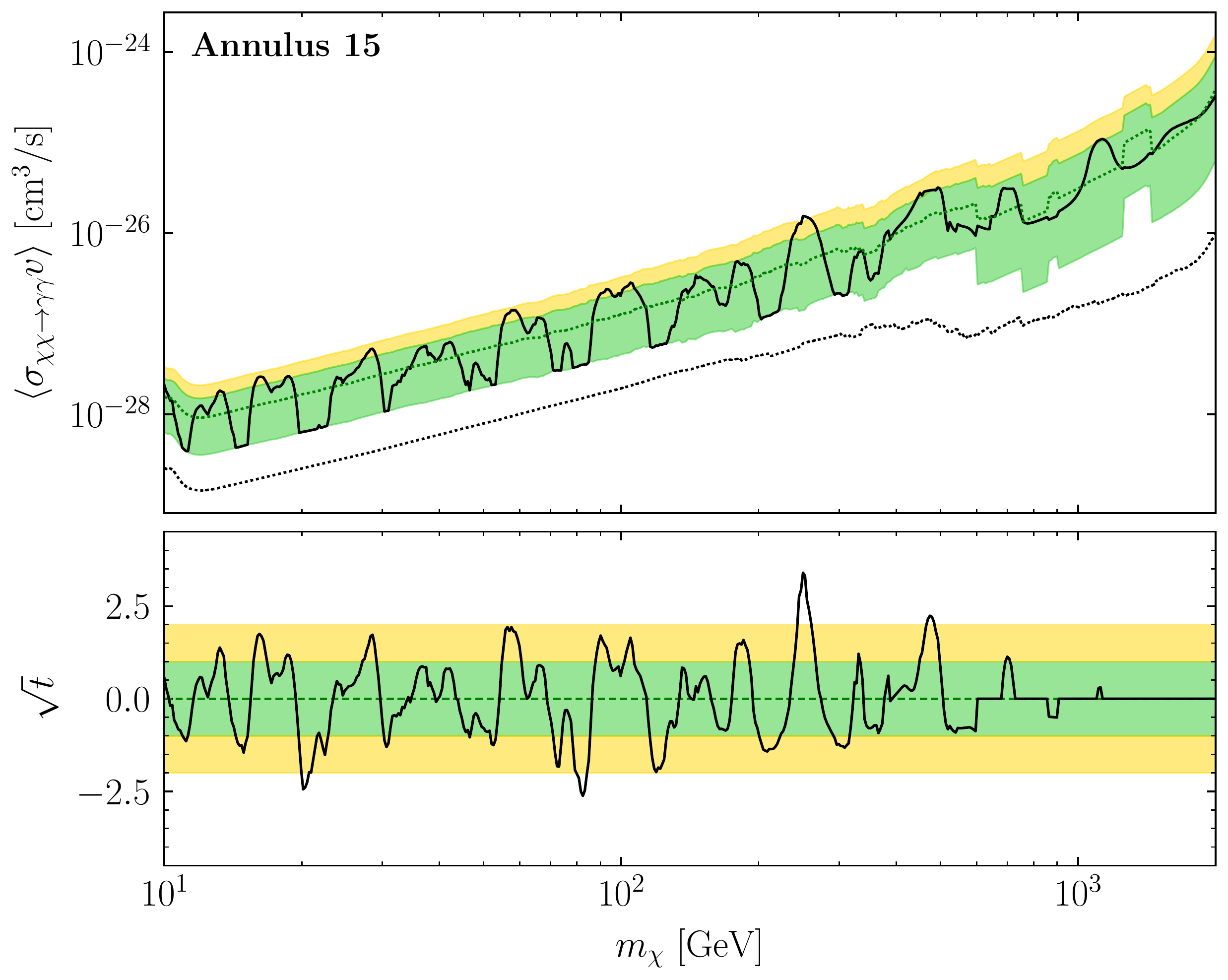}
		\includegraphics[width=0.49\textwidth]{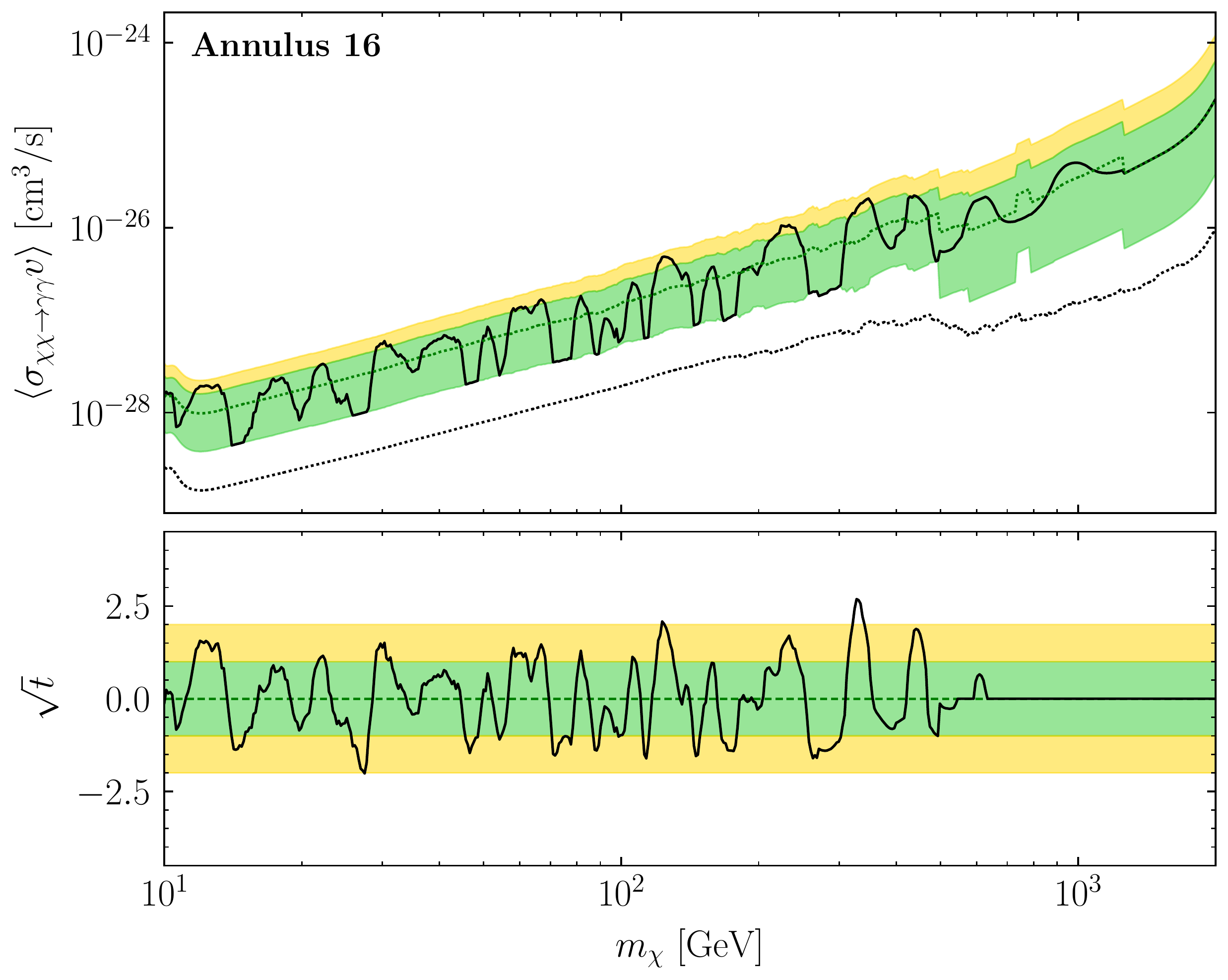}
	\end{center}
	\caption{As in the left panel of Fig.~\ref{fig:Results_NFW_Limits} but for Annulus 15 and Annulus 16.}
	\label{fig:Annuli1516}
\end{figure*}

\begin{figure*}[!htb]
	\begin{center}
		\includegraphics[width=0.49\textwidth]{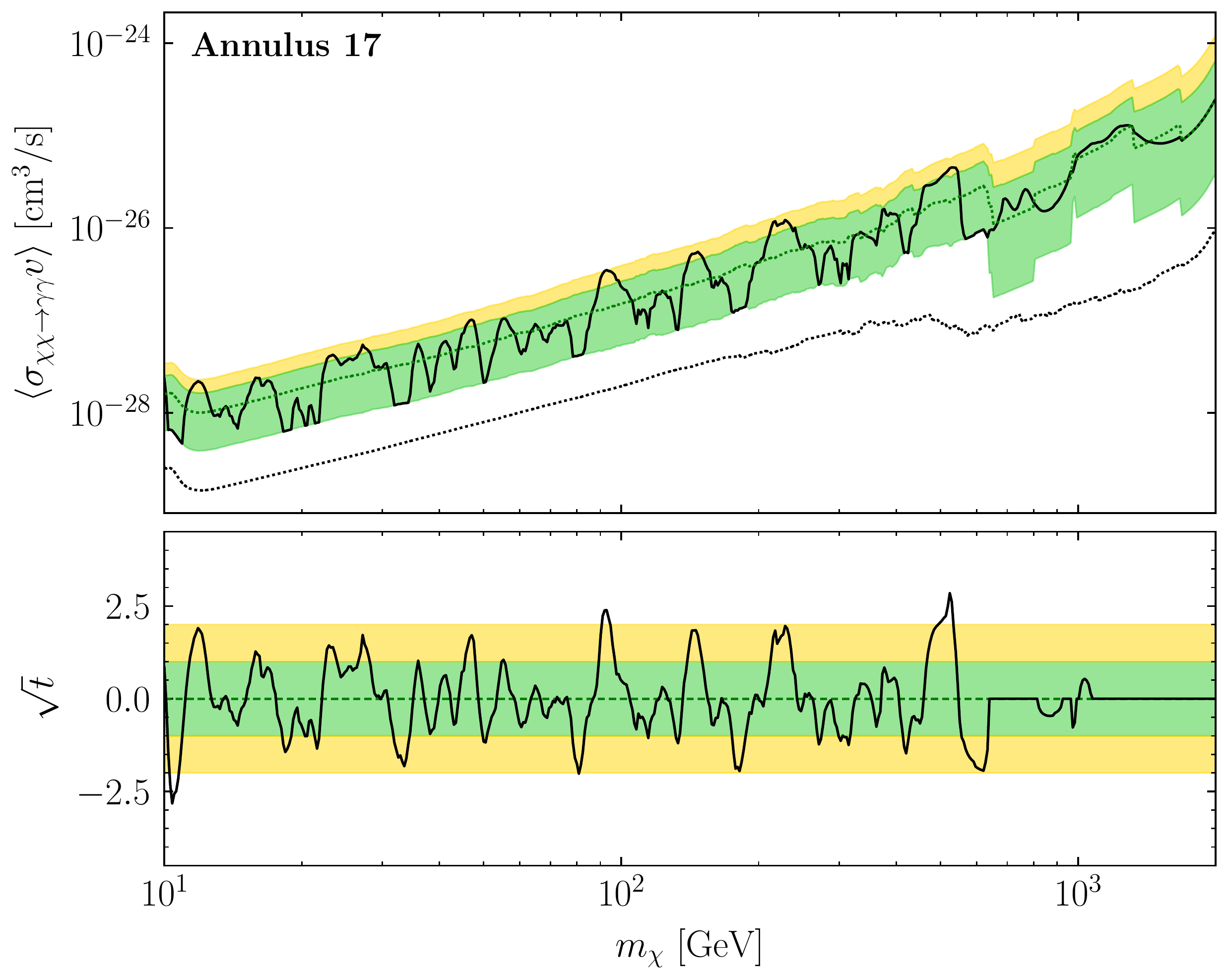}
		\includegraphics[width=0.49\textwidth]{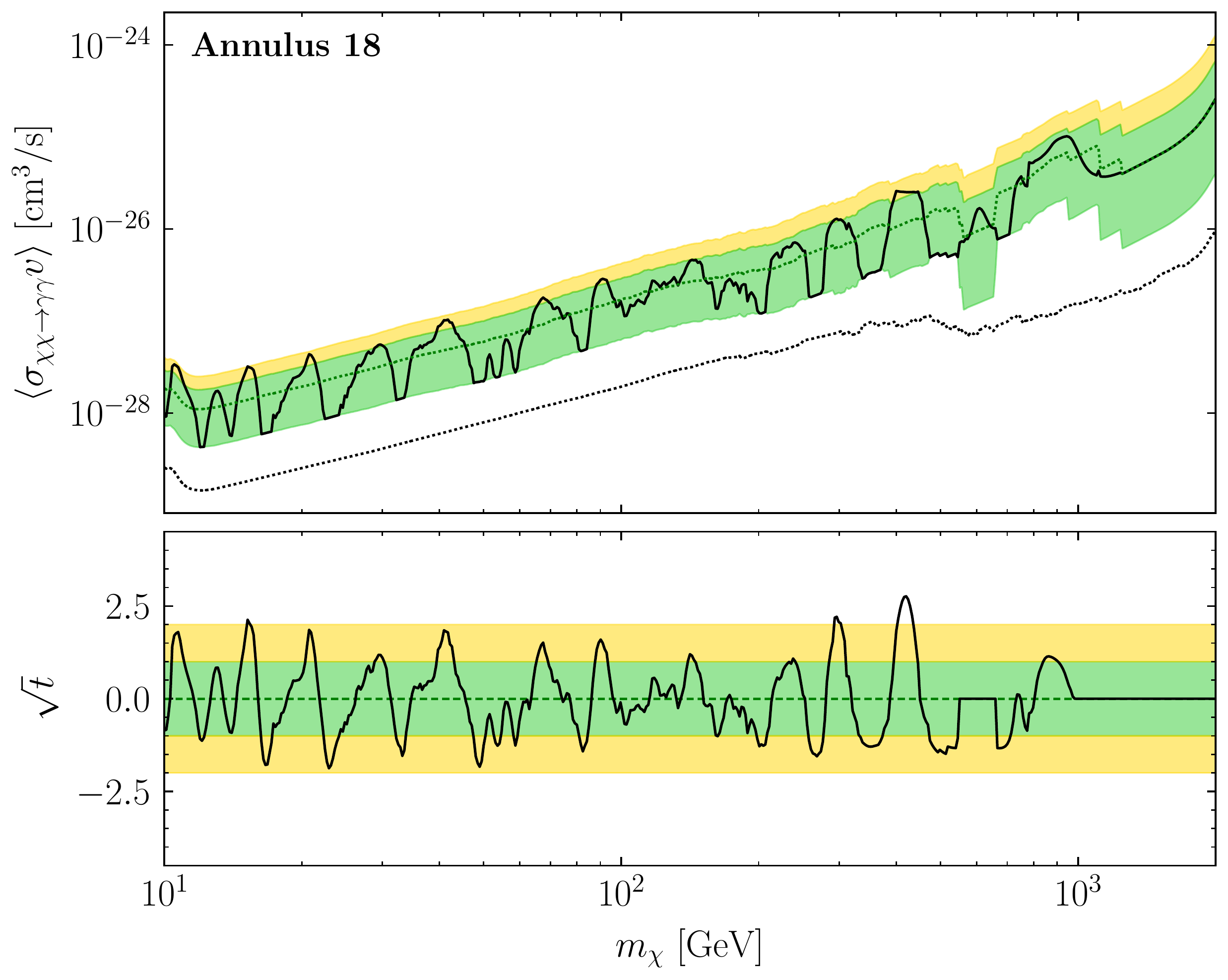}
	\end{center}
	\caption{As in the left panel of Fig.~\ref{fig:Results_NFW_Limits} but for Annulus 17 and Annulus 18.}
	\label{fig:Annuli1718}
\end{figure*}

\begin{figure*}[!htb]
	\begin{center}
		\includegraphics[width=0.49\textwidth]{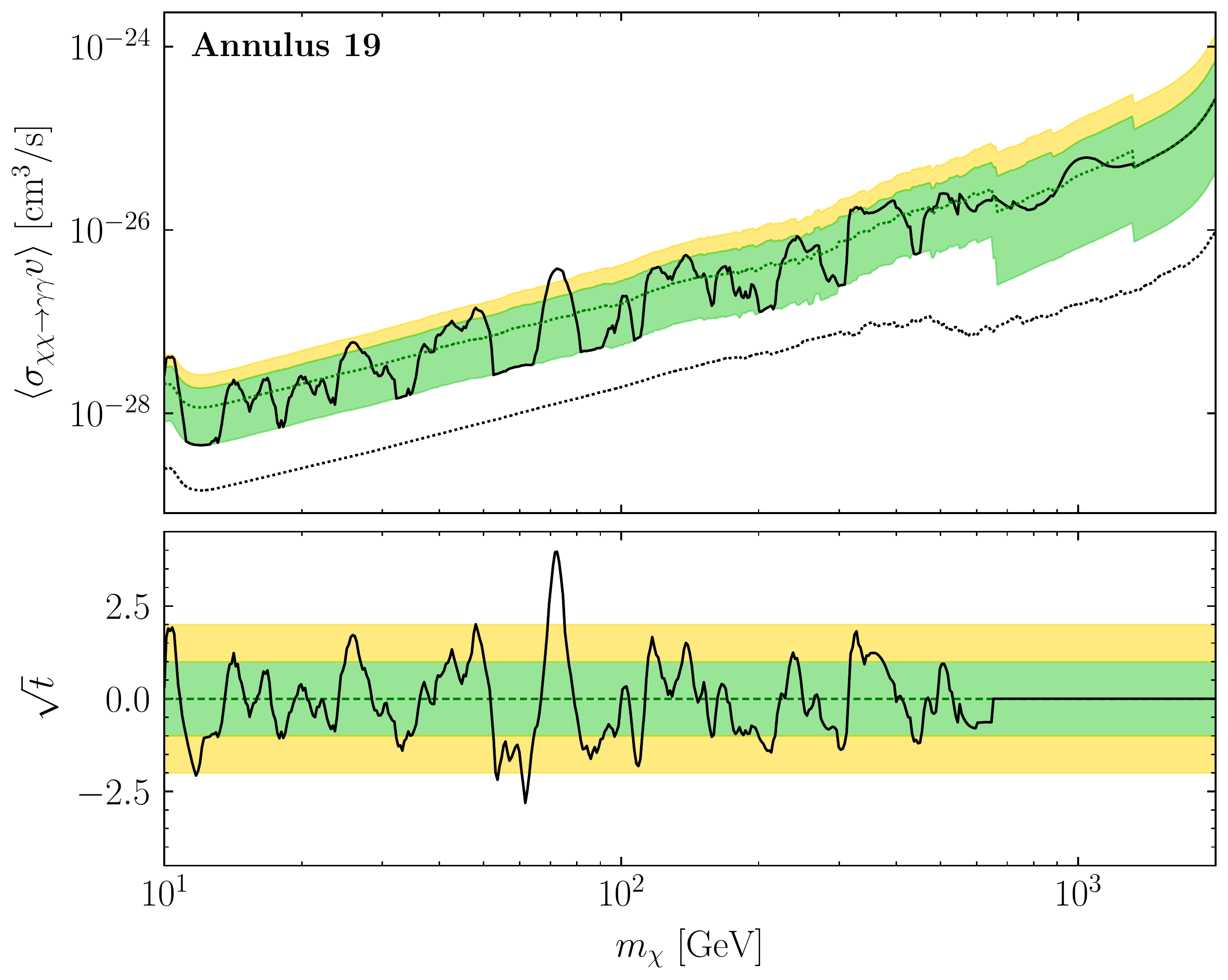}
		\includegraphics[width=0.49\textwidth]{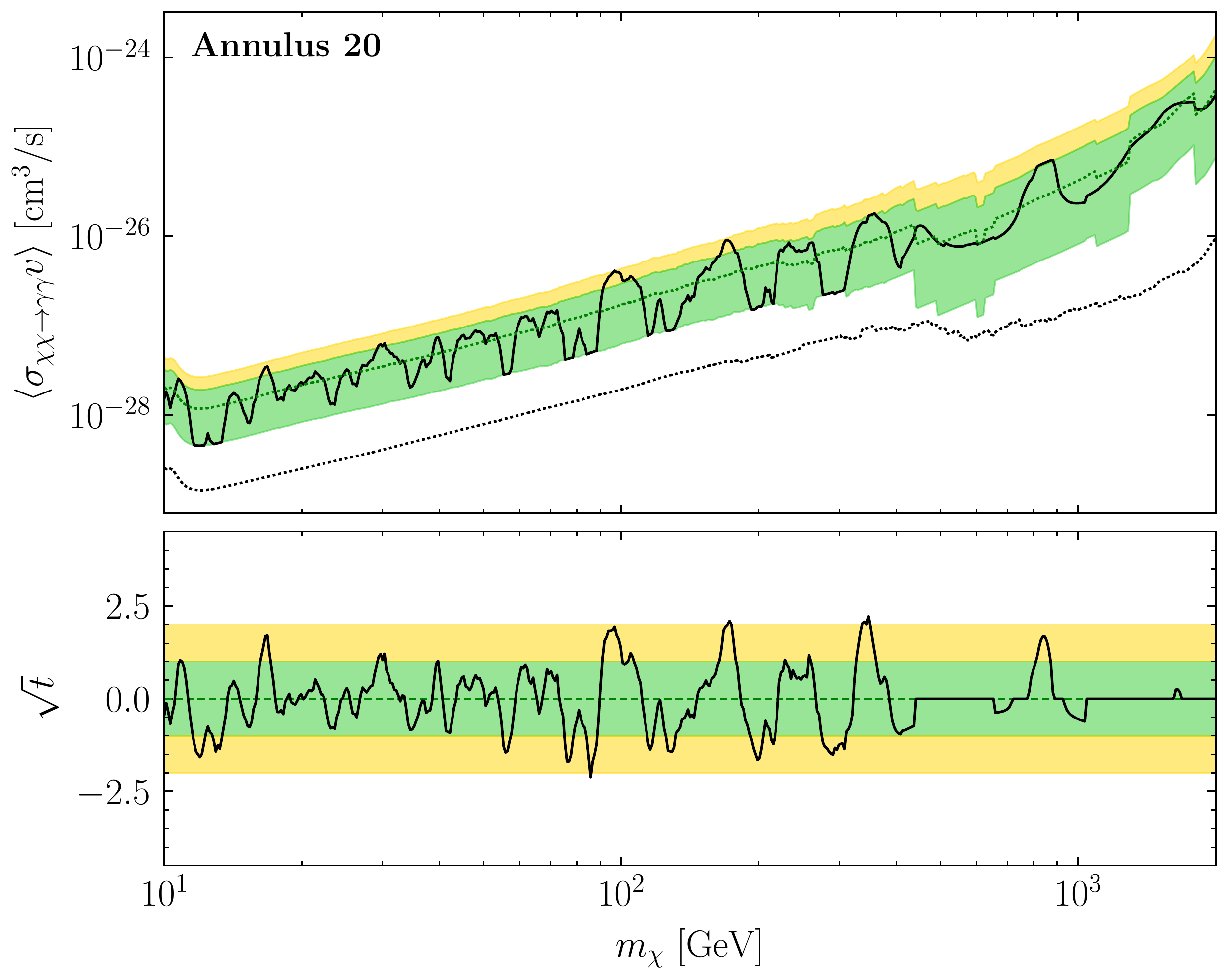}
	\end{center}
	\caption{As in the left panel of Fig.~\ref{fig:Results_NFW_Limits} but for Annulus 19 and Annulus 20.}
	\label{fig:Annuli1920}
\end{figure*}

\begin{figure*}[!htb]
	\begin{center}
		\includegraphics[width=0.49\textwidth]{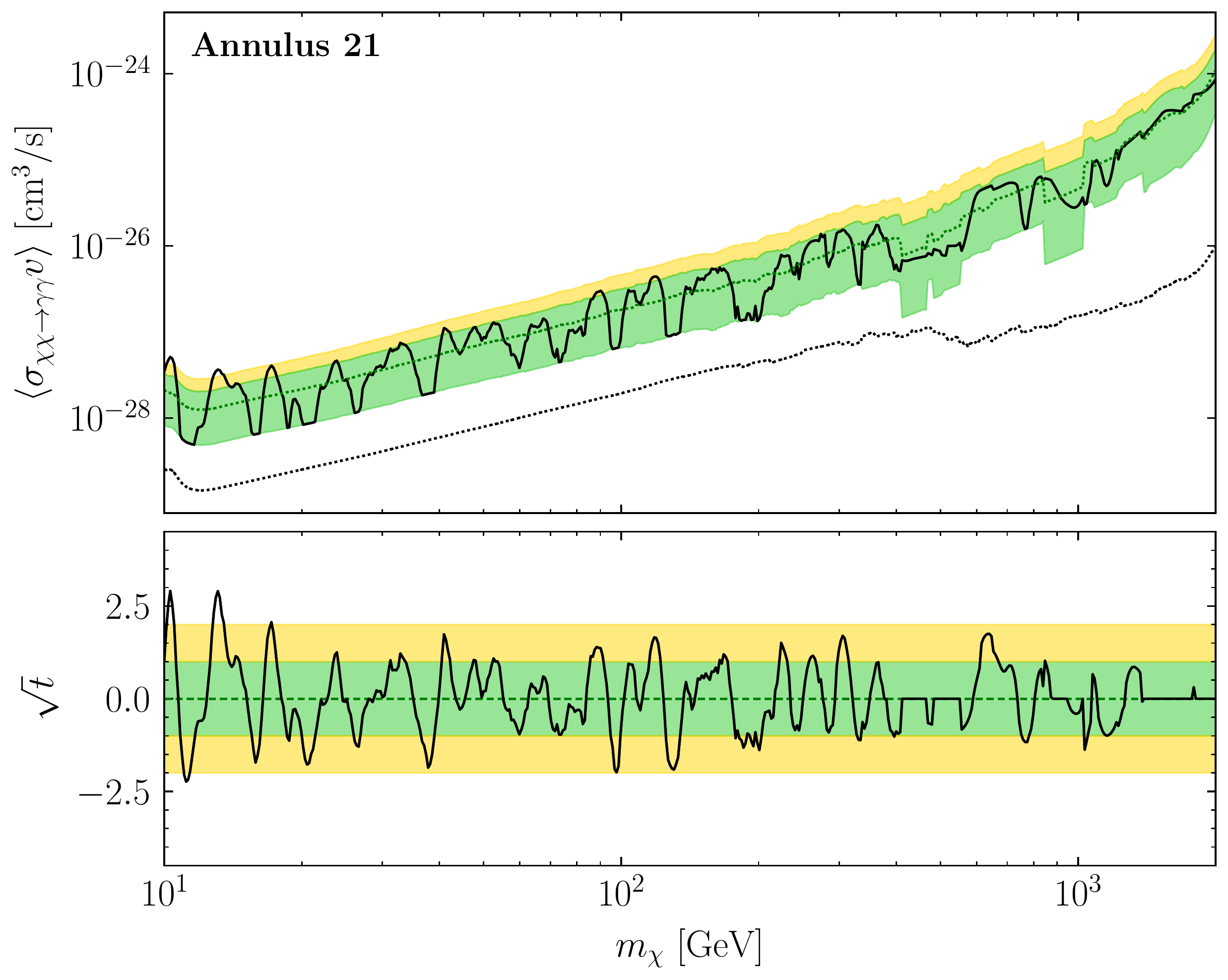}
		\includegraphics[width=0.49\textwidth]{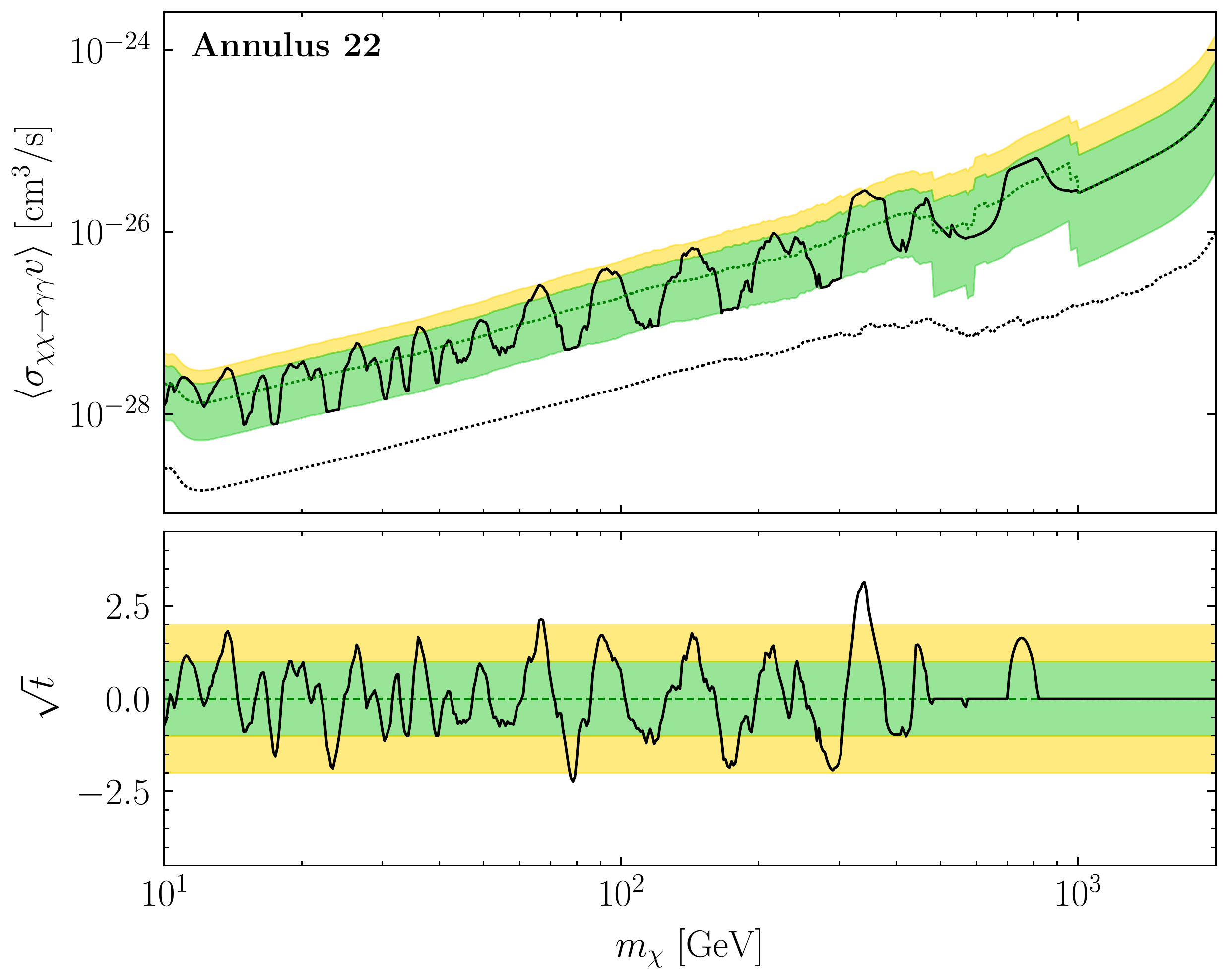}
	\end{center}
	\caption{As in the left panel of Fig.~\ref{fig:Results_NFW_Limits} but for Annulus 21 and Annulus 22.}
	\label{fig:Annuli2122}
\end{figure*}

\begin{figure*}[!htb]
	\begin{center}
		\includegraphics[width=0.49\textwidth]{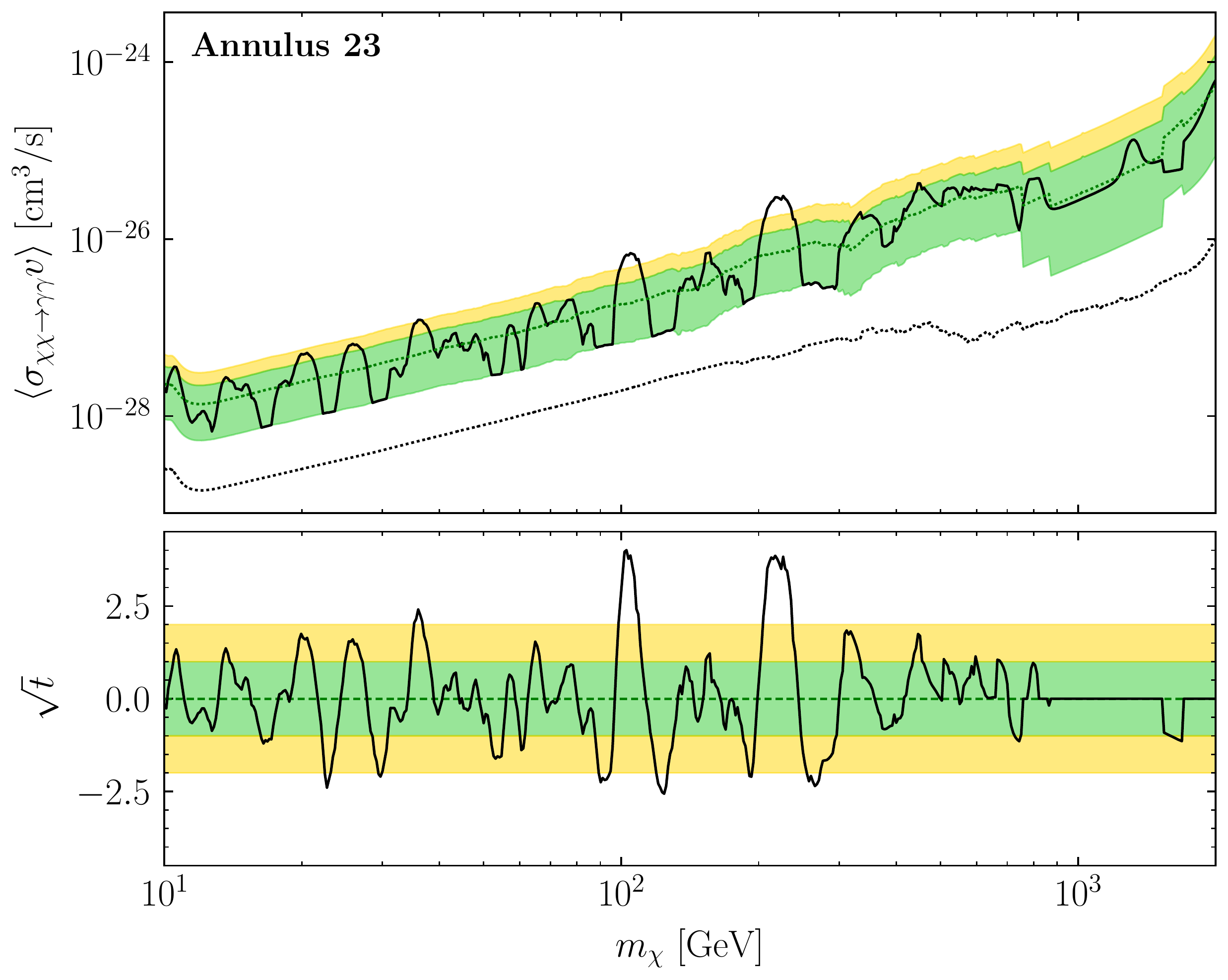}
		\includegraphics[width=0.49\textwidth]{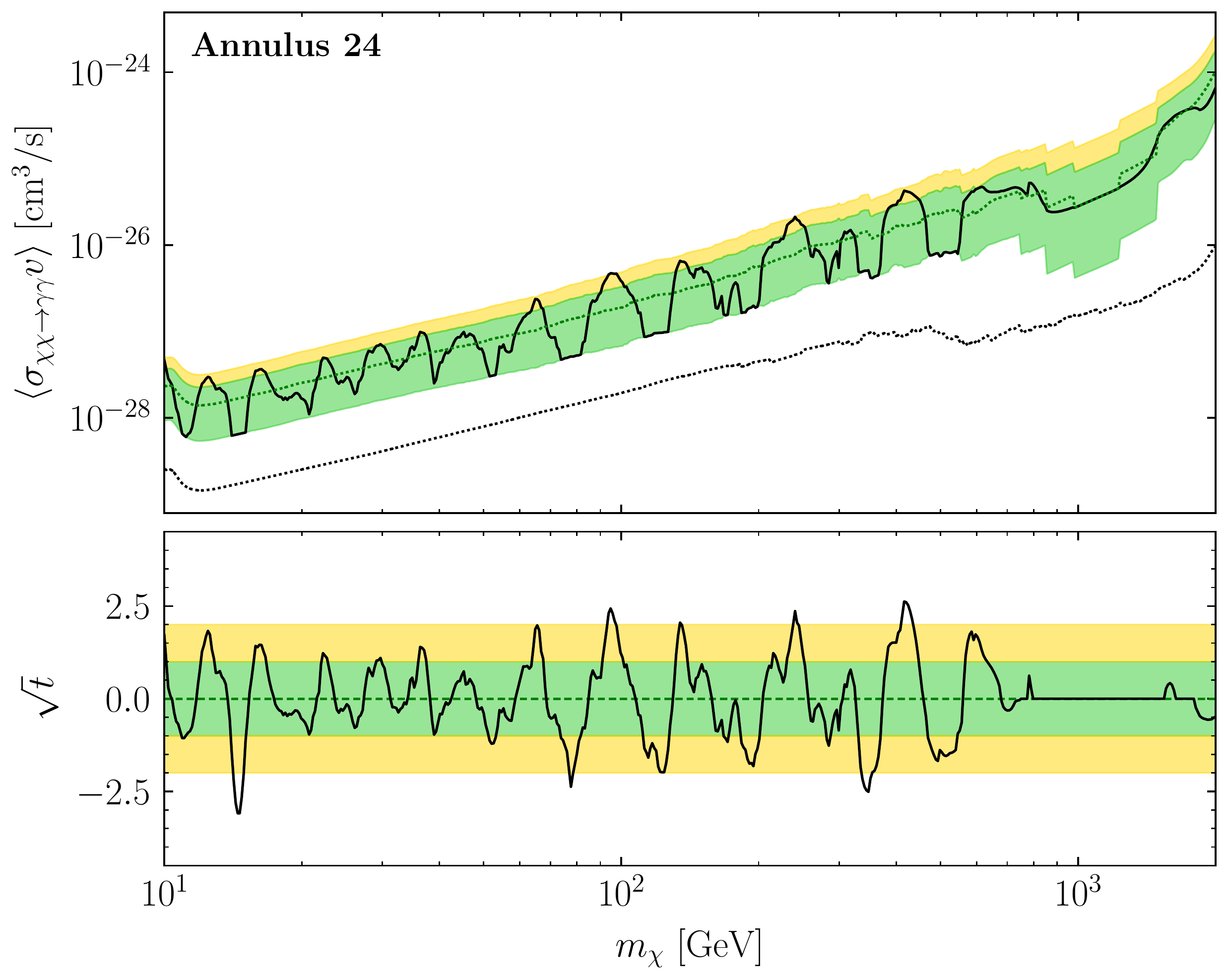}
	\end{center}
	\caption{As in the left panel of Fig.~\ref{fig:Results_NFW_Limits} but for Annulus 23 and Annulus 24.}
	\label{fig:Annuli2324}
\end{figure*}

\begin{figure*}[!htb]
	\begin{center}
		\includegraphics[width=0.49\textwidth]{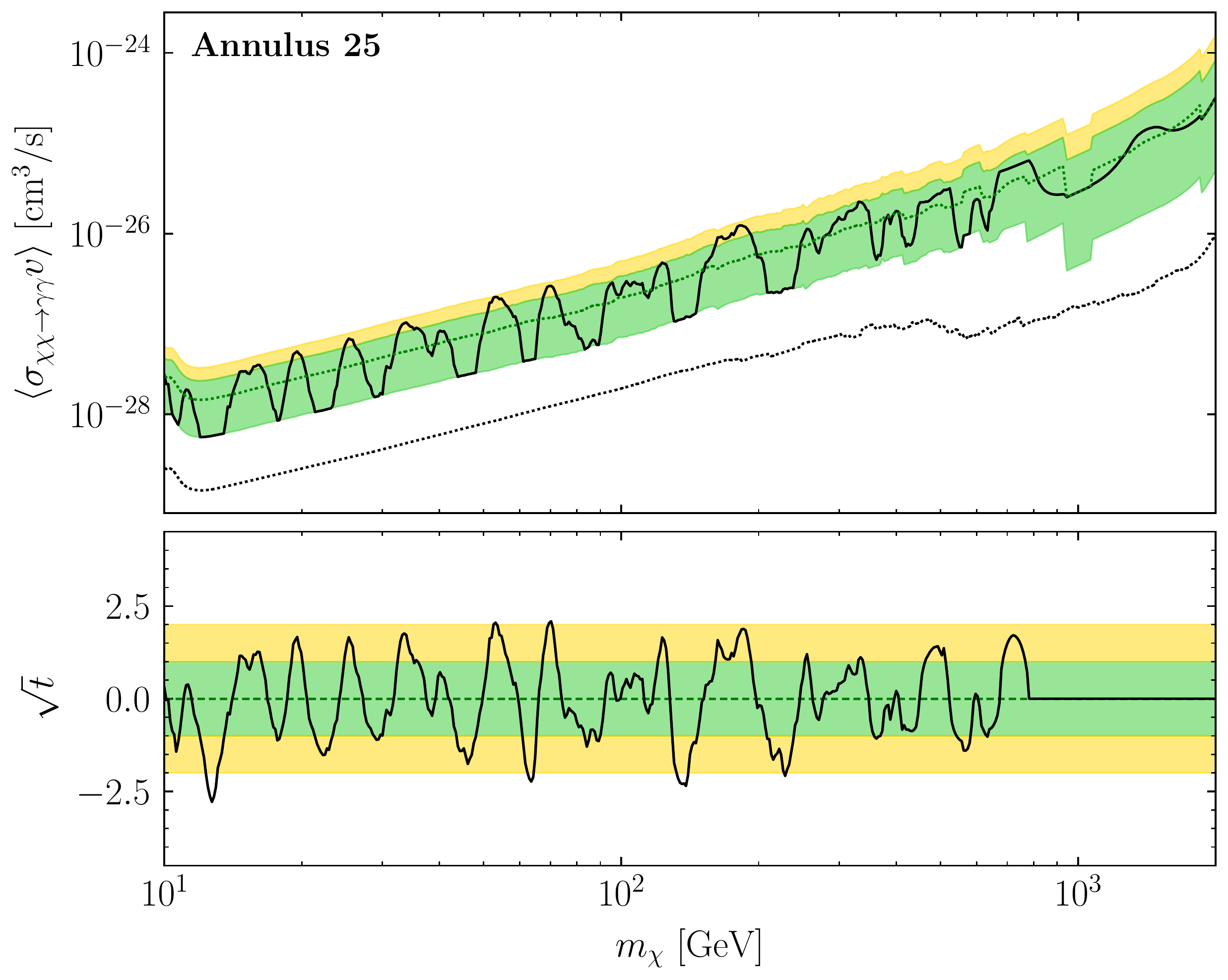}
		\includegraphics[width=0.49\textwidth]{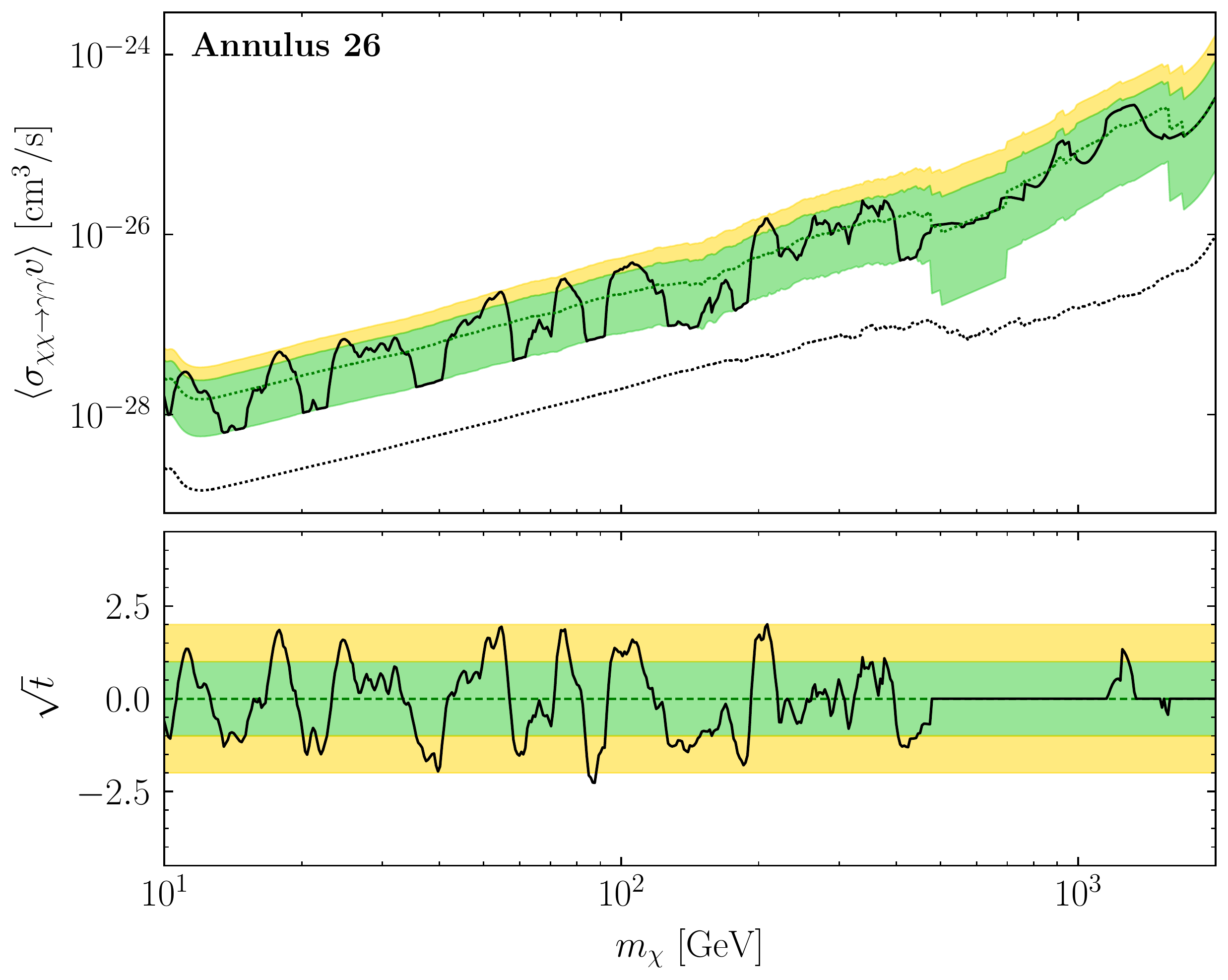}
	\end{center}
	\caption{As in the left panel of Fig.~\ref{fig:Results_NFW_Limits} but for Annulus 25 and Annulus 26.}
	\label{fig:Annuli2526}
\end{figure*}

\begin{figure*}[!htb]
	\begin{center}
		\includegraphics[width=0.49\textwidth]{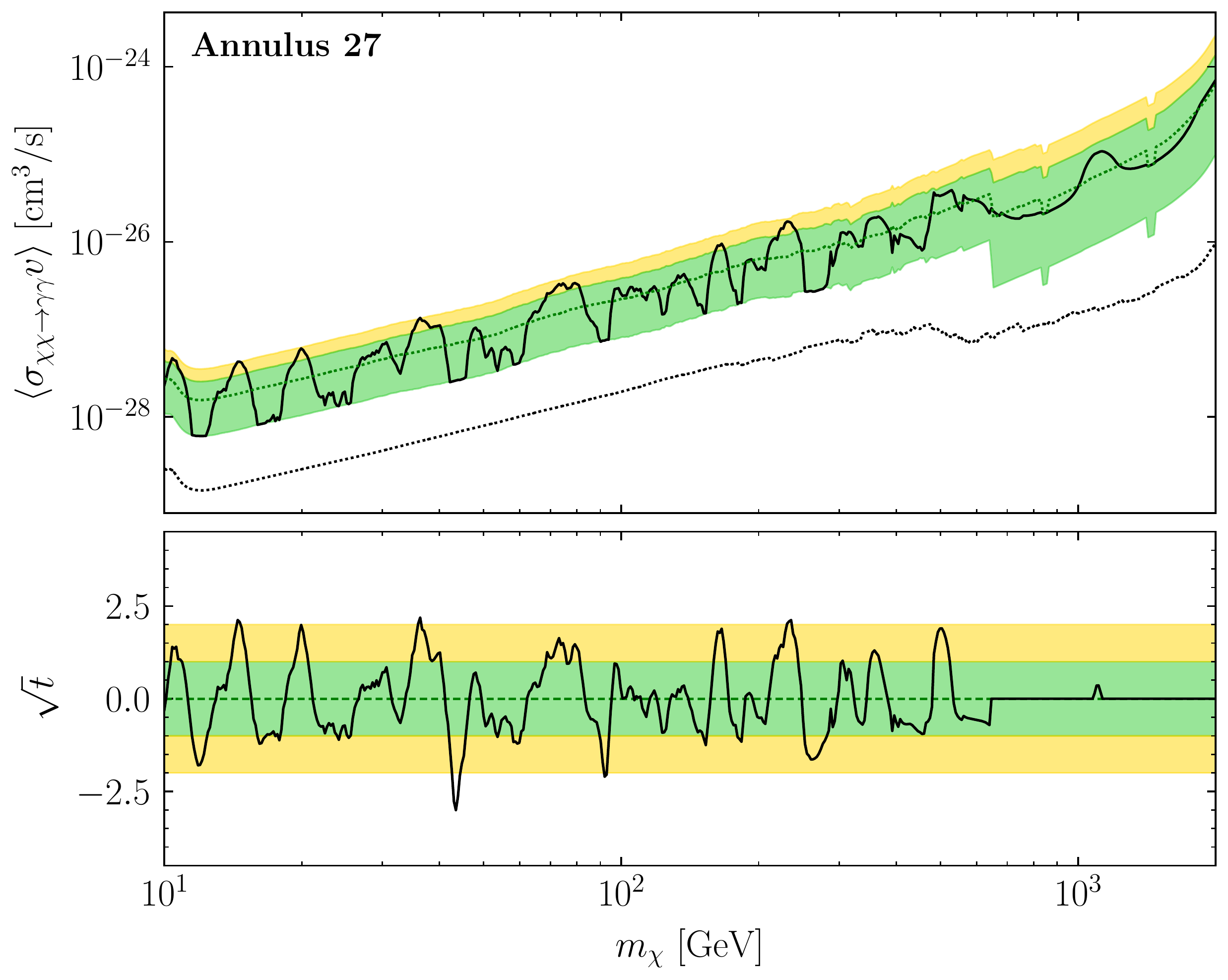}
		\includegraphics[width=0.49\textwidth]{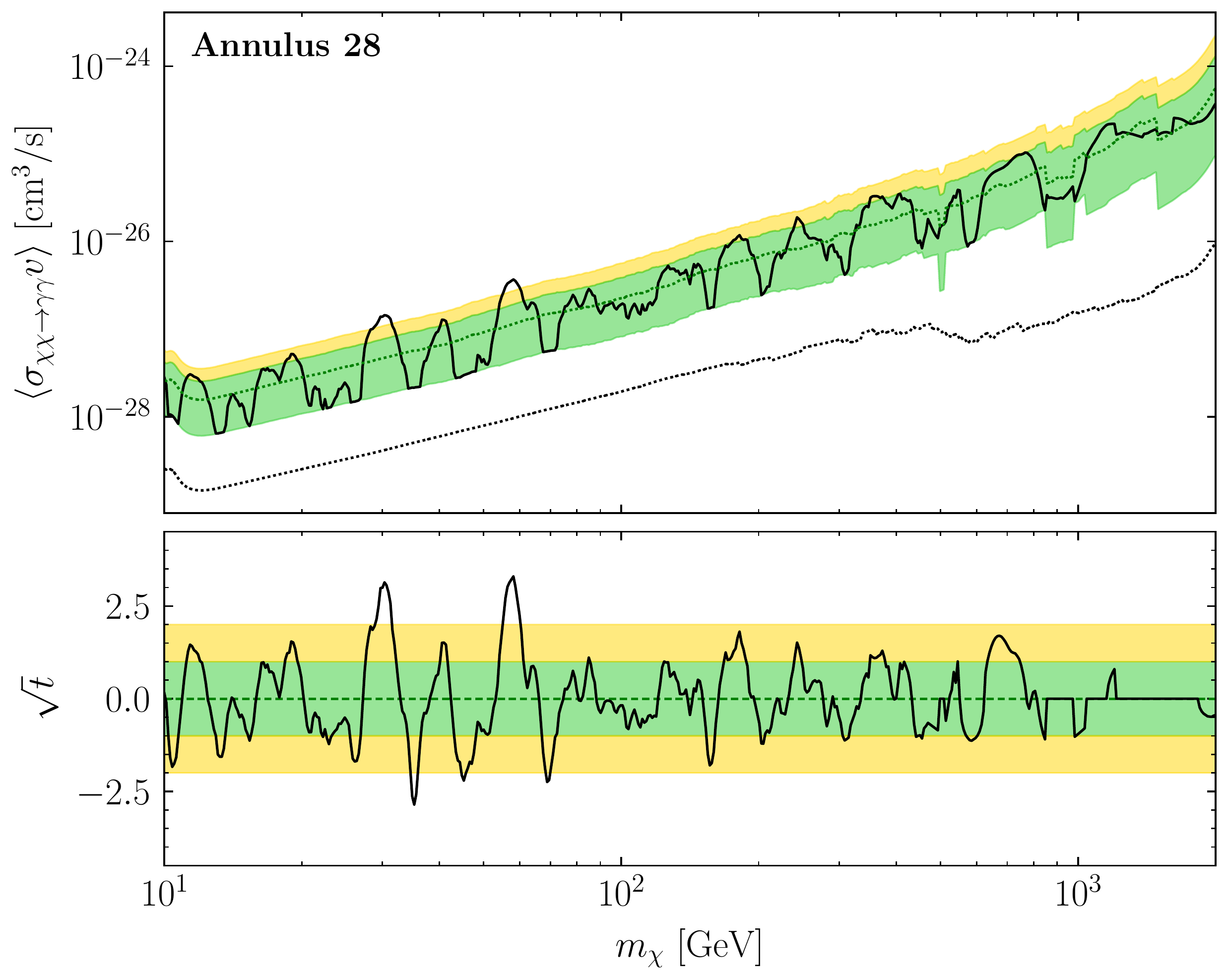}
	\end{center}
	\caption{As in the left panel of Fig.~\ref{fig:Results_NFW_Limits} but for Annulus 27 and Annulus 28.}
	\label{fig:Annuli12728}
\end{figure*}

\begin{figure*}[!htb]
	\begin{center}
		\includegraphics[width=0.49\textwidth]{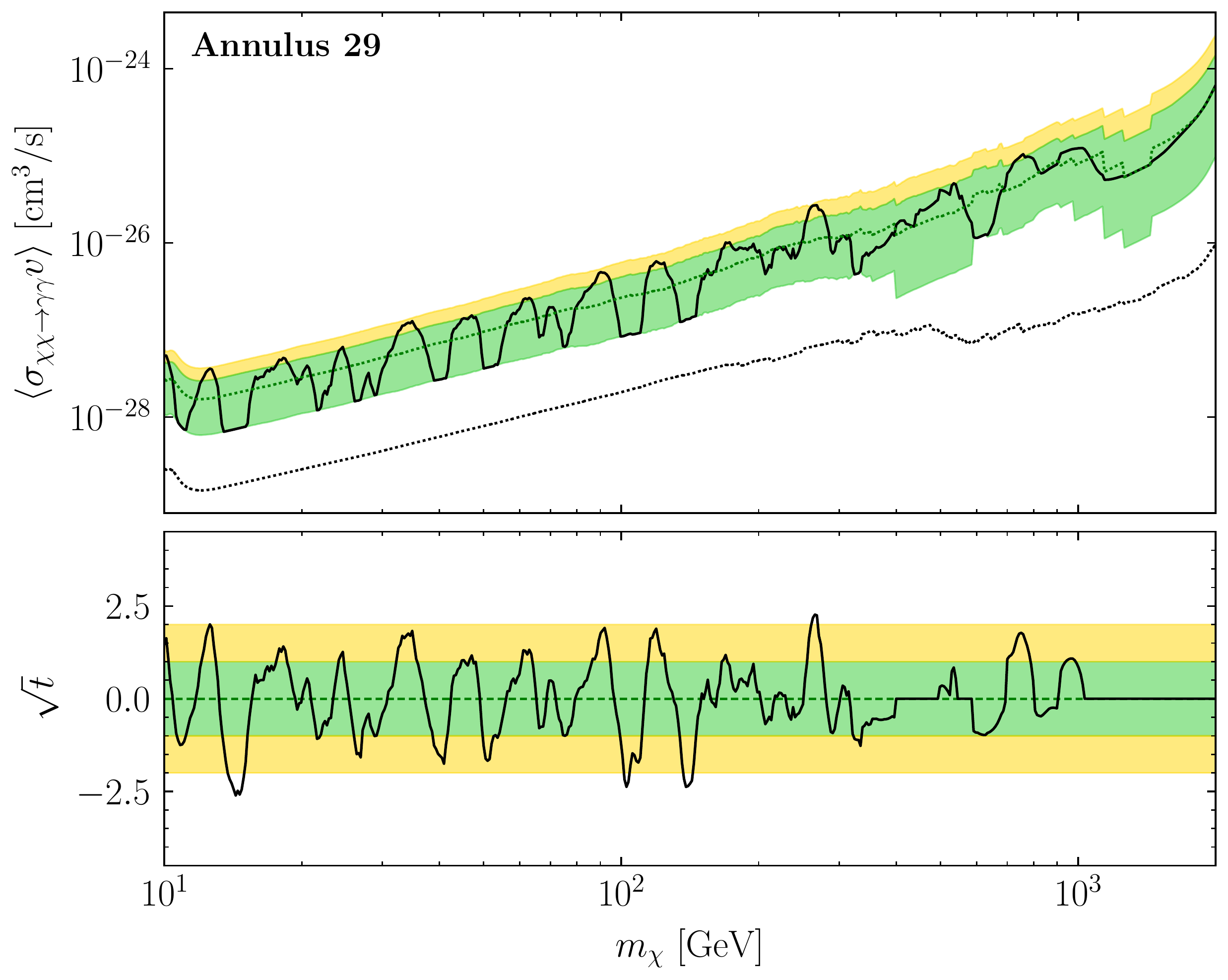}
		\includegraphics[width=0.49\textwidth]{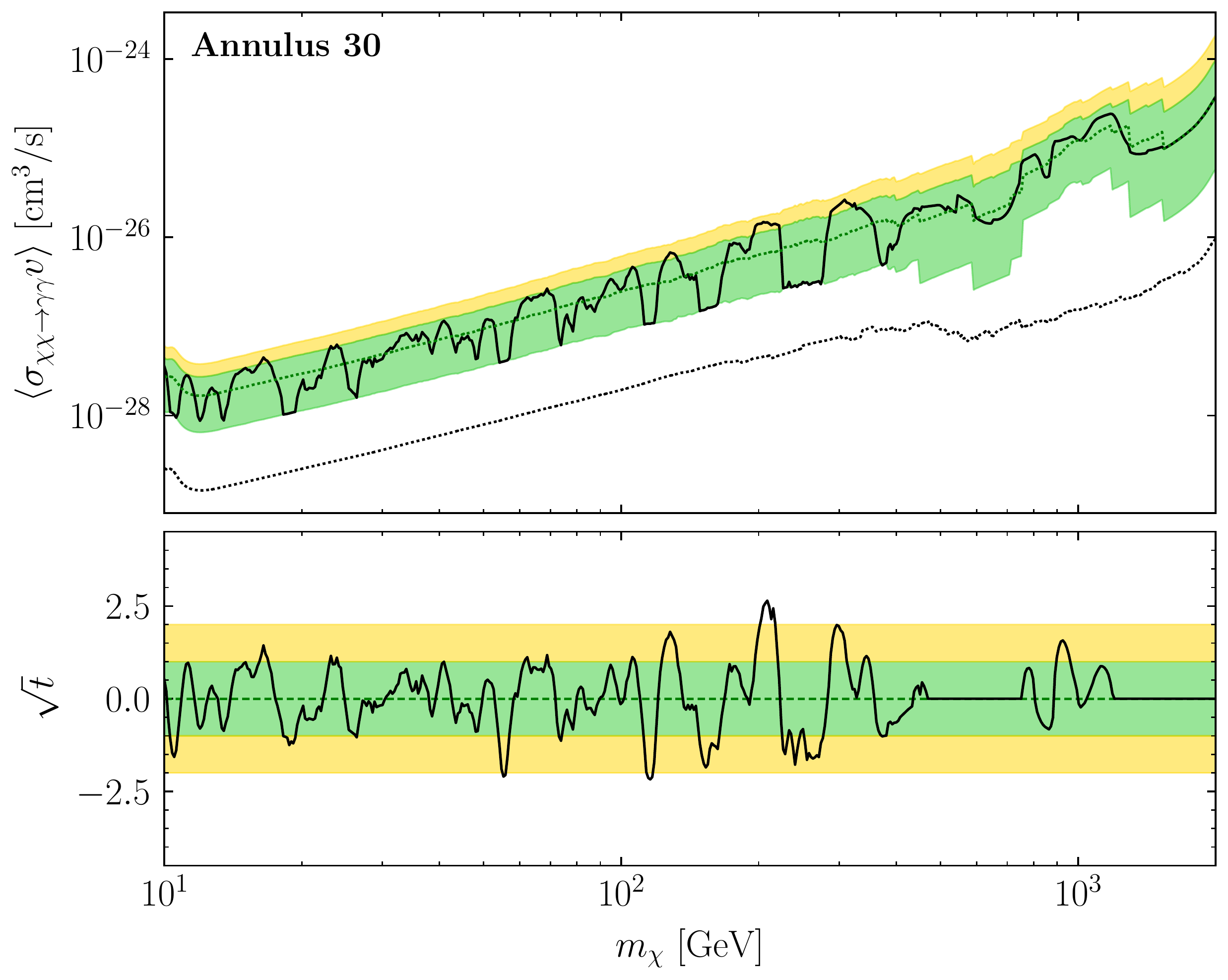}
	\end{center}
	\caption{As in the left panel of Fig.~\ref{fig:Results_NFW_Limits} but for Annulus 29 and Annulus 30.}
	\label{fig:Annuli2930}
\end{figure*}

In this section, we provide the limits and associated detection significances corresponding to a joint analysis over EDISP quartiles for each annuli. We present the results for the NFW annihilation analysis.  The individual annuli results are illustrated in Figs.~\ref{fig:Annuli12}, ~\ref{fig:Annuli34},~\ref{fig:Annuli56},~\ref{fig:Annuli78},~\ref{fig:Annuli910},~\ref{fig:Annuli1112},~\ref{fig:Annuli1314},~\ref{fig:Annuli1516},~\ref{fig:Annuli1718},~\ref{fig:Annuli1920},~\ref{fig:Annuli2122},~\ref{fig:Annuli2324},~\ref{fig:Annuli2526},~\ref{fig:Annuli12728},~\ref{fig:Annuli2930}.
We indicate the expected sensitivity of the full, joint  analysis with a black dotted line.

\section{Independent EDISP Results}
\begin{figure*}[!htb]
	\begin{center}
		\includegraphics[width=0.32\textwidth]{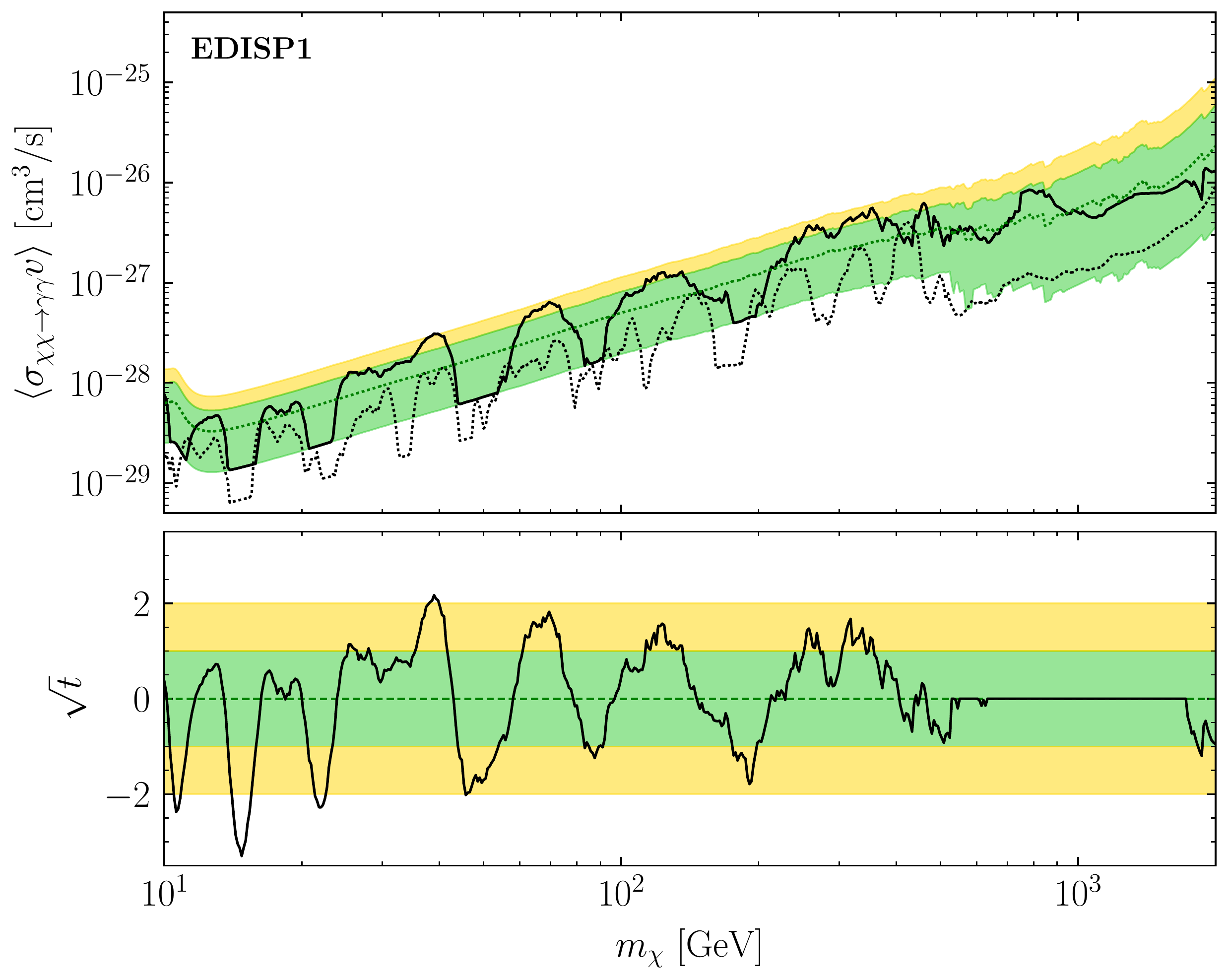}
		\includegraphics[width=0.32\textwidth]{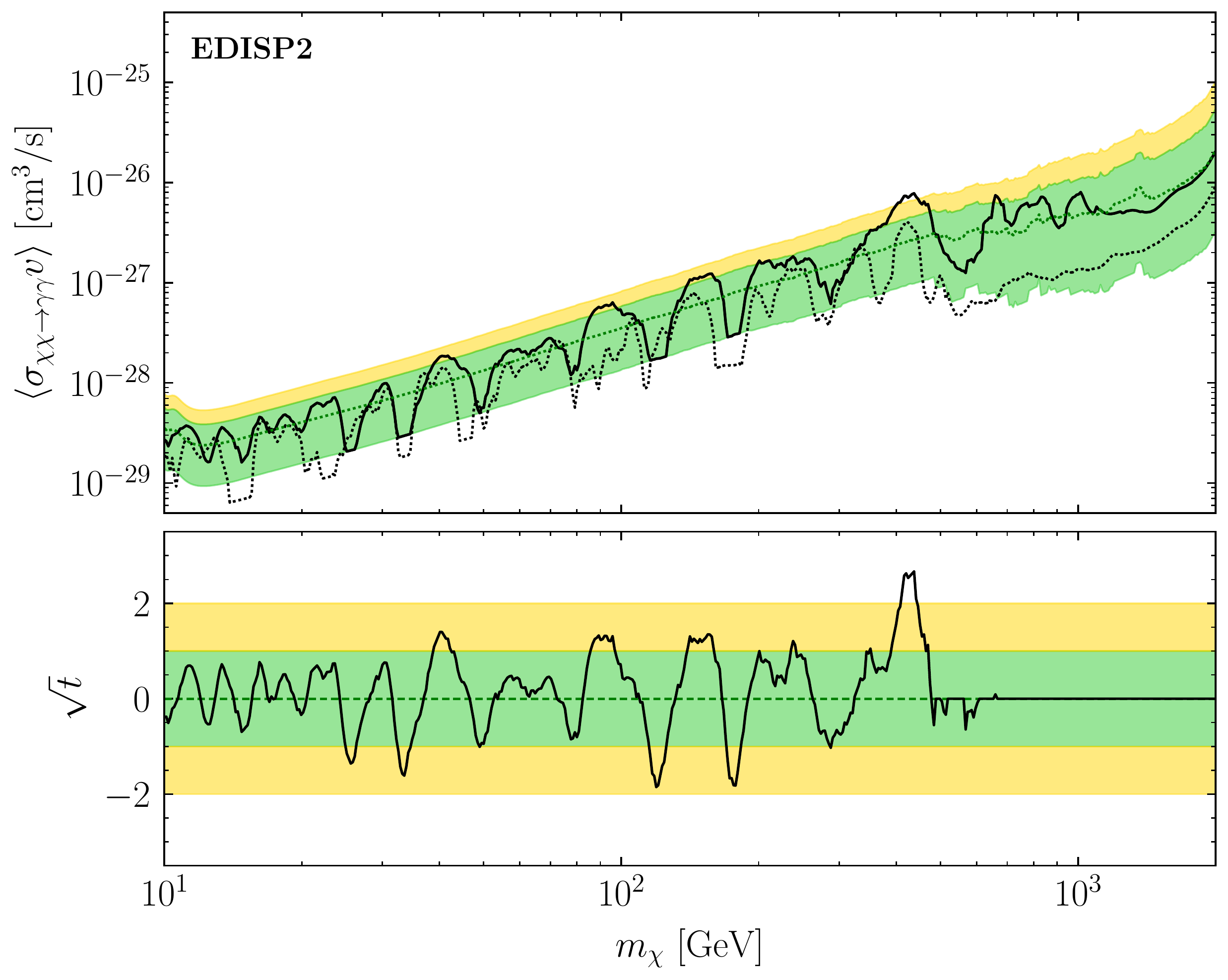}
		\includegraphics[width=0.32\textwidth]{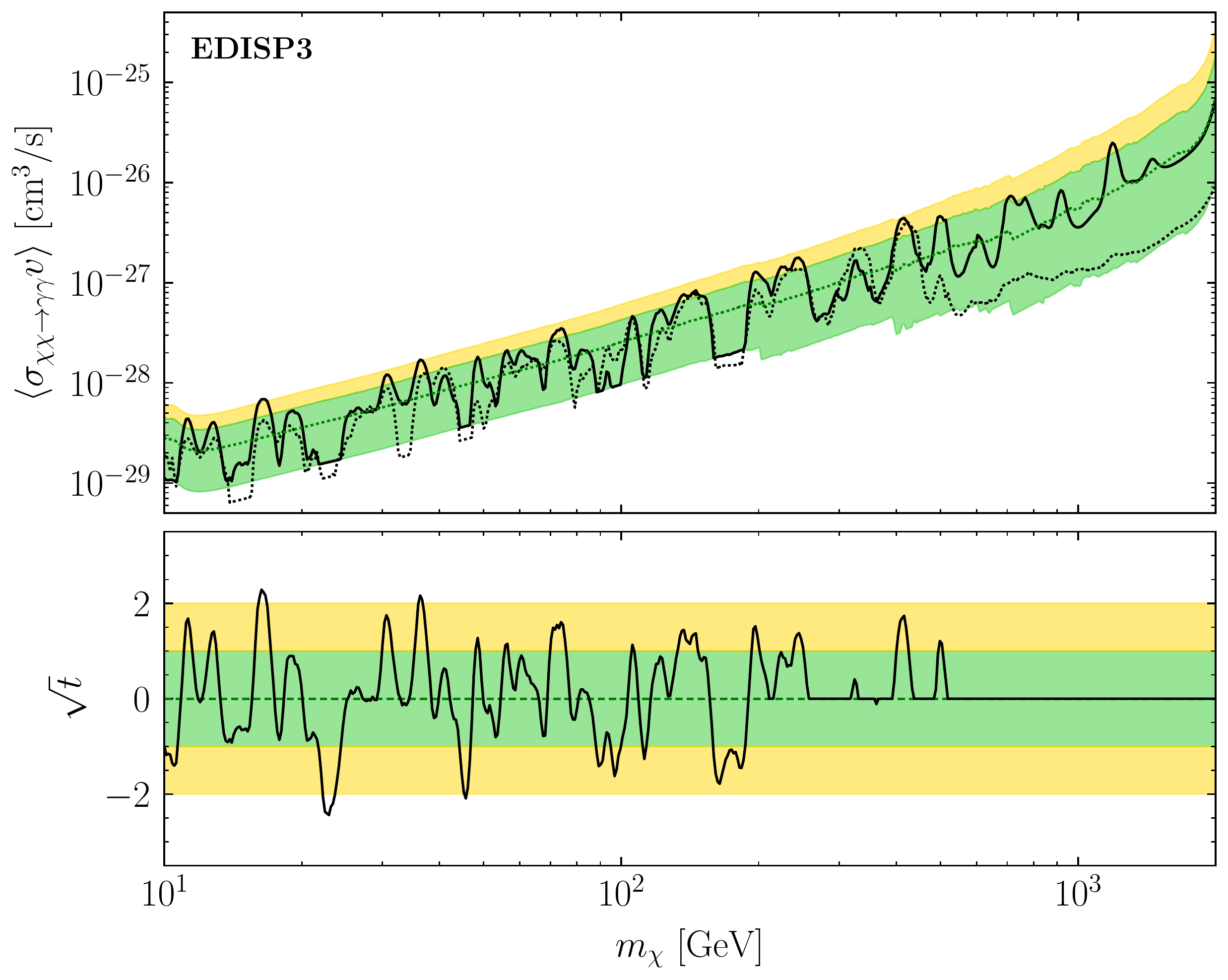}
	\end{center}
	\caption{As in the left panel of Fig.~\ref{fig:Results_NFW_Limits}, but for each EDISP quartile considered independently. We indicate the limits associated with the joint analysis over all three quartiles with a black dotted line.}
	\label{fig:EDISP_Results}
\end{figure*}

In this Appendix, we consider the analysis of each of the energy dispersion quartiles independently, with results presented in Fig.~\ref{fig:EDISP_Results}. As might be expected, the strongest limits are generally achieved by EDISP3, the top quartile of data by energy resolution, followed by EDISP2 and then EDISP1.

\section{Signal Injection Tests}
\label{app:injection_test}
\begin{figure*}[!htb]
	\begin{center}
		\includegraphics[width=0.9\textwidth]{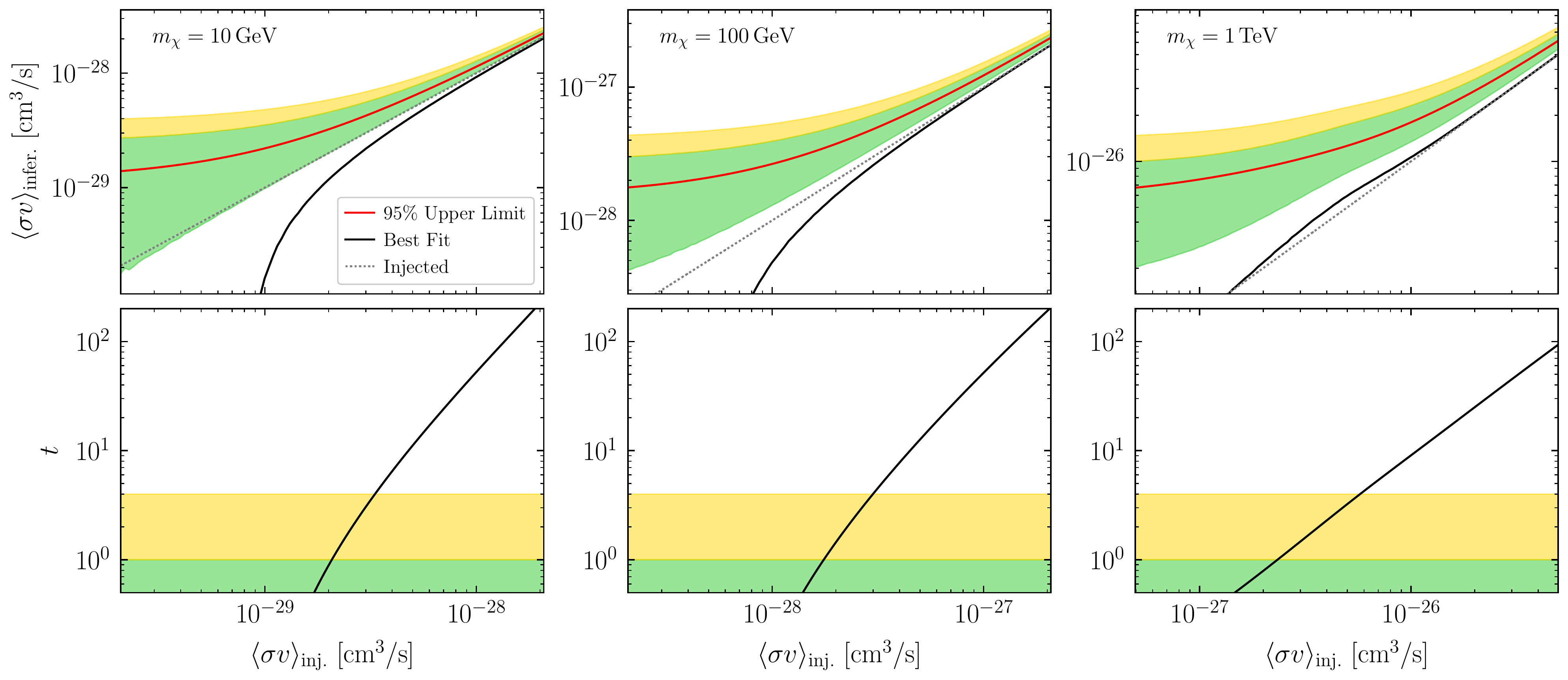}
	\end{center}
	\caption{The results testing an annihilation signal injection for three different candidate DM masses. For more details, see the text.}
	\label{fig:Annihilation_Injection}
\end{figure*}

\begin{figure*}[!htb]
	\begin{center}
		\includegraphics[width=0.9\textwidth]{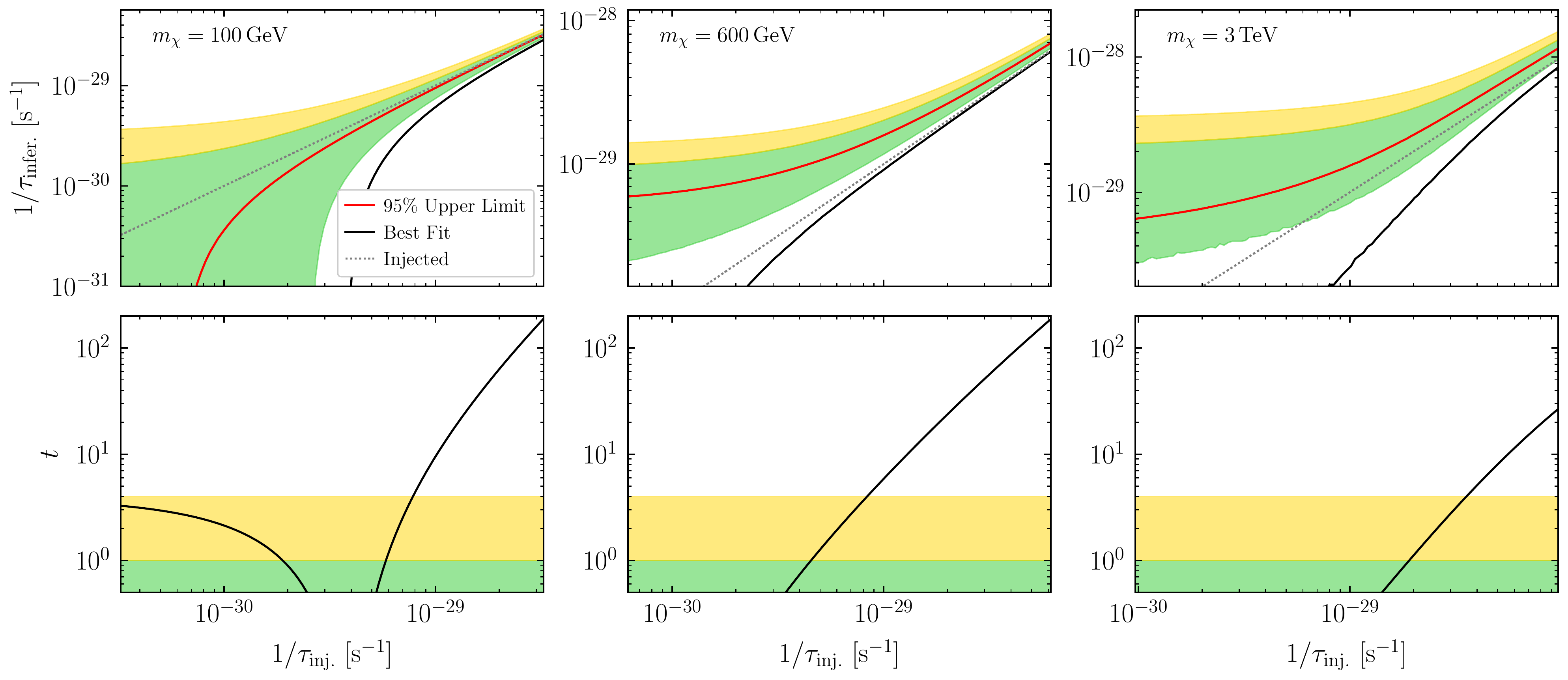}
	\end{center}
	\caption{As in Fig.~\ref{fig:Annihilation_Injection}, but for injected decay signals.}
	\label{fig:Decay_Injection}
\end{figure*}

In this Appendix, we perform signal injection tests under our fiducial analysis scheme. For three masses in the NFW annihilation search (Fig.~\ref{fig:Annihilation_Injection}) and three masses in the NFW decay search (Fig.~\ref{fig:Decay_Injection}), we inject the expected signal at varying signal strengths atop real data, then apply our analysis procedure to the synthetic data.

In the top panel, we compare injected signal strength, which sets the number of photons added on top of the real data, with the maximum likelihood estimate and the associated 95$^\mathrm{th}$ percentile upper limit for the signal strength parameter. Green and yellow bands indicate the $1\sigma$ and $2\sigma$ containment intervals for our upper limit. These figures demonstrate that our limit-setting procedure is accurately estimating the strength of the injected signal strength and placing a limit with appropriate coverage to within statistical uncertainties. 

In the bottom panel, we provide the value of the discovery TS $t$ as a function of injected signal strength, which we compare to the $1\sigma$ and $2\sigma$ thresholds for local significance indicated by the green and yellow bands. In all six cases, sufficiently bright signals result in excesses that produce large TSs, supporting that our analysis is capable of detecting any high-sigificance line-like excesses that may have been (but evidently are not) present in the data.

\clearpage

\bibliography{refs}
\end{document}